\documentclass[12pt]{article}
\usepackage[utf8]{inputenc}
\usepackage[T1]{fontenc}

\usepackage{graphicx}
\graphicspath{ {./plot/} }

\usepackage{authblk}

\usepackage{caption}

\usepackage{float}
\usepackage{subcaption}
\usepackage{amsmath, amssymb, amsthm}
\usepackage{mathtools}
\usepackage{bbm}
\usepackage{bm}
\usepackage{mathrsfs}
\usepackage{tikz}
\usepackage{pgfplots}
\usepgfplotslibrary{dateplot}
\usepackage{yhmath}
\usepackage{multirow} 

\usepackage{hyperref}
\hypersetup{hidelinks}
\hypersetup{
colorlinks=true,
linkcolor=blue,
citecolor=blue
}
\usepackage{xr}

\usepackage{enumerate, enumitem}
\usepackage{fancyhdr, graphicx, proof, comment, multicol}
\usepackage[none]{hyphenat} 

\usepackage{microtype} 
\usepackage{mdframed} 

\usepackage[round]{natbib}
\setcitestyle{authoryear,open={(},close={)}}

\newtheorem{proposition}{Proposition}

\newtheorem{definition}{Definition}

\begin{document}
\def\spacingset#1{\renewcommand{\baselinestretch}%
{#1}\small\normalsize} \spacingset{1}






\title{\bf Flexible Bayesian Modeling for Longitudinal Binary and Ordinal Responses}
\author{Jizhou Kang and Athanasios Kottas\thanks{Jizhou Kang (jkang37@ucsc.edu)
is Ph.D. student, and Athanasios Kottas (thanos@soe.ucsc.edu) is Professor, Department 
of Statistics, University of California, Santa Cruz, CA, USA.} \\
University of California, Santa Cruz, CA, USA\\
}
%
\maketitle

\bigskip
\begin{abstract}


Longitudinal studies with binary or ordinal responses are widely encountered in 
various disciplines, where the primary focus is on the temporal evolution of the 
probability of each response category. Traditional approaches build from  
the generalized mixed effects modeling framework. Even amplified with nonparametric 
priors placed on the fixed or random effects, such models are restrictive due to the 
implied assumptions on the marginal expectation and covariance structure of the responses. 
We tackle the problem from a functional data analysis perspective, treating the 
observations for each subject as realizations from subject-specific stochastic processes 
at the measured times. 
We develop the methodology focusing initially on binary responses, for which we assume 
the stochastic processes have Binomial marginal distributions. Leveraging the logits 
representation, we model the discrete space processes through 
continuous 
space processes. We utilize a hierarchical framework to model the mean and covariance 
kernel of the continuous space processes nonparametrically and simultaneously through 
a Gaussian process prior and an Inverse-Wishart process prior, respectively. The prior 
structure results in flexible inference for the evolution and correlation of binary 
responses, while allowing for borrowing of strength across all subjects. The modeling 
approach can be naturally extended to ordinal responses. Here, the continuation-ratio 
logits factorization of the multinomial distribution is key for efficient modeling 
and inference, including a practical way of dealing with unbalanced longitudinal data. 
The methodology is illustrated with synthetic data examples and an analysis of 
college students' mental health status data.


\end{abstract}

\noindent%
{\it Keywords:} Bayesian hierarchical modeling; 
Continuation-ratio logits; Functional data analysis; Markov chain Monte Carlo; 
Student-t process.

\newpage
\spacingset{1.7} 

\section{Introduction}
\label{sec:intro}

Recent years have witnessed a rapid growth of longitudinal studies with binary and 
ordinal responses in several disciplines, including econometrics, and the health 
and social sciences. In such studies, of primary importance are the probability 
response curves, i.e., the probabilities of the response categories that evolve
dynamically over time.
This article aims at developing a hierarchical framework, 
customized to longitudinal settings, that allows flexible inference for the probability 
response curves. In addition, the defining characteristic of longitudinal data is that 
repeated measurements on the same subject induce dependence. Hence, a further objective 
is to 
flexibly model lead-lag correlations among repeated measurements.

The development of statistical methods for longitudinal binary and ordinal data stems from 
models for longitudinal continuous responses, postulating the generalized linear model framework. 
Analogous to the continuous case, a specific model is formulated under one of three broad
approaches pertaining to marginal models, conditional models, or subject-specific models. 
Marginal models provide alternative modeling options when likelihood-based approaches are 
difficult to implement.
A conditional model describes the distribution of responses conditional on the covariates 
and also on part of the other components of the responses. In a subject-specific model, 
the effects of a subset of covariates are allowed to vary randomly from one individual 
to another. In the absence of predictor variables, functions of the observation time are 
usually used as covariates. We refer to \citet{Molenberghs2006} for a comprehensive review. 
In Section \ref{subsec:literaturereview}, we elaborate on the connection of our 
proposed modeling approach with existing methods.

In this article, we introduce a novel viewpoint for longitudinal binary and ordinal data 
analysis. 
We begin with the model construction for longitudinal binary responses. The critical insight 
that distinguishes our methodology from the majority of the existing literature is functional 
data analysis. We treat the subjects' measurements as stochastic process realizations at the 
corresponding time points. The benefits are twofold. First, the models can incorporate 
unbalanced data from longitudinal studies in a unified scheme; directly inferring the 
stochastic process provides a well-defined probabilistic model for the missing values. 
Secondly, we can exploit the power of Bayesian hierarchical modeling for continuous 
functional data \citep[e.g.,][]{Yang2016}. To that end, we adopt the Binomial distribution 
with the logit link that connects binary responses to continuous signals, which, subject to 
additive measurement error, are then modeled as (conditionally) independent and 
identically distributed (i.i.d.) realizations from a Gaussian process (GP) with 
random mean and covariance function. We place an Inverse-Wishart process (IWP) prior 
on the covariance function, and conditional on it, use a GP prior for the mean function. 
Therefore, the two essential ingredients in longitudinal modeling, the trend and the 
covariance structure, are modeled simultaneously and nonparametrically.

The hierarchical structure allows borrowing of strength across the subjects' trajectories. 
We apply a specific setting of hyperpriors for the GP and IWP priors, such that marginalizing 
over them, the latent continuous functions have a Student-t process (TP) prior. The 
TP enhances the flexibility of the GP \citep[e.g.,][]{Shah2014}. It retains attractive GP 
properties, such as analytic marginal and predictive distributions, and it yields 
predictive covariance that, unlike the GP, explicitly depends on the observed values. 
For inferential purposes, we represent the joint posterior distribution in multivariate form 
through evaluating the functions on the pooled grid, resulting in the common 
normal-inverse-Wishart conditional conjugacy. In conjunction with the Pólya-Gamma data 
augmentation technique \citep{Polson2013}, we develop a relatively simple and effective posterior 
simulation algorithm, circumventing the need for specialized techniques or tuning of 
Metropolis-Hastings steps.

To extend the model for ordinal responses, we utilize the continuation-ratio logits 
representation of the multinomial distribution. Such representation features an 
encoding of an ordinal response with $C$ categories as a sequence of $C-1$ binary 
indicators, in which the $j$-th indicator signifies whether the ordinal response belongs 
to the $j$-th category or to one of the higher categories. 
We show that fitting a multinomial model for the ordinal responses is equivalent 
to fitting separately the aforementioned model on the binary indicators. 
Hence, we can conduct posterior simulation for each response category in a parallel fashion, 
leading to significant computational efficiency gains in model implementation.

In modern longitudinal studies, it is common that the complete vector of repeated measurements 
is not collected on all subjects. As a specific example, in ecological momentary assessment 
(EMA) studies, emotions and behaviors are repeatedly measured for a cohort of participants, 
through wearable electronic devices \citep{EMABook2018}. For instance, in the \textit{StudentLife} 
study \citep{StudentLife2014}, researchers monitored the students' mental status through 
pop-up questionnaires on their smartphones that prompted multiple times at pseudorandom 
intervals during the study period. Since the data collection process is based on the 
participants' conscious responding to prompted questions several times a day, non-response 
is inevitable. Missing values are typically considered to be a nuisance rather than a 
characteristic of EMA time series. Parametric and nonparametric Bayesian methods have been 
developed to handle longitudinal data with missingness; see \citet{Daniels2020} for a review. 
The common issue is that one has to bear the drawbacks of making either structured or 
unstructured assumptions to manage missingness. The unstructured approach leads to flexibility, 
yet it may result in difficulties due to a large number of parameters relative to the sample 
size. Besides, the majority of the existing literature on longitudinal studies with missingness 
focuses on the scenario with continuous responses, and the extension to discrete responses is not trivial.

Accordingly, our contributions can be summarized as follows: (i) we model the mean and 
covariance jointly and nonparametrically, avoiding potential biases caused by a pre-specified 
model structure; (ii) we unify the toolbox for balanced and unbalanced longitudinal studies; 
(iii) the model encourages borrowing of strength, preserving systematic patterns that are 
common across all subject responses; (iv) we develop a computationally efficient posterior 
simulation method by taking advantage of conditional conjugacy; (v) the model facilitates 
applications for ordinal responses with a moderate to large number of categories. 

The rest of the paper is organized as follows. 
Section \ref{sec:binarymodel} develops the methodology for binary responses, including
model formulation, study of model properties, and the computational approach to 
inference and prediction.
Section \ref{sec:realapp} illustrates the modeling approach through an EMA study that 
focuses on analyzing students' mental health through binary outcomes. The modeling extension 
for longitudinal ordinal responses is presented in Section \ref{sec:polyordinalmodel}, 
including an illustration involving an ordinal outcome from the same EMA study. Finally, 
Section \ref{sec:summary} concludes with discussion.

%
\section{The modeling approach for binary responses}
\label{sec:binarymodel}

Here, we develop the methodology for longitudinal binary responses.
The data consist of repeated binary responses on $n$ subjects, 
with the observation on subject $i$ at time $\tau_{it}$ denoted by $Y_{it}$. The set of 
repeated outcomes for the $i$-th subject is collected into a $T_i$-dimensional vector 
$\mathbf{Y}_i=$ $(Y_{i1},\ldots,Y_{iT_i})^\top$. The hierarchical model construction 
is presented in Section \ref{subsec:standardmodel}. In Section \ref{subsec:modprop}, 
we discuss model properties related to our inference objectives. Bayesian inference and 
prediction is developed in Section \ref{subsec:modelapply}. In Section \ref{subsec:simstudy},
we outline the findings from simulation studies, the details of which are included in 
the Supplementary Material. Finally, to place our contribution within the literature, we 
discuss in Section \ref{subsec:literaturereview} the proposed model in the context of 
relevant Bayesian nonparametric approaches.

Regarding notation under our functional data analysis modeling approach, 
we use the regular letter and its bold form to distinguish between the trajectory of 
responses over time and its evaluation on a number of time points. We use similar notation 
for other functional variables, possibly including the time input(s) inside parentheses
or as a subscript. Moreover, $\tau$ and $\boldsymbol{\tau}$ denote the generic time input 
and a grid of times, respectively.

\subsection{Model specification}
\label{subsec:standardmodel}

We examine the data from a functional data analysis perspective, treating each observed 
data vector $\mathbf{Y}_i$ as the evaluation of trajectory $Y_i(\tau)$ on grid 
$\boldsymbol{\tau}_i=$ $(\tau_{i1},\ldots,\tau_{iT_i})^{\top}$, for $i=1,\ldots,n$. 
The $n$ trajectories are assumed to be (conditionally) independent realizations from 
a continuous-time stochastic process. The prior probability model is built on the stochastic 
process. This approach avoids strong pre-determined assumptions on the transition 
mechanism within the sequence of subject-specific responses in $\mathbf{Y}_i$, while it 
is suitable to accommodate repeated measurements regardless of their observational pattern.  

%
%

The functional data analysis view of longitudinal data dates back at least to 
\citet{Zhao2004}, 
where it is suggested that functional data analysis tools, 
such as principal component analysis, can be used to capture 
periodic structure in longitudinal data. Indeed, \citet{Yao2005} study functional 
principal component analysis (FPCA) for sparse longitudinal data, a method that can 
provide effective recovery of the entire individual trajectories from fragmental data. 
FPCA has been applied in finance \citep{Ingrassia2005}, biomechanics \citep{Dona2009}, 
and demographic studies \citep{Shamshoian2020}. Its extension to examine sequences 
of discrete data is studied in \citet{Hall2008}.

Our methodology builds from a GP-based hierarchical model for continuous functional 
data \citep{Yang2016}. Regarding mean-covariance estimation, the model in \cite{Yang2016} 
can be considered as a Bayesian counterpart of \citet{Yao2005}. 
The hierarchical scheme enables a natural extension to studies with binary responses. 
We assume that, subject to measurement error, the $i$-th subject's responses, 
$Y_{it} \equiv$ $Y_i(\tau_{it})$, depend on the $i$-th trajectory of the underlying 
process, evaluated at times $\tau_{it}$, through the following model
\begin{equation*}
Y_i(\tau_{it}) \mid Z_i(\tau_{it}),\epsilon_{it} \, \stackrel{ind.}{\sim} \,
Bin(1, \varphi(Z_i(\tau_{it})+\epsilon_{it})),\quad t=1,\ldots,T_i,\quad i=1,\ldots,n,
\label{eq:bernmodel}
\end{equation*}
where $\varphi(x) = $ $\exp(x)/\{ 1 + \exp(x) \}$ denotes the expit function. 
The error terms are i.i.d. from a white noise process, that is, 
$\epsilon_{it} \mid \sigma^2_{\epsilon} \stackrel{i.i.d.}{\sim} N(0,\sigma^2_{\epsilon})$, 
and independent of the process realizations $Z_i(\cdot)$. The main building block 
for the model construction is a hierarchical GP prior for the $Z_i(\cdot)$.
In particular, given random mean function $\mu(\cdot)$ and covariance kernel 
$\Sigma(\cdot,\cdot)$, the $Z_i(\cdot)$ are i.i.d. GP realizations, denoted by 
$Z_i \mid \mu,\Sigma \stackrel{i.i.d.}{\sim} GP(\mu,\Sigma)$, for $i=1,\ldots,n$.
The hierarchical GP prior model is completed with nonparametric priors for the 
mean function and covariance kernel:
\begin{equation}
\mu \mid \Sigma \sim GP(\mu_0,\Sigma/\kappa), \quad 
\Sigma\sim IWP(\nu,\Psi_{\boldsymbol{\phi}}) ,
\label{eq:gpiwpprior}
\end{equation}
where $GP(\cdot,\cdot)$ and $IWP(\cdot,\cdot)$ denote the GP and IWP prior, 
respectively. The nonparametric prior reflects the intuition that parametric forms 
will generally not be sufficiently flexible for the mean and covariance functions.

We adopt an IWP prior for the covariance kernel, defined such that, on any finite 
grid $\boldsymbol{\tau}=(\tau_1,\ldots,\tau_T)$ with $|\boldsymbol{\tau}|$ points, 
the projection $\Sigma(\boldsymbol{\tau},\boldsymbol{\tau})$ follows an inverse-Wishart 
distribution with mean 
$\Psi_{\boldsymbol{\phi}}(\boldsymbol{\tau},\boldsymbol{\tau}) / (\nu-2)$, denoted by 
$IW(\nu,\Psi_{\boldsymbol{\phi}}(\boldsymbol{\tau},\boldsymbol{\tau}))$.
Here, $\Psi_{\boldsymbol{\phi}}(\cdot,\cdot)$ is a non-negative definite function 
with parameters $\boldsymbol{\phi}$. Note that we use the parameterization from 
\citet{Dawid1981} for the inverse-Wishart distribution, in particular, $\nu$ is the 
shape parameter and $\nu+|\boldsymbol{\tau}|-1$ is the degrees of freedom parameter in the
more common parameterization. \citet{Yang2016} validate that this parameterization defines 
an infinite dimensional probability measure whose finite dimensional projection on grid 
$\boldsymbol{\tau}$ coincides with the inverse-Wishart distribution 
$IW(\nu,\Psi_{\boldsymbol{\phi}}(\boldsymbol{\tau},\boldsymbol{\tau}))$.

The model formulation is completed with prior specification for the hyperparameters. 
The error variance is assigned an inverse gamma prior, 
$\sigma_{\epsilon}^2\sim IG(a_{\epsilon},b_{\epsilon})$. We focus primarily on 
stationary specifications under the prior structure in (\ref{eq:gpiwpprior}). 
In particular, we work with mean function, $\mu_0(\tau) \equiv \mu_0$, and isotropic 
covariance function, $\Psi_{\boldsymbol{\phi}}$, within the Matérn class, a widely 
used class of covariance functions \citep{Rasmussen2006}. In general, 
the Matérn covariance function is specified by a scale parameter $\sigma^2$, a range 
parameter $\rho$, and a smoothness parameter $\iota$. 
To encourage 
smoothness in the probability response curves, we set 
$\iota = 5/2$, such that the covariance kernel is given by
\begin{equation*}
\Psi_{\boldsymbol{\phi}}(\tau,\tau^{\prime}) \, = \, 
\sigma^2 \, \left( 1+\frac{\sqrt{5}|\tau-\tau^{\prime}|}{\rho}+
\frac{5|\tau-\tau^{\prime}|^2}{3\rho^2} \right) \,
\exp\left( -\frac{\sqrt{5}|\tau-\tau^{\prime}|}{\rho} \right),
    \label{eq:matern52covfun}
\end{equation*}
where $\boldsymbol{\phi}=\{\sigma^2,\rho\}$. 
For hyperparameters $\mu_0$, $\sigma^2$, $\rho$, we take the commonly used choice, 
\begin{equation*}
\mu_0\sim N(a_{\mu},b_{\mu}),\quad \sigma^2\sim \text{Gamma}(a_{\sigma},b_{\sigma}),\quad 
\rho\sim Unif(a_{\rho},b_{\rho}).
    \label{eq:hyperprior}
\end{equation*}
Finally, we set $\kappa = (\nu-3)^{-1}$, such that the continuous-time process for 
the $Z_i(\cdot)$ is a TP when $\mu$ and $\Sigma$ are marginalized out 
(see Section \ref{subsec:modprop} for details). As a consequence, parameter $\nu$ controls 
the tail heaviness of the marginal process, with smaller values of $\nu$ corresponding to 
heavier tails. We place a uniform prior on $\nu$, $\nu\sim Unif(a_{\nu},b_{\nu})$, 
with $a_{\nu}>3$ to ensure positive definiteness of $\Sigma/\kappa$.

As discussed in \citet{Diggle1988}, the correlation of repeated measurements on the same 
subject commonly has the following patterns. First, it should decrease with respect to 
the measurements' separation in time, while remaining positive to indicate the measurements 
are from the same subject. This feature is encapsulated by the form of the covariance 
kernel $\Psi_{\boldsymbol{\phi}}$. The IWP prior elicits realizations for which this property 
holds a priori, while enabling a flexible estimate of the covariance structure with information 
from the data a posteriori. Second, measurements that are made arbitrarily close in time are 
subject to imperfect correlation, possibly caused by subsampling of each subject. This feature 
is represented by the error term in our model. Moreover, the motivation for adding the error 
term arises from the fact that measurement error is introduced in the estimation of a 
continuous-time function based on data collected at discrete time points.

In addition to the aforementioned methodological considerations, adding the error term is 
practically important. Effectively, the error term serves as a nugget to the covariance matrix. 
It can alleviate numerical problems that may arise from its inversion, a calculation required
in the posterior inference procedure. Moreover, adding the error term is common practice in 
other areas involving GP-based models, including spatial statistics \citep[e.g.,][]{Carmack2012} 
and computer model emulation \citep[e.g.,][]{Andrianakis2012}.

Although the probability model is formulated through stochastic process realizations, 
posterior simulation is based on the corresponding finite dimensional distributions (f.d.d.s.). 
Consequently, to write the model for the data, we need to represent the likelihood and prior 
in multivariate forms through evaluating the functions on finite grids. 
Denoting $Y_i(\boldsymbol{\tau}_i)$ by $\mathbf{Y}_i$, $Z_i(\boldsymbol{\tau}_i)$ by 
$\mathbf{Z}_i$, and $\boldsymbol{\epsilon}_i=$
$(\epsilon_{i1},\ldots,\epsilon_{iT_i})^{\top}$, the 
model for the data can be written as 
\begin{equation}
\begin{split}
&\mathbf{Y}_i\mid \mathbf{Z}_i,\boldsymbol{\epsilon}_i \, \stackrel{ind.}{\sim} \,
\prod^{T_i}_{t=1}Bin(1, \varphi(Z_{it}+\epsilon_{it})),\quad i=1,\ldots,n,\\ 
&\mathbf{Z}_i\mid \mu(\boldsymbol{\tau}_i),\Sigma(\boldsymbol{\tau}_i,\boldsymbol{\tau}_i) \,
\stackrel{ind.}{\sim} \, N(\mu(\boldsymbol{\tau}_i),\Sigma(\boldsymbol{\tau}_i,\boldsymbol{\tau}_i)),
\quad \boldsymbol{\epsilon}_i\mid \sigma_{\epsilon}^2
\, \stackrel{ind.}{\sim} \, N(\mathbf{0},\sigma_{\epsilon}^2 \, \mathbf{I}).
\end{split}
\label{eq:finiterepbinmodel}
\end{equation}
Notice that the grids $\{\boldsymbol{\tau}_i:i=1,\ldots,n\}$ are not necessarily the same
for all subjects. Therefore, the shared GP and IWP prior in (\ref{eq:gpiwpprior}) need to 
be evaluated on the pooled grid $\boldsymbol{\tau}=\cup_{i=1}^n\boldsymbol{\tau}_i$. 
If $\boldsymbol{\mu}$, $\boldsymbol{\Sigma}$, and $\boldsymbol{\Psi}_{\boldsymbol{\phi}}$ 
denote $\mu(\boldsymbol{\tau})$, $\Sigma(\boldsymbol{\tau},\boldsymbol{\tau})$, and $\Psi_{\boldsymbol{\phi}}(\boldsymbol{\tau},\boldsymbol{\tau})$, respectively, then
\begin{equation}
\boldsymbol{\mu} \mid \boldsymbol{\Sigma},\mu_{0}, \nu 
\, \sim \, N(\mu_0\mathbf{1}, (\nu-3) \boldsymbol{\Sigma}),\quad 
\boldsymbol{\Sigma} \mid \nu, \boldsymbol{\phi} 
\, \sim \, IW(\nu,\boldsymbol{\Psi}_{\boldsymbol{\phi}}).
\label{eq:multigpiwppool}
\end{equation}
The hierarchical model formulation for the data in (\ref{eq:finiterepbinmodel}) 
and (\ref{eq:multigpiwppool}) forms the basis for the posterior simulation algorithm, 
which is discussed in detail in Section \ref{subsec:modelapply}.

\subsection{Model properties}
\label{subsec:modprop}

To fix ideas for the following discussion, we refer to $Z_i(\tau)$ as the signal process of 
the binary process $Y_i(\tau)$, and to $\mathcal{Z}_i(\tau)=Z_i(\tau)+\epsilon_i(\tau)$ as 
the latent process of $Y_i(\tau)$. Since the stochastic process is characterized by its 
f.d.d.s., we shall investigate the random vectors $\mathbf{Y}_{\boldsymbol{\tau}}=$
$Y_i(\boldsymbol{\tau})$, $\boldsymbol{\mathcal{Z}}_{\boldsymbol{\tau}}=$
$\mathcal{Z}_{i}(\boldsymbol{\tau})$, and $\mathbf{Z}_{\boldsymbol{\tau}}=$
$Z_{i}(\boldsymbol{\tau})$, for a generic grid vector $\boldsymbol{\tau}=$
$(\tau_1,\ldots,\tau_T)^{\top}$. We surpass the subject index $i$ because the subject 
trajectories are identically distributed. The Supplementary Material includes 
proofs for the propositions included in this section.

Among the various inference goals in a study that involves longitudinal binary data, 
estimating the probability response curve and the covariance structure of the repeated 
measurements are the most important ones.  In Proposition \ref{prop:meancovcondsignal}, 
we derive the probability response curves and covariance matrix of the binary 
vector $\mathbf{Y}_{\boldsymbol{\tau}}$, conditional on the signal vector
$\mathbf{Z}_{\boldsymbol{\tau}}$ and error variance $\sigma^2_{\epsilon}$. 
The probability response curve can be defined generically 
as $\mathbf{P}_{\mathbf{y}\boldsymbol{\tau}} =$
$(\text{Pr}(Y_{\tau_1}=y_{\tau_1}\mid \mathbf{Z}_{\boldsymbol{\tau}},\sigma_{\epsilon}^2),
\ldots,\text{Pr}(Y_{\tau_T}=y_{\tau_T}\mid \mathbf{Z}_{\boldsymbol{\tau}},
\sigma_{\epsilon}^2) )^{\top}$, where $y_{\tau_t}$ is either 0 or 1. 
Without loss of generality, we focus on $\mathbf{P}_{\mathbf{1}\boldsymbol{\tau}}$.


\begin{proposition}
\label{prop:meancovcondsignal}
The probability response curve is given by
$\mathbf{P}_{\mathbf{1}\boldsymbol{\tau}} =$
$E(\boldsymbol{\pi}(\boldsymbol{\mathcal{Z}}_{\boldsymbol{\tau}})
\mid \mathbf{Z}_{\boldsymbol{\tau}},\sigma_{\epsilon}^2)$, 
where $\boldsymbol{\pi}(\mathbf{x})$ denotes the vector operator that applies 
the expit function to every entry of $\mathbf{x}$. Regarding the covariance matrix, 
for $\tau\in\boldsymbol{\tau}$, 
$Var(Y_{\tau}\mid \mathbf{Z}_{\boldsymbol{\tau}},\sigma_{\epsilon}^2) =$
$E(\varphi(\mathcal{Z}_{\tau})\mid \mathbf{Z}_{\boldsymbol{\tau}},\sigma_{\epsilon}^2) - E^2(\varphi(\mathcal{Z}_{\tau})\mid \mathbf{Z}_{\boldsymbol{\tau}},\sigma_{\epsilon}^2)$, 
and for $\tau, \tau^{\prime}\in\boldsymbol{\tau}$, with $\tau^{\prime} \neq \tau$, $Cov(Y_{\tau},Y_{\tau^{\prime}}\mid\mathbf{Z}_{\boldsymbol{\tau}},\sigma_{\epsilon}^2)=$
$Cov(\varphi(\mathcal{Z}_{\tau}),\varphi(\mathcal{Z}_{\tau^{\prime}})\mid \mathbf{Z}_{\boldsymbol{\tau}},\sigma_{\epsilon}^2)$. 
The conditional expectations in all of the above expressions are with respect to distribution,
$\boldsymbol{\mathcal{Z}}_{\boldsymbol{\tau}}\mid \mathbf{Z}_{\boldsymbol{\tau}},
\sigma_{\epsilon}^2 \sim N(\mathbf{Z}_{\boldsymbol{\tau}},\sigma_{\epsilon}^2 \, \mathbf{I})$.
\end{proposition}


The practical utility of Proposition \ref{prop:meancovcondsignal} lies on performing 
posterior inference for the probability response curve and the covariance structure of 
the binary process, conditioning on the signal process and the noise. With posterior 
samples of $\mathbf{Z}_{\boldsymbol{\tau}}$ and $\sigma^2_{\epsilon}$, we can 
simulate $\boldsymbol{\mathcal{Z}}_{\boldsymbol{\tau}}$ 
from $N(\mathbf{Z}_{\boldsymbol{\tau}},\sigma^2_{\epsilon}\mathbf{I})$ and numerically 
compute the corresponding moments in Proposition \ref{prop:meancovcondsignal}. 
The entries of $\boldsymbol{\mathcal{Z}}_{\boldsymbol{\tau}}$ are independent, given 
$\mathbf{Z}_{\boldsymbol{\tau}}$, and thus simulating 
$\boldsymbol{\mathcal{Z}}_{\boldsymbol{\tau}}$ is not computationally demanding, 
even when $|\boldsymbol{\tau}|$ is large. 


We next establish a closer connection between the binary process and the signal process.  
Proposition \ref{prop:deltaapproxlogitnormal} reveals that the evolution of the binary process 
over time can be (approximately) expressed as a function of the expectation of the signal 
process and the total variance. Moreover, the covariance of the binary process is approximately 
the covariance of the signal process scaled by a factor related to the expectation of the signal.

\begin{proposition}
\label{prop:deltaapproxlogitnormal}
Consider the proposed model as described in (\ref{eq:finiterepbinmodel}) and 
denote $\mu(\boldsymbol{\tau})=\boldsymbol{\mu}$, 
and $\Sigma(\boldsymbol{\tau},\boldsymbol{\tau})=\boldsymbol{\Sigma}$. Then, 
\begin{align*}
&\text{Pr}(Y_{\tau}=1\mid \boldsymbol{\mu},\boldsymbol{\Sigma},\sigma_{\epsilon}^2)\approx \varphi(\text{E}(Z_{\tau}\mid \boldsymbol{\mu},\boldsymbol{\Sigma}))+\frac{\text{Var}(Z_{\tau}\mid \boldsymbol{\mu},\boldsymbol{\Sigma})+\sigma_{\epsilon}^2}{2}\varphi^{\prime\prime}(\text{E}(Z_{\tau}\mid \boldsymbol{\mu},\boldsymbol{\Sigma})),\,\, \forall \tau\in\boldsymbol{\tau}, \nonumber\\
&\text{Cov}(Y_{\tau},Y_{\tau^{\prime}}\mid\boldsymbol{\mu},\boldsymbol{\Sigma},\sigma_{\epsilon}^2)
\approx 
\varphi^{\prime}(\text{E}(Z_{\tau}\mid \boldsymbol{\mu},\boldsymbol{\Sigma}))\varphi^{\prime}(\text{E}(Z_{\tau^{\prime}}\mid \boldsymbol{\mu},\boldsymbol{\Sigma})) \, 
\text{Cov}(Z_{\tau},Z_{\tau^{\prime}}\mid\boldsymbol{\mu},\boldsymbol{\Sigma})\\
& -\frac{1}{4}[\text{Var}(Z_{\tau}\mid \boldsymbol{\mu},\boldsymbol{\Sigma})+\sigma_{\epsilon}^2][\text{Var}(Z_{\tau^{\prime}}\mid\boldsymbol{\mu},\boldsymbol{\Sigma})+\sigma_{\epsilon}^2]\varphi^{\prime\prime}(\text{E}(Z_{\tau}\mid \boldsymbol{\mu},\boldsymbol{\Sigma}))\varphi^{\prime\prime}(\text{E}(Z_{\tau^{\prime}}\mid \boldsymbol{\mu},\boldsymbol{\Sigma})),\,\, \forall \tau,\tau^{\prime}\in\boldsymbol{\tau}.
\end{align*}
Here, $\varphi^{\prime}(x)=\frac{d\varphi(x)}{dx}=\varphi(x)[1-\varphi(x)]$ 
and $\varphi^{\prime\prime}(x)=$ $\frac{d^2\varphi(x)}{dx^2}=$ 
$\varphi(x)[1-\varphi(x)][1-2\varphi(x)]$.
\end{proposition}

Our inference results are based on exact expressions, such as the ones in 
Proposition \ref{prop:meancovcondsignal}. Nonetheless, the approximate expressions 
derived in Proposition \ref{prop:deltaapproxlogitnormal} are practically useful to 
gain more insight on properties of the binary process, as well as for prior specification. 
Note that exploring properties of the binary process is not trivial due to the lack 
of general analytical forms for moments of logit-normal distributions. Hence, a 
connection with properties of the signal process is useful. For instance, if we 
specify the covariance for the signal process to decrease as a function of separation 
in time, 
an analogous structure will hold (approximately) for the binary process. 

The previous discussion focuses on studying the f.d.d.s of the binary process 
given the signal process. Therefore, it is important to investigate 
the marginal f.d.d.s of the signal process. We show that, under the specification 
$\kappa =$ $(\nu-3)^{-1}$, the f.d.d.s. of the signal process correspond to a 
multivariate Student-t (MVT) distribution, and thus the signal process is a TP. 
We first state the definition of the MVT distribution and the TP
\citep[see, e.g.,][]{Shah2014}. Notice that we use the covariance matrix as a 
parameter for the MVT distribution, instead of the more common parameterization 
based on a scale matrix.

\begin{definition}
\label{def:defmvtandtp}
The random vector $\mathbf{Z}\in\mathbb{R}^n$ is MVT distributed, denoted 
$\mathbf{Z}\sim MVT(\nu,\boldsymbol{\mu},\boldsymbol{\Psi})$, 
if it has density
\begin{equation*}
\frac{\Gamma(\frac{\nu+n}{2})}{[(\nu-2)\pi]^{n/2}
\Gamma(\frac{\nu}{2})}|\boldsymbol{\Psi}|^{-1/2}
\left(
1 + \frac{(\mathbf{Z}-\boldsymbol{\mu})^T\boldsymbol{\Psi}^{-1}
(\mathbf{Z}-\boldsymbol{\mu})}{\nu-2} \right)^{-\frac{\nu+n}{2}}
\label{eq:defmvt}
\end{equation*}
where $\nu > 2$ is the degrees of freedom parameter, $\boldsymbol{\mu}\in\mathbb{R}^n$,
and $\boldsymbol{\Psi}$ is an $n\times n$ symmetric, positive definite matrix.
Under this parameterization, $E(\mathbf{Z})=\boldsymbol{\mu}$ 
and $Cov(\mathbf{Z})=\boldsymbol{\Psi}$. 

Consider a process $Z(\tau)$ formulated through mean function $\mu(\tau)$, a 
non-negative kernel function $\Psi(\tau,\tau)$, and parameter $\nu>2$, such that its 
f.d.d.s correspond to the MVT distribution with mean vector and covariance matrix induced 
by $\mu(\tau)$ and $\Psi(\tau,\tau)$, respectively. Then, $Z(\tau)$ follows a TP, denoted 
by $Z(\tau)\sim TP(\nu,\mu(\tau),\Psi(\tau,\tau))$.
\end{definition}


Marginalizing over $\boldsymbol{\mu}$ and $\boldsymbol{\Sigma}$
in (\ref{eq:finiterepbinmodel}) and (\ref{eq:multigpiwppool}),
the implied distribution for $\mathbf{Z}_{\boldsymbol{\tau}}$ is MVT, with degrees of 
freedom parameter $\nu$ (with $\nu > 3$ in our context), mean vector $\mu_0 \mathbf{1}$,
and covariance matrix $\boldsymbol{\Psi}_{\boldsymbol{\phi}} =$
$\Psi_{\boldsymbol{\phi}}(\boldsymbol{\tau},\boldsymbol{\tau})$. We thus obtain
the following result for the signal process.

\begin{proposition}
\label{prop:marginalsignal}
Under the model formulation in (\ref{eq:finiterepbinmodel}) and (\ref{eq:multigpiwppool}),
the signal process follows marginally a TP, that is, 
$Z\sim TP(\nu,\mu_0,\Psi_{\boldsymbol{\phi}})$.
\end{proposition}



Proposition \ref{prop:marginalsignal} is beneficial 
in terms of both computation and interpretation. Without a constraint on $\kappa$, as 
in \citet{Yang2016}, the marginal distribution of $\mathbf{Z}_{\boldsymbol{\tau}}$ does 
not have analytical form. Hence, for prediction at new time points, one has to sample 
from an IWP and a GP, which is computationally intensive, especially for a dense grid. 
In contrast, we can utilize the analytical form of the TP predictive distribution to 
develop a predictive inference scheme that resembles that of GP-based models
(see Section \ref{subsec:modelapply}). 
Moreover, the result highlights the model property that the degrees of freedom 
parameter $\nu$ controls how heavy tailed the process is. Smaller values of $\nu$ 
correspond to heavier tails. As $\nu$ gets larger, the tails resemble Gaussian tails. 
Additionally, $\nu$ controls the dependence between $Z_{\tau}$ and $Z_{\tau^{\prime}}$, 
which are jointly MVT distributed, with smaller values indicating higher dependence. 
Such interpretation of parameter $\nu$ facilitates the choice of its hyperprior.

The local behavior of stochastic process realizations is crucial for interpolation. 
Under the longitudinal setting, continuous, or perhaps differentiable, signal process 
trajectories are typically anticipated. Evidently, the observed
data can not visually inform the smoothness of signal process realizations. Rather, 
such smoothness should be captured in the prior specification that incorporates 
information about the data generating mechanism. For weakly stationary processes, 
mean square continuity is equivalent to the covariance function being continuous at 
the origin \citep{Stein1999}. And, the process is $\iota$-times mean square differentiable 
if and only if the $2\iota$-times derivative of the covariance function at the origin 
exists and is finite. Under our model, the signal process follows a TP marginally. 
Its covariance structure is specified by the Matérn covariance function with smoothness 
parameter $\iota$. Referring to the behavior of the Matérn class of covariance functions 
at the origin, we obtain the following result for the mean square continuity and 
differentiability of the signal process.

\begin{proposition}
\label{prop:smoothsignal}
Consider the proposed model with marginal signal process 
$Z\sim TP(\nu,\mu_0,\Psi_{\boldsymbol{\phi}})$, where $\Psi_{\boldsymbol{\phi}}$ belongs to 
the Matérn family of covariance functions with smoothness parameter $\iota$. Then, 
the signal process is mean square continuous and $\lfloor\iota\rfloor$-times mean 
square differentiable.   
\end{proposition}


%
%

The results in this section study several properties that are useful in model
implementation. Indeed, the practical utility of such model properties with respect 
to prior specification and posterior inference is discussed in the next section.

\subsection{Prior specification and posterior inference}
\label{subsec:modelapply}


The model described in Section \ref{subsec:standardmodel} contains parameters 
$\{\sigma_{\epsilon}^2,\mu_0,\sigma^2,\rho,\nu\}$ whose prior hyperparameters need 
to be specified. We develop a default specification strategy that relies on the 
model properties explored in Section \ref{subsec:modprop}.

First, we set the prior for $\mu_0$ such that the prior expected probability 
response curve does not favor any category, and the corresponding prior uncertainty 
bands span a significant portion of the unit interval. For instance, this can 
be achieved with prior $\mu_0\sim N(0,100)$ which yields prior expected probability
of positive response of about $1/2$ across $\tau$.
In general, we would not expect to have available prior information about the 
variance and correlation structure of the unobserved signal process, which are 
controlled by parameters $\sigma^2$ and $\rho$. However, 
Proposition \ref{prop:deltaapproxlogitnormal} suggests an approximate relationship 
between the covariance structure of the binary process and the signal process, 
and we can thus specify the corresponding priors similarly to GP-based models. 
In particular, we select the uniform prior for the range parameter $\rho$ such that 
the correlation between $Z_{\tau}$ and $Z_{\tau^{\prime}}$ decreases to $0.05$ when 
the difference between $\tau$ and $\tau^{\prime}$ is within a pre-specified subset 
of the observation time window. For instance, for the data analysis in Section 
\ref{sec:realapp} where the total observation window comprises 72 days, we used a
$Unif(3,12)$ prior for $\rho$, which implies that the aforementioned correlation 
decreases to $0.05$ when the time difference ranges from 7 to 31 days. 
The hyperprior for $\nu$ is $Unif(a_{\nu},b_{\nu})$. We specify $a_{\nu}>3$ to 
reflect the constraint for $\Sigma/(\nu-3)$ to be a well-defined covariance matrix, 
and $b_{\nu}$ large enough such that the tail behavior of the marginal TP is hard 
to distinguish from that of a GP. For instance, a default choice is $a_{\nu}=4$ 
and $b_{\nu}=30$.

We follow \citet{Fong2010} to specify the prior for 
$\sigma_{\epsilon}^2\sim IG(a_{\epsilon},b_{\epsilon})$. Integrating 
out $\sigma_{\epsilon}^2$, the measurement error $\epsilon$ is marginally 
distributed as a univariate Student-t distribution with location parameter 0, scale 
parameter $b_{\epsilon}/a_{\epsilon}$, and degrees of freedom parameter $2a_{\epsilon}$. 
For a predetermined measurement error range $(-R,R)$ with degree of freedom $\upsilon$, 
we can use the relationship $\pm t_{1-(1-q)/2}^{\upsilon}\sqrt{b_{\epsilon}/a_{\epsilon}}=\pm R$ 
to obtain $a_{\epsilon}=$ $\upsilon/2$ and $b_{\epsilon}=$
$R^2\upsilon/[2(t_{1-(1-q)/2}^{\upsilon})^2]$, where $t^{\upsilon}_q$ is the 
$q$-th percentile of a Student-t distribution with $\upsilon$ degrees of freedom.


Proceeding to posterior inference, we develop an MCMC algorithm based on 
(\ref{eq:finiterepbinmodel}) and (\ref{eq:multigpiwppool}). We introduce layers of 
latent variables,  beginning with $\xi_{it}\sim PG(1,0)$ for every observation $Y_{it}$, 
where $PG(a,b)$ denotes the Pólya-Gamma distribution with shape parameter $a$ and 
tilting parameter $b$ \citep{Polson2013}. Denote the collection of Pólya-Gamma variables 
for each subject by $\boldsymbol{\xi}_i=(\xi_{i1},\ldots,\xi_{iT_i})^{\top}$. Also, 
introduce $\mathcal{Z}_{it}=Z_{it}+\epsilon_{it}$, and let $\boldsymbol{\mathcal{Z}}_i=$
$(\mathcal{Z}_{i1},\ldots,\mathcal{Z}_{iT_i})^{\top}$. Recall 
that $\boldsymbol{\tau}=\cup_{i=1}^n\boldsymbol{\tau}_i$ is the pooled grid. Denote 
the evaluations on the pooled grid by $\tilde{\mathbf{Z}}_i=Z_i(\boldsymbol{\tau})$ 
and let $\mathbf{Z}_i^*=\tilde{\mathbf{Z}}_i\setminus\mathbf{Z}_i$. 
That is, $\mathbf{Z}_i^*=Z_i(\boldsymbol{\tau}_i^*)$,
where $\boldsymbol{\tau}_i^*=\boldsymbol{\tau}\setminus\boldsymbol{\tau}_i$ is the set 
of grid points at which the $i$-th trajectory misses observations. Then, the hierarchical 
model for the data $\{Y_{it}: t=1,\ldots,T_i, \, i=1,\ldots,n\}$ can be expressed as 
\begin{equation*}
    \begin{split}
        &Y_{it}\mid\mathcal{Z}_{it},\xi_{it}\stackrel{ind.}{\sim} \mathcal{B}(\mathcal{Z}_{it},\xi_{it}),\,\,\xi_{it}\stackrel{i.i.d.}{\sim} PG(1,0),\,\, t=1,\ldots,T_i,\\
        &\boldsymbol{\mathcal{Z}}_i\mid \mathbf{Z}_i,\sigma_{\epsilon}^2\stackrel{ind.}{\sim} N(\mathbf{Z}_i,\sigma_{\epsilon}^2\mathbf{I}_{T_i}),\,\, \tilde{\mathbf{Z}}_i=(\mathbf{Z}_i,\mathbf{Z}_i^*)^{\top}\mid\boldsymbol{\mu},\boldsymbol{\Sigma}\stackrel{i.i.d.}{\sim}N(\boldsymbol{\mu},\boldsymbol{\Sigma}),\,\, i=1,\ldots,n,\\
        &\sigma_{\epsilon}^2\sim IG(a_{\epsilon},b_{\epsilon}),\,\,\boldsymbol{\mu}\mid \mu_0,\boldsymbol{\Sigma},\nu\sim N(\mu_0\mathbf{1},(\nu-3)\boldsymbol{\Sigma}),\,\,\mu_0\sim N(a_{\mu},b_{\mu}),\\
        & \boldsymbol{\Sigma}\mid \nu,\boldsymbol{\Psi}_{\boldsymbol{\phi}}\sim IW(\nu,\boldsymbol{\Psi}_{\boldsymbol{\phi}}),\,\,\boldsymbol{\Psi}_{\boldsymbol{\phi}}=\Psi_{\boldsymbol{\phi}}(\boldsymbol{\tau},\boldsymbol{\tau}),\,\, \boldsymbol{\phi}=\{\sigma^2,\rho\},\\
        & \sigma^2\sim \text{Gamma}(a_{\sigma},b_{\sigma}),\,\,\rho\sim Unif(a_{\rho},b_{\rho}),\,\,\nu\sim Unif(a_{\nu},b_{\nu}).
    \end{split}
    \label{eq:hiermodelbin}
\end{equation*}
Here, $\mathcal{B}(\mathcal{Z}_{it},\xi_{it}) \propto$ 
$\exp\{(Y_{it}-0.5) \, \mathcal{Z}_{it}-0.5\,\xi_{it}\mathcal{Z}_{it}^2\}$
denotes the probability mass function of $Y_{it}$ conditional on both sets of latent variables,
$\mathcal{Z}_{it}$ and $\xi_{it}$.
Based on Theorem 1 from \cite{Polson2013}, marginalizing out $\xi_{it}$ from 
$\mathcal{B}(\mathcal{Z}_{it},\xi_{it})$, we obtain the Binomial distribution for $Y_{it}$
conditional on latent variable $\mathcal{Z}_{it}$, that is, 
$Y_{it} \mid \mathcal{Z}_{it} \stackrel{ind.}{\sim} Bin(1,\varphi(\mathcal{Z}_{it}))$.
Hence, the joint posterior density of all model parameters can be written as
\begin{equation}
    \begin{split}
        p(\{\boldsymbol{\mathcal{Z}}_i\}_{i=1}^n,&\{\boldsymbol{\xi}_i\}_{i=1}^n,\{\tilde{\mathbf{Z}}_i\}_{i=1}^n,
        \boldsymbol{\mu},\boldsymbol{\Sigma},\sigma_{\epsilon}^2,\mu_0,\sigma^2,\rho,\nu\mid \{\mathbf{Y}_i\}_{i=1}^n)\\
        &\propto \prod_{i=1}^n\{p(\mathbf{Y}_i\mid \boldsymbol{\mathcal{Z}}_i,\boldsymbol{\xi}_i)p(\boldsymbol{\xi}_i)p(\boldsymbol{\mathcal{Z}}_i\mid \mathbf{Z}_i,\sigma_{\epsilon}^2)p(\mathbf{Z}_i^*\mid\mathbf{Z}_i,\boldsymbol{\mu},\boldsymbol{\Sigma})p(\mathbf{Z}_i\mid \boldsymbol{\mu},\boldsymbol{\Sigma})\}\\
        &\times p(\boldsymbol{\mu}\mid \mu_0,\boldsymbol{\Sigma},\nu)p(\boldsymbol{\Sigma}\mid \sigma^2,\rho,\nu)p(\sigma_{\epsilon}^2)p(\mu_0)p(\sigma^2)p(\rho)p(\nu).
    \end{split}
    \label{eq:jointpostmodelbin}
\end{equation}

The introduction of the latent variables enables a Gibbs sampling scheme with conditionally 
conjugate updates. Denote generically by $p(\boldsymbol{\theta}\mid -)$ 
the posterior full conditional for parameter $\boldsymbol{\theta}$. Notice 
that $p(\boldsymbol{\mathcal{Z}}_i,\boldsymbol{\xi}_i\mid -)\propto $
$p(\mathbf{Y}_i \mid \boldsymbol{\mathcal{Z}}_i,\boldsymbol{\xi}_i) p(\boldsymbol{\xi}_i) p(\boldsymbol{\mathcal{Z}}_i\mid \mathbf{Z}_i,\sigma_{\epsilon}^2)$, 
which matches the Bayesian logistic regression structure in \citet{Polson2013}. 
Therefore, $p(\boldsymbol{\mathcal{Z}}_i\mid -)$ and $p(\boldsymbol{\xi}_i\mid -)$ 
can be sampled directly. 
Factorizing the prior of $\tilde{\mathbf{Z}}_i$ as $p(\tilde{\mathbf{Z}}_i|\boldsymbol{\mu},\boldsymbol{\Sigma})=$
$p(\mathbf{Z}_i^*\mid\mathbf{Z}_i,\boldsymbol{\mu},\boldsymbol{\Sigma})p(\mathbf{Z}_i\mid \boldsymbol{\mu},\boldsymbol{\Sigma})$, 
results in $p(\mathbf{Z}_i^*,\mathbf{Z}_i| -)\propto$
$ p(\mathbf{Z}_i^*\mid\mathbf{Z}_i,\boldsymbol{\mu},\boldsymbol{\Sigma})
p(\mathbf{Z}_i\mid \boldsymbol{\mu},\boldsymbol{\Sigma})
p(\boldsymbol{\mathcal{Z}}_i\mid \mathbf{Z}_i,\sigma_{\epsilon}^2)$. This forms yields
ready updates for $\mathbf{Z}_i^*$ and $\mathbf{Z}_i$ using GP-based predictive sampling.
All other model parameters can be sampled using standard updates. The details of the 
MCMC algorithm are given in the Supplementary Material.

We have linked the probability response curve and covariance structure of the 
binary process $Y_i(\tau)$ to the corresponding signal process $Z_i(\tau)$. 
To estimate the signal process, we obtain posterior samples for 
$\mathbf{Z}_i^+=Z_i(\boldsymbol{\tau}^+)$, where 
$\boldsymbol{\tau}^+\supset \boldsymbol{\tau}$ is a finer grid than the pooled grid.
Denote $\check{\boldsymbol{\tau}}=\boldsymbol{\tau}^+\setminus\boldsymbol{\tau}$ 
as the time points where none of the subjects have observations, and 
let $\check{\mathbf{Z}}_i=Z_i(\check{\boldsymbol{\tau}})$. Using the marginal TP 
result from Proposition \ref{prop:marginalsignal}, 
\begin{equation*}
    \begin{pmatrix} \tilde{\mathbf{Z}}_i \\ \check{\mathbf{Z}}_i \end{pmatrix}\sim 
    MVT\left( \nu,
    \begin{pmatrix} \boldsymbol{\mu}_{0\boldsymbol{\tau}} \\ \boldsymbol{\mu}_{0\check{\boldsymbol{\tau}}} \end{pmatrix}, \begin{pmatrix} \boldsymbol{\Psi}_{\boldsymbol{\tau},\boldsymbol{\tau}} & \boldsymbol{\Psi}_{\boldsymbol{\tau},\check{\boldsymbol{\tau}}} \\ \boldsymbol{\Psi}_{\check{\boldsymbol{\tau}},\boldsymbol{\tau}} & 
    \boldsymbol{\Psi}_{\check{\boldsymbol{\tau}},\check{\boldsymbol{\tau}}}
    \end{pmatrix} \right),
    \label{eq:jointpostpred}
\end{equation*}
where $\boldsymbol{\mu}_{0\cdot}=\mu_0\mathbf{1}_{|\cdot|}$, and 
$\boldsymbol{\Psi}_{\cdot,\cdot}$ denotes the covariance function evaluation 
$\Psi_{\boldsymbol{\phi}}(\cdot,\cdot)$. Next, based on the conditionals of the 
MVT distribution \citep{Shah2014},
\begin{equation}
    \check{\mathbf{Z}}_i\mid \tilde{\mathbf{Z}}_i \sim 
    MVT \left( \nu+|\boldsymbol{\tau}|,\check{\boldsymbol{\mu}}_{i\check{\boldsymbol{\tau}}},\frac{\nu+S_{i\boldsymbol{\tau}}-2}{\nu+|\boldsymbol{\tau}|-2}\check{\boldsymbol{\Psi}}_{\check{\boldsymbol{\tau}},\check{\boldsymbol{\tau}}}\right),
    \label{eq:condpostpred}
\end{equation}
with $\check{\boldsymbol{\mu}}_{i\check{\boldsymbol{\tau}}}=\boldsymbol{\Psi}_{\check{\boldsymbol{\tau}},\boldsymbol{\tau}}\boldsymbol{\Psi}_{\boldsymbol{\tau},\boldsymbol{\tau}}^{-1}(\tilde{\mathbf{Z}}_i- \boldsymbol{\mu}_{0\boldsymbol{\tau}})+ \boldsymbol{\mu}_{0\check{\boldsymbol{\tau}}}$, $S_{i\boldsymbol{\tau}}=(\tilde{\mathbf{Z}}_i- \boldsymbol{\mu}_{0\boldsymbol{\tau}})^{\top}\boldsymbol{\Psi}_{\boldsymbol{\tau},\boldsymbol{\tau}}^{-1}(\tilde{\mathbf{Z}}_i- \boldsymbol{\mu}_{0\boldsymbol{\tau}})$ and $\check{\boldsymbol{\Psi}}_{\check{\boldsymbol{\tau}},\check{\boldsymbol{\tau}}}=\boldsymbol{\Psi}_{\check{\boldsymbol{\tau}},\check{\boldsymbol{\tau}}}-\boldsymbol{\Psi}_{\check{\boldsymbol{\tau}},\boldsymbol{\tau}}\boldsymbol{\Psi}_{\boldsymbol{\tau},\boldsymbol{\tau}}^{-1}\boldsymbol{\Psi}_{\boldsymbol{\tau},\check{\boldsymbol{\tau}}}$. 
Using (\ref{eq:condpostpred}), given each posterior sample for $\tilde{\mathbf{Z}}_i$, 
$\mu_0$, $\boldsymbol{\phi}$ and $\nu$, we can complete the posterior realization for the 
signal process over the finer grid. As discussed in Section \ref{subsec:modprop}, we can 
then obtain full posterior inference for functionals of the binary process.

The predictive distribution of the signal process also illustrates the information 
borrowed across subjects. For the $i$-th subject, the grid, $\boldsymbol{\tau}^+$,
where predictions are made can be 
partitioned as $\boldsymbol{\tau}_i\cup\boldsymbol{\tau}_i^*\cup\check{\boldsymbol{\tau}}$, 
where $\boldsymbol{\tau}_i^*=\boldsymbol{\tau}\setminus\boldsymbol{\tau}_i$ represents 
the grid points where subject $i$ does not have observations, while at least one of 
the other subjects have observations. Then, we first predict $Z_i(\boldsymbol{\tau}_i^*)$ 
conditioning on $Z_i(\boldsymbol{\tau}_i)$ by the GP predictive distribution, and next 
predict $Z_i(\check{\boldsymbol{\tau}})$ conditioning on $Z_i(\boldsymbol{\tau}_i)$ 
and $Z_i(\boldsymbol{\tau}_i^*)$ by the TP predictive distribution. 
Comparing with the GP, (\ref{eq:condpostpred}) suggests the TP is scaling the predictive 
covariance by the factor $\frac{\nu+S_{i\boldsymbol{\tau}}-2}{\nu+|\boldsymbol{\tau}|-2}$. 
Note that $S_{i\boldsymbol{\tau}}$ is distributed as the sum of squares of 
$|\boldsymbol{\tau}|$ independent $MVT_1(\nu,0,1)$ random variables and hence 
$\text{E}(S_{i\boldsymbol{\tau}})=|\boldsymbol{\tau}|$. Accordingly, if we have made 
good interpolation prediction, the predictive covariance for extrapolation of
$Z_i(\check{\boldsymbol{\tau}})$ is expected to scale down and vice versa. 
Comparing with predicting both $Z_i(\boldsymbol{\tau}_i^*)$ and 
$Z_i(\check{\boldsymbol{\tau}})$ conditioning on $Z_i(\boldsymbol{\tau}_i)$ 
through the GP predictive distribution, our model allows using information across 
subjects to adjust the individual 
trajectory's credible interval.

Another crucial benefit of modeling the signal process as a TP emerges when we consider 
making predictions at $\check{\boldsymbol{\tau}}$, the grid points where none 
of the subjects have observations. Under the hierarchical GP prior in \citet{Yang2016}, 
for which the marginal is not generally a TP, such predictions would require the 
conditional distribution $\boldsymbol{\Sigma}_{\check{\boldsymbol{\tau}},\check{\boldsymbol{\tau}}}
\mid \boldsymbol{\Sigma}_{\boldsymbol{\tau},\boldsymbol{\tau}}$ from their joint 
inverse-Wishart distribution, which is not analytically available. We circumvent this 
issue by marginalizing out $\mu$ and $\Sigma$. The predictions are then based on the 
conditional $\check{\mathbf{Z}}_i\mid \tilde{\mathbf{Z}}_i$ from their joint multivariate 
t distribution, which is the MVT distribution in (\ref{eq:condpostpred}). Hence, for prediction 
on a grid denser than the pooled grid $\boldsymbol{\tau}$, the marginal TP specification 
for the signal process is a practically important model feature.

\subsection{Synthetic data examples}
\label{subsec:simstudy}

We assess the model by applying it to carefully designed simulation scenarios that reflect 
our main contributions. The full details are provided in the Supplementary Material. Here, 
we briefly discuss the simulation study setting and summarize the main findings.

For the {\color{blue} three} sets of simulation studies we considered, the 
longitudinal binary responses are generated from the following generic process:
\begin{equation}
    \begin{split}
    &
    Y_i(\boldsymbol{\tau}_{i}) \mid \mathcal{Z}_i(\boldsymbol{\tau}_{i}) \stackrel{ind.}{\sim} 
    Bin(1,\eta(\mathcal{Z}_i(\boldsymbol{\tau}_{i}))), \quad \boldsymbol{\tau}_i=(\tau_{i1},\ldots,\tau_{iT_i}),\quad i=1,\ldots,n,\\
    & \mathcal{Z}_i(\boldsymbol{\tau}_{i}) 
    =f(\boldsymbol{\tau}_i)+\boldsymbol{\omega}_i+\boldsymbol{\epsilon}_i\quad \boldsymbol{\epsilon}_i\stackrel{i.i.d.}{\sim} N(\mathbf{0},\sigma_{\epsilon}^2\mathbf{I}),
    \end{split}   
    \label{eq:simgendata}
\end{equation}
where $\eta(\cdot)$ is a link function mapping $\mathbb{R}$ to $(0,1)$, $f(\tau)$ 
is a signal function, and $\boldsymbol{\omega}_i$ is a realization from a mean zero 
continuous stochastic process that depicts the temporal covariance within the $i$-th subject.

The first set of simulation studies focuses on evaluating the effectiveness of the 
proposed model in capturing the fluctuation of the temporal trend. We consider 
different link function, signal function, and temporal covariance structure combinations, 
and we simulate unbalanced data with different sparsity levels. The results demonstrate 
that, despite the data generating process and the sparsity level, the model can 
recover not only the subject's probability response curve, but also the underlying 
continuous signal function.

The objective of the second set of simulation studies is to explore the performance 
of the proposed model in estimating the within subject covariance structure. To this end, 
we examine a number of possible choices for generating the $\boldsymbol{\omega}_i$ 
in (\ref{eq:simgendata}), which imply covariance structures that are not of the 
same form as the covariance kernel of the model. The results reveal that the model 
can recover the true covariance between the signal variables, 
$(Z_i(\tau_{it}),Z_i(\tau_{i t^{\prime}}))$, and the binary 
responses, $(Y_i(\tau_{it}),Y_i(\tau_{i t^{\prime}}))$, thus providing empirical 
evidence for the robustness of the covariance kernel choice.

In both cases, we examine simplified versions of the model for comparison. 
The simplified models are constructed by modeling either the mean structure or 
the covariance structure parametrically in the two sets of simulation studies, 
respectively. Demonstrating that the proposed model outperforms its parametric 
backbones, we highlight the practical utility of the nonparametric modeling 
for the mean and covariance structure.  

To further illustrate the practical benefits of the functional 
data analysis perspective, in the third simulation study, we consider a scenario 
where observations are made irregularly. Through formal model comparison, we show 
that the proposed model outperforms a traditional approach under the GLMM framework.

\subsection{Connections with existing literature}
\label{subsec:literaturereview}


Our methodology is broadly related with certain Bayesian nonparametric methods. 
The proposed model is related to a particular class of conditional models, known as 
transition models, which induce the aging effect by allowing past values to explicitly 
affect the present observation, usually through autoregressive dynamics. 
\citet{DiLucca2013} studied a class of non-Gaussian autoregression models
for continuous responses, which can be extended to handle binary longitudinal outcomes by 
treating them as a discretized version of the continuous outcomes. 
\citet{MariaJASA2018} developed a nonparametric density regression model for ordinal 
regression relationships that evolve in discrete time. 
Compared with the proposed methodology, these models are more flexible in terms of the 
binary response distribution. However, it is demanding to handle higher than first-order
dynamics, and there is no natural way to treat missing data under a discrete time 
autoregressive framework, hindering applications for unbalanced longitudinal studies.

The proposed model is more closely related to subject-specific models, where the 
responses are assumed to be independent conditioning on subject-specific effects. 
The main approach has been to construct models for longitudinal binary responses
building from Bayesian nonparametric models for longitudinal continuous data, 
developed under the mixed effects framework, utilizing Dirichlet 
process mixture models \citep[e.g.,][]{LiLinMuller2010,Ghosh2010,Quintana2016} 
or additive GP models \citep[e.g.,][]{Cheng2019}.
For instance, embedding a Dirichlet process mixture of normals prior as the probability model 
for the latent variables, \citet{Jara2007} and \citet{TangDuan2012} consider binary responses,
and \citet{Kunihama2019} handle mixed-scale data comprising continuous and binary responses. 
The proposed model differs in the way of treating subject-specific effects, and it arguably 
offers benefits in terms of computational efficiency. 

There is a growing trend of adopting functional data analysis tools in longitudinal data 
modeling. These methods specify observations as linear combinations of functional principal 
components (FPCs), with the FPCs represented as expansions of a pre-specified basis. 
Bayesian methods include \citet{Jiang2020} for continuous responses, and 
\citet{vanDerLinde2009} for binary and count responses. Challenges include inference 
which is sensitive to the basis choice, and a complex orthogonality constraint on the FPCs.
Recently, \citet{JamseAmy2022} proposed 
an approach that can serve as foundation for generalized FPC analysis of sparse and irregular 
binary responses. Nonetheless, our model involves a more parsimonious formulation, 
including the structure with the GP and TP predictive distributions. 
\section{Application with binary responses: \textit{Studentlife} data}
\label{sec:realapp}



\subsection{Data for analysis}
\label{subsec:datarealapp}


\textit{Studentlife} \citep{StudentLife2014} is a study that integrates automatic 
sensing data and an EMA component to probe students' mental health status and to study 
its relationship with students' academic performance and behavior trends. The data were 
collected by a smartphone app carried by 48 students over a 10-week term at Dartmouth 
College. The dataset 
is available from the R package ``studentlife'' \citep{Studentlifepackage}. 


We focus on a subset of the data that corresponds to assessing the students' emotional status. 
In the \textit{Studentlife} study, the assessment of emotion is conducted by the Photographic 
Affect Meter (PAM), a tool for measuring affect in which users select from a wide variety of 
photos the one which best suits their current mood \citep{Pollak2011}. The PAM survey is 
deployed to the mobile app and prompts everyday during the study period. The participants 
either respond to the survey, or ignore it, introducing missingness. The outcome of the survey 
contains two attributes, the PAM valence and the PAM arousal. They are scores of -2 to 2 
(excluding 0) that measure the subject's extent of displeasure to pleasure or state of 
activation ranging from low to high, respectively. We dichotomize the valence and arousal 
scores by their sign, representing the positive values by 1. In this section, we focus on 
analyzing the change of binary valence and arousal responses to evaluate students' affects 
as the term progresses.    


The data were collected during the spring 2013 term at Dartmouth college. We set the study 
period according to the official academic calendar, from the first day of classes (March 25, 2013) 
to the end of the final exam period (June 4, 2013), resulting in a total of 72 days. We exclude 
subjects with less than 12 responses, resulting in 45 students. The longitudinal recordings of 
valence or arousal of the $i$-th student are denoted by $Y_i(\boldsymbol{\tau}_i)$, for
$i=1,\ldots,45$, where the student-specific grid points are a subset of $\boldsymbol{\tau}=$
$(0,1,\ldots,71)^{\top}$, representing the days on which the measurements are recorded. 
Several special events occurred during the study period, and we are particularly interested 
in investigating the change of students' affects on the time intervals around these events. 
Specifically, the events and corresponding periods are: (i) Days following the Boston marathon 
bombing (April 15, 2013 to April 17, 2013); (ii) The Green Key (a spring festival at Dartmouth) 
period (May 17, 2013 to May 18, 2013); (iii) The Memorial Day long weekend (May 25, 2013 to 
May 27, 2013); (iv) The final examination period (May 31, 2013 to June 3, 2013).       


We retrieve the data for the specific responses and study period from the R package 
``studentlife'' that contains the database for the entire study. Over all observations, 
the percentage of missing values is 31.1\%. There are slightly more missing responses at 
the beginning and toward the end of the study.
In light of the structure of EMA studies, the missing pattern for each 
subject can be viewed as random. We elaborate on the missing-at-random assumption in the 
Supplementary Material.

\begin{figure}[t!]
    \centering
    \begin{subfigure}[b]{0.48\textwidth}
            \includegraphics[width=\textwidth,height=5cm]{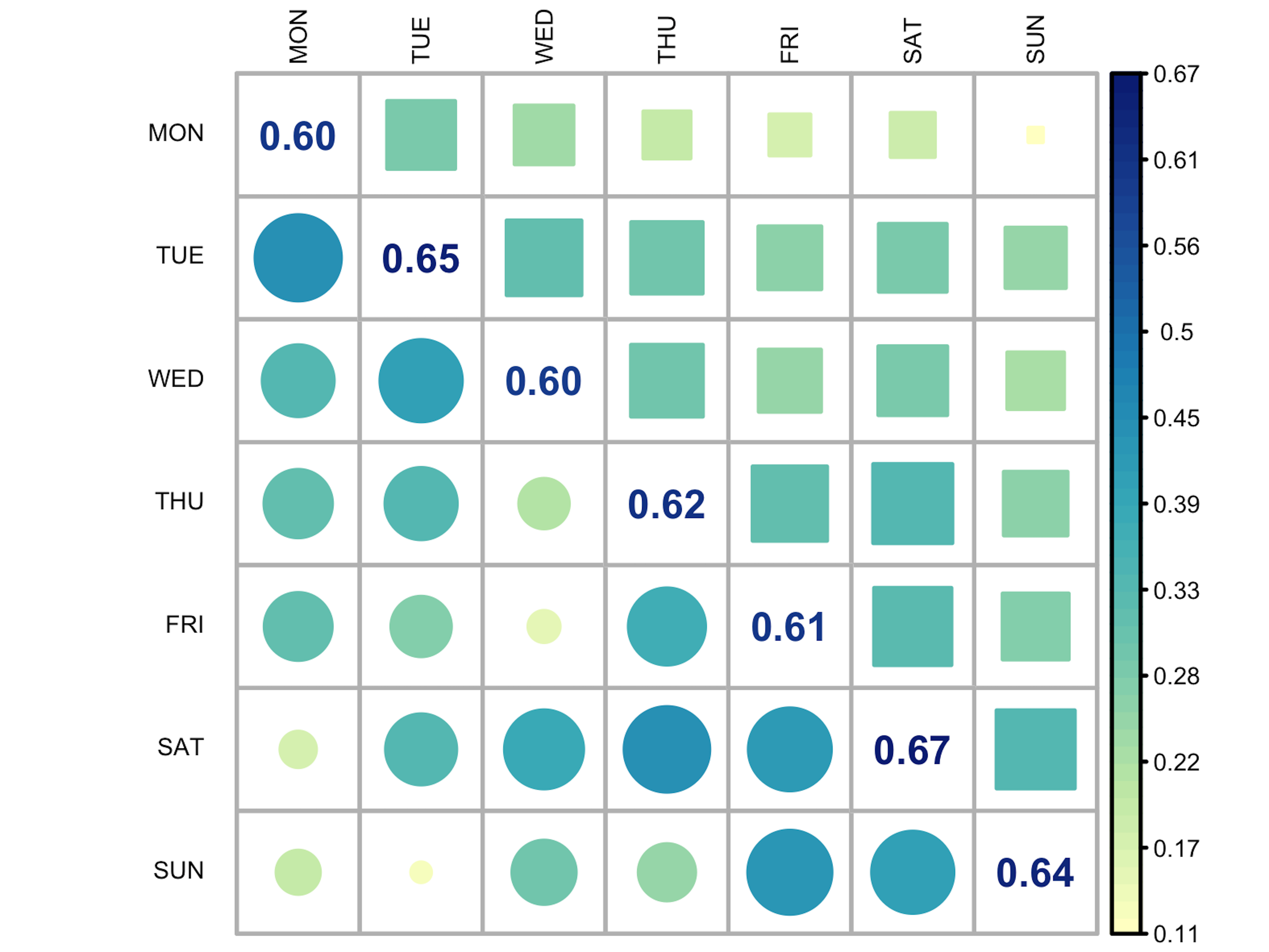}
            \caption{{\footnotesize Valence.}}
    \end{subfigure}
    \begin{subfigure}[b]{0.48\textwidth}
            \includegraphics[width=\textwidth,height=5cm]{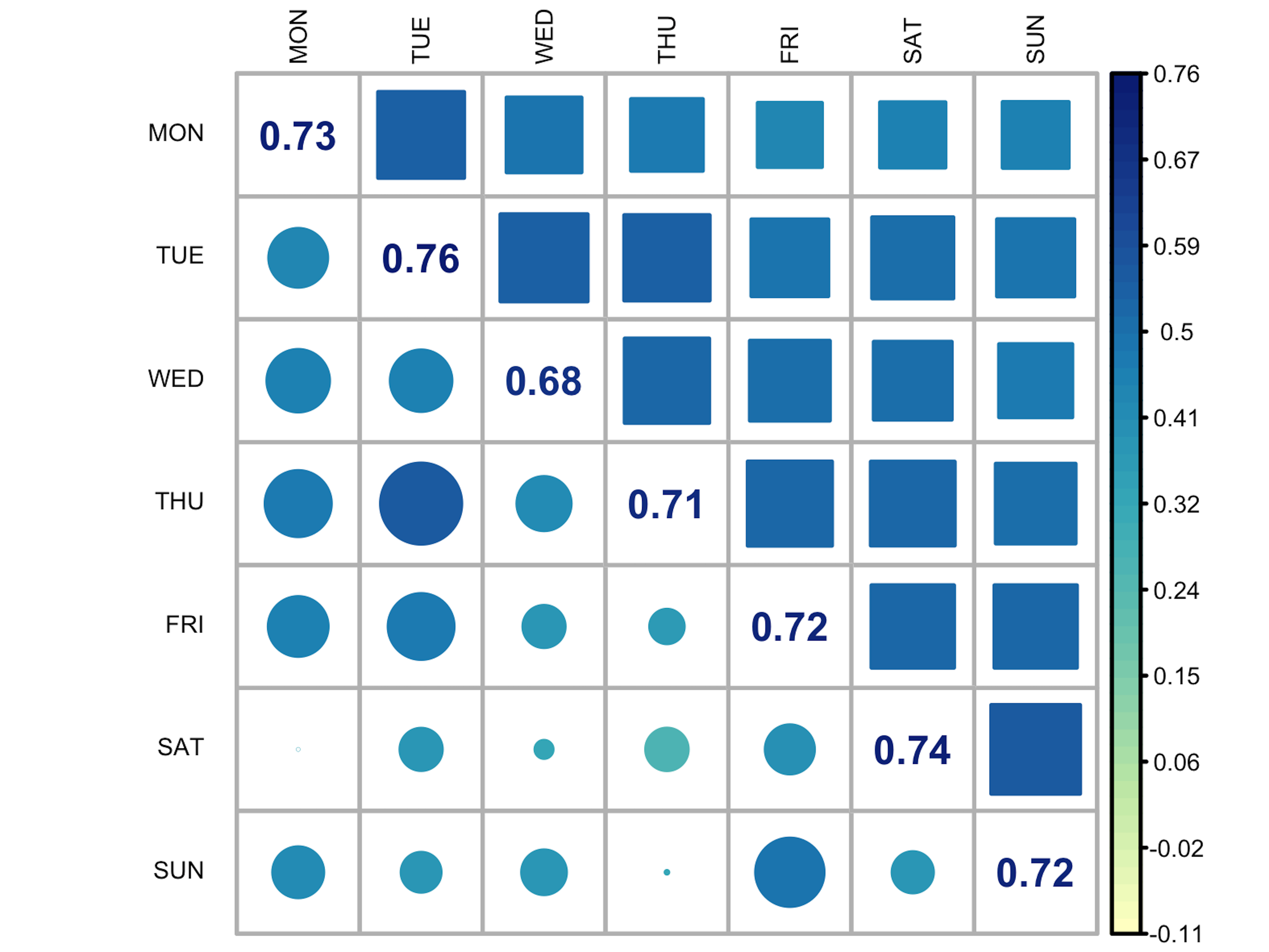}
            \caption{{\footnotesize Arousal.}}
    \end{subfigure}
\caption{\textit{Studentlife} data. Empirical estimate of the correlation coefficients 
between binary responses within a week. In each panel, the upper triangle and the lower 
triangle are for the Pearson and the tetrachoric correlation coefficient, respectively. }
    \label{fig:datacorr}
\end{figure}

We further explore the correlations between the binary responses within a week. We split the 
whole observation sequence into batches representing a week, and empirically calculate the 
Pearson and the tetrachoric correlation coefficient for each pair of time and distance 
combinations. The results, plotted in Figure \ref{fig:datacorr}, suggest that the correlation 
of the students' response to valence and arousal decreases slowly in time. 
%
%
%

\subsection{Analysis and results}
\label{subsec:resultsrealapp}

We fit the proposed model for the binary valence and arousal responses separately. 
We specify the prior for the model parameters by the procedure mentioned in 
Section \ref{subsec:modelapply}. (Results from prior sensitivity analysis are 
presented in the Supplementary Material.) Posterior inference results are based on 
5000 MCMC samples obtained every 4 iterations from a chain of 50000 iterations with 
a 30000 burn-in period (which is conservative).

\begin{figure}[t!]
\centering
\includegraphics[width=16cm,height=5cm]{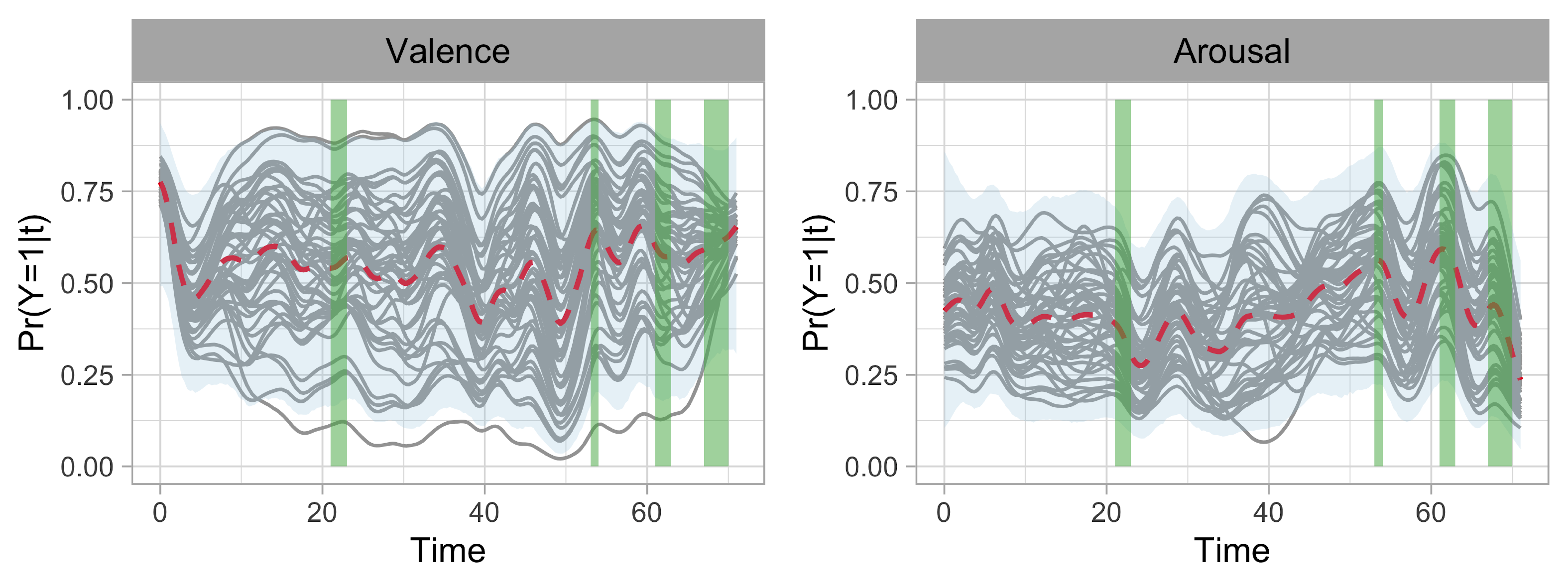}
\caption{\textit{Studentlife} data. Posterior mean (dashed line) and 95\% interval estimate 
(shaded region) of the probability response curve for an out-of-sample subject. The posterior 
mean estimates of probability response curves for in-sample subjects are given by the solid 
lines. The vertical shaded regions correspond to the four special time periods 
(see Section \ref{subsec:datarealapp}).}
\label{fig:realappprobcurve}
\end{figure}

We first examine in Figure \ref{fig:realappprobcurve} the probability response curves, 
defined as the probability of obtaining positive valence or arousal as a function of time.
For the valence, the happiness level drops as the term begins and increases when the term 
ends. The Boston marathon bombing may have had a minor effect on the valence. We observe 
local peaks around the Green Key festival and the Memorial Day holiday. As the students 
finish their exams, there is a trend toward happiness. As for arousal, it is relatively 
stable at the beginning of the term, and fluctuates as the term progresses. There is a 
drop in activation level after the Boston marathon bombing and during the final exam 
period, while the activation level reaches a local maximum at around the Green Key 
festival and the Memorial Day holiday. 
 
\begin{figure}[t!]
    \centering
    \begin{subfigure}[b]{0.24\textwidth}
            \includegraphics[width=\textwidth,height=4cm]{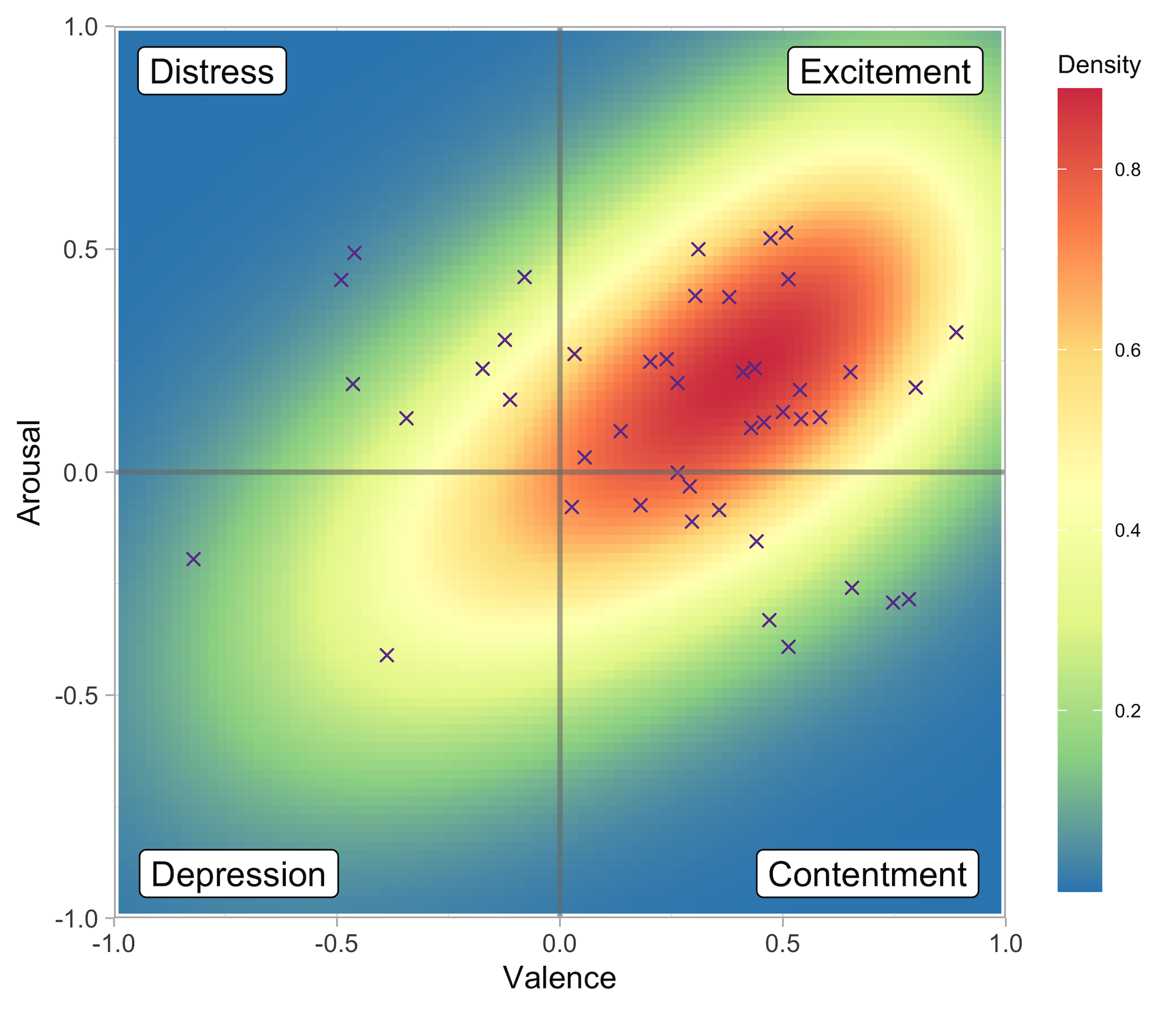}
            \caption{{\footnotesize Green Key}}
    \end{subfigure}
    \begin{subfigure}[b]{0.24\textwidth}
            \includegraphics[width=\textwidth,height=4cm]{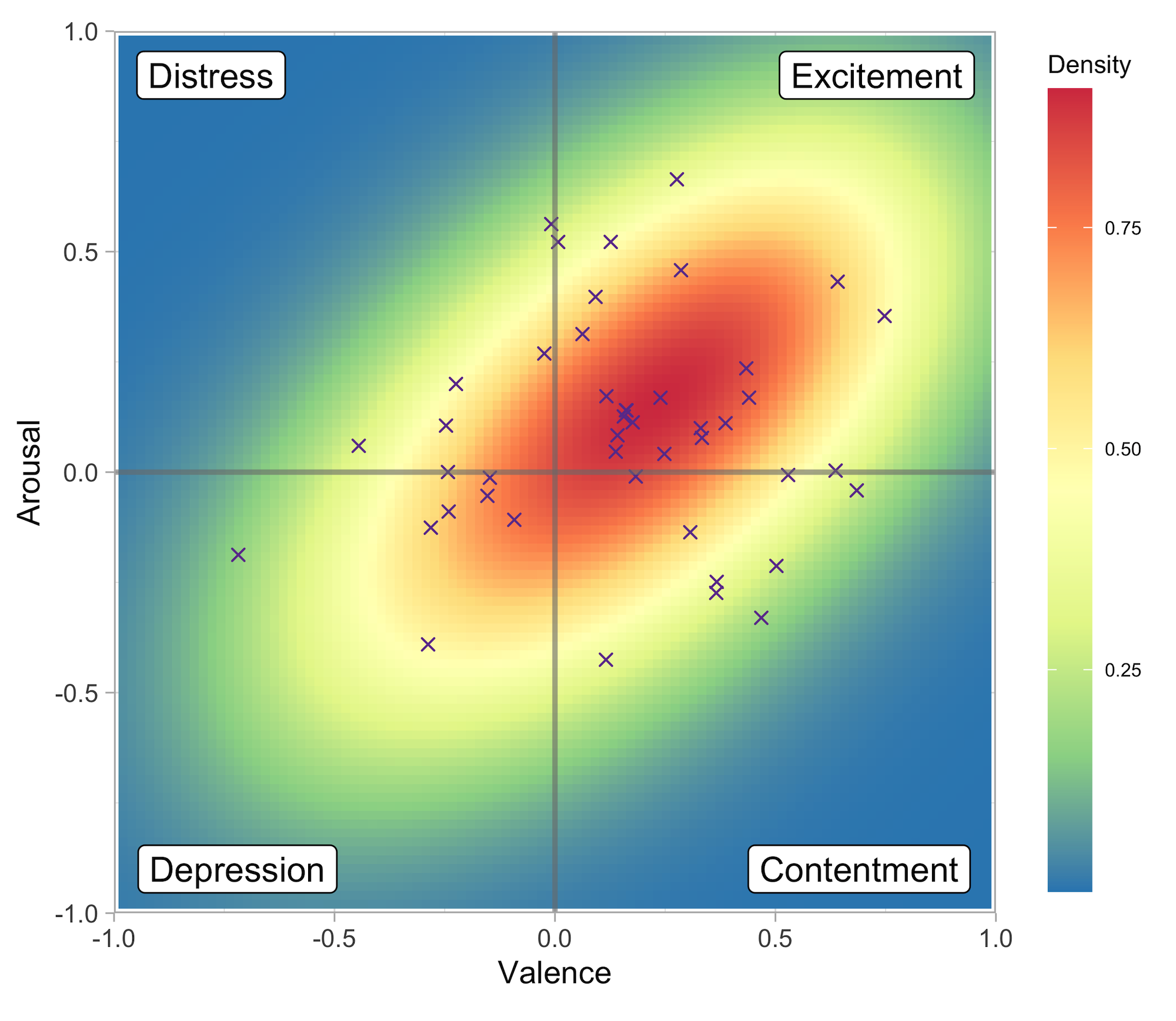}
            \caption{{\footnotesize Memorial Day}}
    \end{subfigure}
    \begin{subfigure}[b]{0.24\textwidth}
            \includegraphics[width=\textwidth,height=4cm]{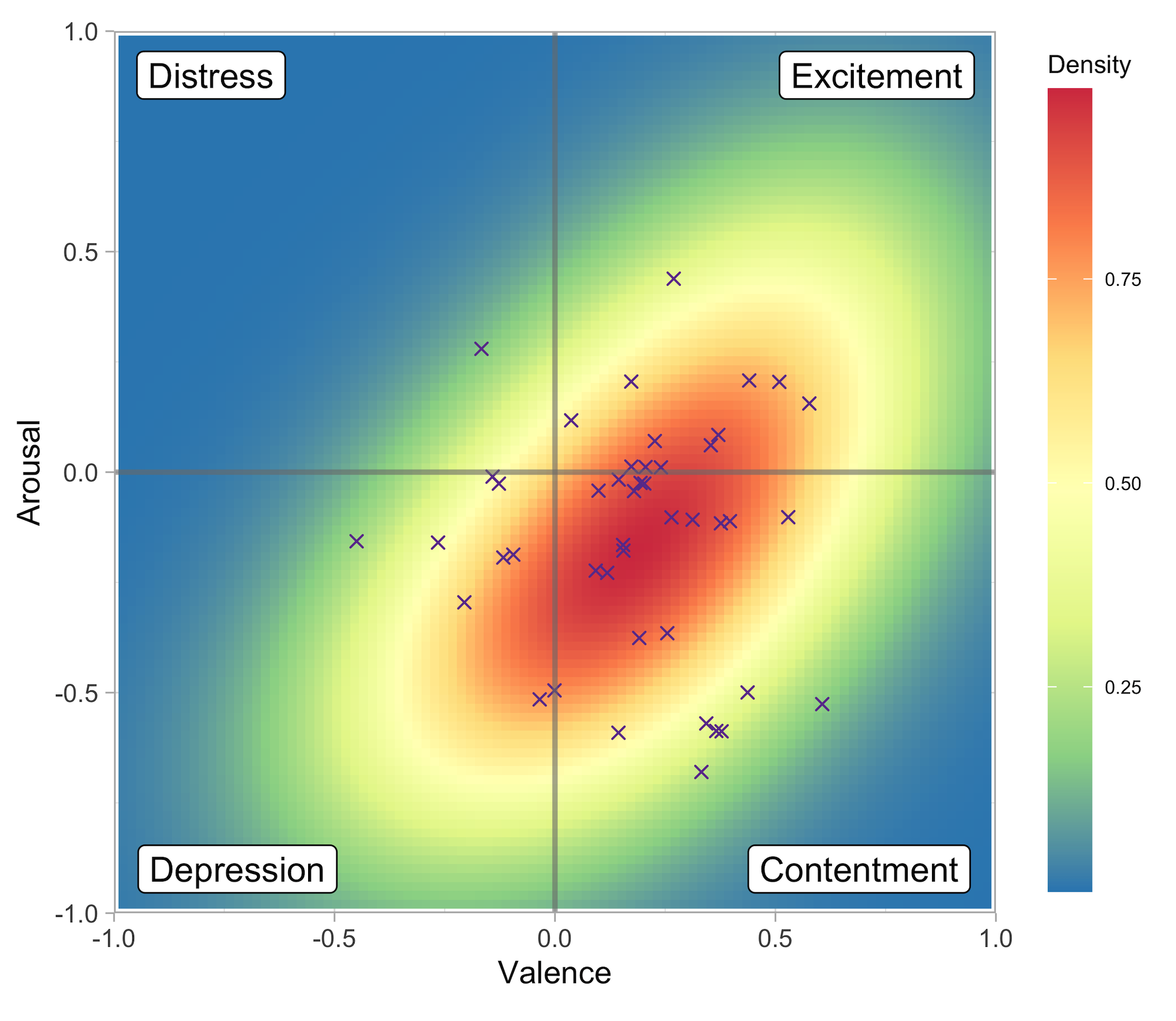}
            \caption{{\footnotesize Final exams begin}}
    \end{subfigure}
    \begin{subfigure}[b]{0.24\textwidth}
            \includegraphics[width=\textwidth,height=4cm]{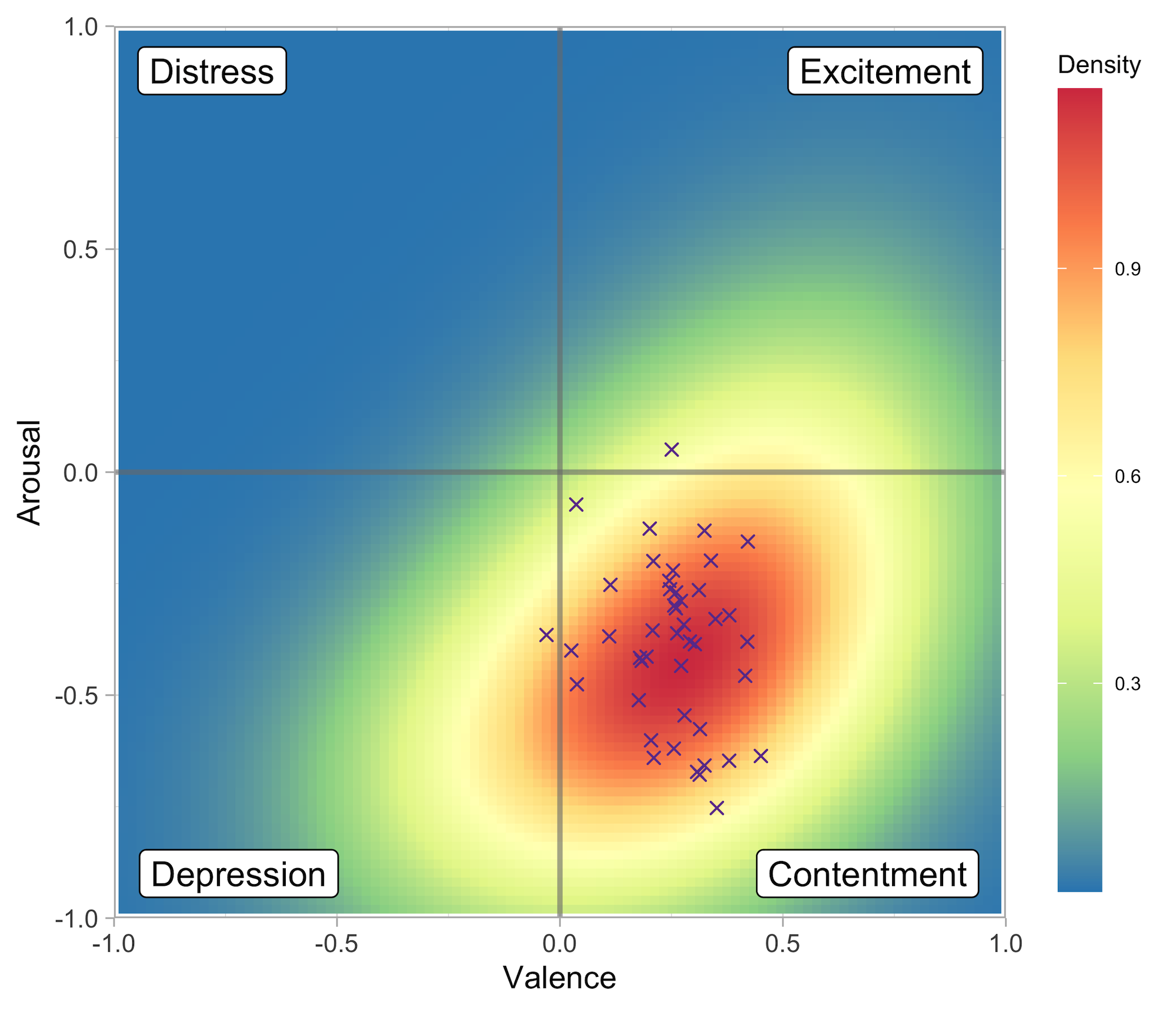}
            \caption{{\footnotesize Final exams end}}
    \end{subfigure}
\caption{\textit{Studentlife} data. Posterior density estimate of an out-of-sample 
subject's valence and arousal probability over the mood coordinate space on four 
specific days. In each panel, the crosses represent the posterior means of the 
in-sample subjects' valence and arousal probability mapped to the mood coordinate space.}
    \label{fig:densityday}
\end{figure}

Moreover, we assess the student's emotional status on specific days. According to 
\citet{Russell1980}, various states of emotional status can be represented by points 
located at the two dimensional mood coordinate space spanned by valence for the horizontal 
dimension and arousal for the vertical dimension. Moods such as excitement, distress, 
depression, and contentment, are represented by points in the quadrants of the space. 
For each observation, we can map the corresponding pairs of probabilities for positive 
valence and arousal onto the unit square in the mood space. In Figure \ref{fig:densityday}, 
the density heatmap is obtained by the posterior samples of positive probabilities for 
a new student of the same cohort, while the posterior means of the in-sample positive 
probabilities are marked by crosses. Panels (a) and (b) suggest the students are mostly 
excited at the festival and holiday. Moving from panel (c) to panel (d), we observe that 
the happiness level increases and the activation level decreases towards the end of the 
exam period. 

\begin{figure}[t!]
\centering
\includegraphics[width=16cm,height=4cm]{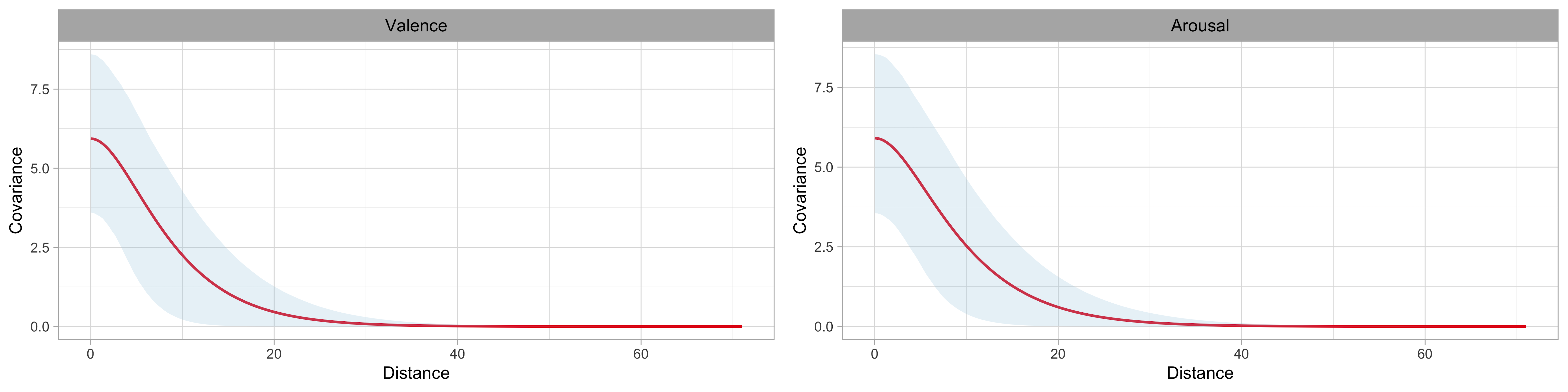}
\caption{\textit{Studentlife} data. Posterior mean (solid line) and 95\% interval 
estimate of the signal process covariance kernel. }
\label{fig:valarocov}
\end{figure}

We also obtain the posterior point and 95\% interval estimate for the covariance kernel 
of the signal process, which is displayed in Figure \ref{fig:valarocov}. 
It is noteworthy that there is a similar decreasing trend for the two distinct 
binary responses of valence and arousal. The practical range, defined as the distance 
at which the correlation is 0.05, has an estimated mean of 20.99 for valence 
and 22.97 for arousal. 

\subsection{Performance comparisons}
\label{subsec:comparerealapp}

For comparison with a traditional approach, we consider an analysis of the data under
the GLMM setting. In particular, we assume the model
\begin{equation*}
		Y_{it}\mid \mathcal{Z}_{it} \stackrel{ind.}{\sim} Bin(1,\varphi(\mathcal{Z}_{it})), \,\,
		\mathcal{Z}_{it}=\tilde{\boldsymbol{\tau}}_{it}^{\top}\boldsymbol{\beta}+\sum_{k=1}^KS_{itk}b_{k}+\mu_i+\epsilon_{it},\,\, 
		t=1,\ldots,T_i, \,\, i=1,\ldots, n,
\end{equation*}  
where $\tilde{\boldsymbol{\tau}}_{it}=(1,\tau_{it})^{\top}$, $\boldsymbol{\beta}$ is the 
vector of fixed effects, and $\epsilon_{it}\stackrel{i.i.d.}{\sim}N(0,\sigma^2_{\epsilon})$ 
is the measurement error. To allow flexibility in modeling the time effect, we consider 
cubic B-spline basis functions with $K=9$ knots that separate naturally the observed interval
by week; $S_{itk}$ is the $k$-th basis associated with time, with parameter 
$b_k\stackrel{i.i.d.}{\sim}N(0,\sigma^2_b)$. Finally, $\mu_i\stackrel{i.i.d.}{\sim}N(0,\sigma_{\mu}^2)$ 
are subject-specific random effects. The model is implemented using the integrated nested Laplace 
approximation (INLA) approach \citep{Rue2009} with the ``INLA'' package in R \citep{Rue2017}. 
We used the default choices provided by the R package for the prior on $\boldsymbol{\beta}$
(a flat prior), and for the values of the variance terms, $\sigma^2_{\epsilon}$, $\sigma^2_{b}$, 
and $\sigma^2_{\mu}$.


\begin{table}[t!] \centering
\small
\caption{\textit{Studentlife} data. Summary of comparison between the proposed model 
and the generalized linear mixed effects model (GLMM) using two different criteria. 
The values in bold correspond to the model favored by the particular criterion.} 
\label{tab:comprealapp}
\begin{tabular}{cccccc}
\hline
\hline
\multirow{2}{*}{Response} & \multirow{2}{*}{Model} & \multicolumn{3}{c}{Posterior predictive loss} & \multirow{2}{*}{CRPS} \\
\cline{3-5} & & $G(\mathcal{M})$ & $P(\mathcal{M})$ & $G(\mathcal{M})+P(\mathcal{M})$ & \\
\hline
\hline
\multirow{2}{*}{Valence} & Proposed & \textbf{428.09} & \textbf{475.31} & \textbf{903.40} & \textbf{0.19}\\
& GLMM & 456.09 & 475.83 & 931.92 & 0.20 \\
\hline
\multirow{2}{*}{Arousal} & Proposed & \textbf{457.62} & 496.63 & \textbf{954.25} & \textbf{0.20}\\
& GLMM & 476.17 & \textbf{492.28} & 968.45 & 0.21 \\
\hline
\hline
\end{tabular}
\end{table}

We perform model comparison using two different metrics: the posterior predictive loss 
criterion which combines a goodness-of-fit term, $G(\mathcal{M})$, and a penalty term, 
$P(\mathcal{M})$, for model complexity \citep{GelfandGhosh1998}; and, the continuous 
ranked probability score (CRPS), defined in terms of predictive cumulative distribution 
functions \citep{Gneiting2007}. Both criteria can be calculated from the posterior samples 
for model parameters, and both favor the model with a smaller value.        
Table \ref{tab:comprealapp} summarizes the results. For the 
valence response, both criteria favor the proposed model. As for the arousal response, 
the proposed model provides a more accurate fit to the data, while being penalized more 
than the GLMM with respect to model complexity. Nonetheless, our model is favored in terms 
of total posterior predictive loss, as well as by the CRPS criterion.

\section{Model for ordinal responses}
\label{sec:polyordinalmodel}

\subsection{The extended model}
\label{subsec:ordinalmodel}

We extend the model developed in Section \ref{subsec:standardmodel} to handle ordinal responses. 
Suppose the observation on subject $i$ at time $\tau_{it}$, denoted by $Y_{it}$, takes $C$ possible 
categories. We can equivalently encode the response as a vector with binary entries 
$\mathbf{Y}_{it}=(Y_{i1t},\ldots,Y_{iCt})$, such that $Y_{it}=j$ is equivalent to $Y_{ijt}=1$ 
and $Y_{ikt}=0$ for any $k\neq j$. We assume a multinomial response distribution for 
$\mathbf{Y}_{it}$, factorized in terms of binomial distributions,
\begin{equation}
	Mult(\mathbf{Y}_{it}\mid m_{it},\omega_{i1t},\ldots,\omega_{iCt})=\prod_{j=1}^{C-1}Bin(Y_{ijt}\mid m_{ijt},\varphi(Z_{ijt}+\epsilon_{ijt}))
	\label{eq:factmulti}
\end{equation}
where $m_{it}=\sum_{j=1}^CY_{ijt}\equiv 1$, $m_{i1t}=m_{it}$, and $m_{ijt}=m_{it}-\sum_{k=1}^{j-1}Y_{ikt}$. 
This factorization bridges the gap between binary and ordinal responses. Similar to the model 
for binary responses, we adopt a functional data analysis perspective on $\{Z_{ijt}\}$, modeling 
them separately through the hierarchical framework developed in Section \ref{subsec:standardmodel}. 
That is, $Z_{ij}(\tau)\mid \mu_j,\Sigma_j\stackrel{i.i.d.}{\sim} GP(\mu_j,\Sigma_j)$, 
for $i=1,\ldots,n$, and 
$\mu_j \mid \Sigma_j\stackrel{ind.}{\sim} GP(\mu_{0j}, (\nu_j -3) \Sigma_j)$, 
$\Sigma_j\stackrel{ind.}{\sim}IWP(\nu_j,\Psi_{\boldsymbol{\phi}_j})$, where 
$\boldsymbol{\phi}_j=\{\sigma^2_j,\rho_j\}$, for $j=1,\ldots,C-1$. The error terms are modeled as 
$\epsilon_{ijt} \mid \sigma^2_{\epsilon j} \stackrel{ind.}{\sim}N(0,\sigma^2_{\epsilon j})$. 
Hence, the hierarchical model for the data can be expressed as 
\begin{equation}
    \begin{split}
        &\mathbf{Y}_i|\{\mathbf{Z}_{ij}\},\{\boldsymbol{\epsilon}_{ij}\}\stackrel{ind.}{\sim} \prod_{t=1}^{T_i}\prod_{j=1}^{C-1}Bin(Y_{ijt}\mid m_{ijt},\varphi(Z_{ijt}+\epsilon_{ijt})),\quad i=1,\ldots,n,\\
        &\mathbf{Z}_{ij}\mid \mu_j(\boldsymbol{\tau}_i),\Sigma_j(\boldsymbol{\tau}_i,\boldsymbol{\tau}_i)\stackrel{ind.}{\sim} N(\mu_j(\boldsymbol{\tau}_i),\Sigma_j(\boldsymbol{\tau}_i,\boldsymbol{\tau}_i)),\quad \boldsymbol{\epsilon}_{ij}\mid \sigma_{\epsilon j}^2\stackrel{ind.}{\sim} 
        N(\mathbf{0},\sigma_{\epsilon j}^2 \, \mathbf{I}),\\
        & \boldsymbol{\mu}_j\mid\mu_{0j},\boldsymbol{\Sigma}_j,\nu_j \stackrel{ind.}{\sim} 
        N(\mu_{0j}\mathbf{1}, (\nu_j - 3) \boldsymbol{\Sigma}_j); \, \boldsymbol{\Sigma}_j\mid\nu_j,\boldsymbol{\Psi}_j\stackrel{ind.}{\sim} 
        IW(\nu_j,\boldsymbol{\Psi}_j), \, j=1,\ldots,C-1
    \end{split}
    \label{eq:fddsmulti}
\end{equation}
where $\mathbf{Y}_{i}=(\mathbf{Y}_{i1},\ldots,\mathbf{Y}_{iT_i})^{\top}$, 
$\mathbf{Z}_{ij}=(Z_{ij1},\ldots,Z_{ijT_i})^{\top}$, 
$\boldsymbol{\epsilon}_{ij}=(\epsilon_{ij1},\ldots,\epsilon_{ijT_i})^{\top}$, 
and the collection of the functional evaluations on the pooled grid $\boldsymbol{\tau}$ 
are denoted by the corresponding bold letter.

The structure in (\ref{eq:factmulti}) is referred to as the continuation-ratio logits representation 
of the multinomial distribution \citep{Tutz1991}. In the context of Bayesian nonparametric modeling,
it has been used as the kernel of nonparametric mixture models for cross-sectional ordinal 
regression \citep{KangKottas2022}.

Examining model properties reveals the practical utility of the continuation-ratio logits 
structure. The factorization in (\ref{eq:factmulti}) allows us to examine the probability response 
curves and the within subject covariance structure in the same fashion as for binary responses. 
Specifically, the continuation-ratio logit for response category $j$ is the logit of the conditional 
probability of response $j$, given that the response is $j$ or higher. As a consequence, for any 
finite grid $\boldsymbol{\tau}=(\tau_1,\ldots,\tau_T)^{\top}$, the probability response 
curves are given by
\begin{equation}
\begin{split}
\mathbf{P}_{\mathbf{j}\boldsymbol{\tau}}& = 
(\text{Pr}(Y_{\tau_1}=j\mid \mathbf{Z}_{\boldsymbol{\tau}},\sigma_{\epsilon}^2),
\ldots,\text{Pr}(Y_{\tau_T}=j\mid \mathbf{Z}_{\boldsymbol{\tau}},
\sigma_{\epsilon}^2))^{\top}\\
&=\text{E}\left( \boldsymbol{\pi}_{j \boldsymbol{\tau}}\mid \mathbf{Z}_{j 
\boldsymbol{\tau}},\sigma_{\epsilon j}^2 \right)
\prod_{k=1}^{j-1} \text{E}\left( (1-\boldsymbol{\pi}_{k \boldsymbol{\tau}})\mid \mathbf{Z}_{k \boldsymbol{\tau}},\sigma_{\epsilon k}^2 \right),
\end{split}
	\label{eq:probmult}
\end{equation}
where 
$\boldsymbol{\pi}_{j\boldsymbol{\tau}}=(\varphi(\boldsymbol{\mathcal{Z}}_{j1}),\ldots,\varphi(\boldsymbol{\mathcal{Z}}_{jT}))^{\top}$ 
and $\boldsymbol{\mathcal{Z}}_{j \boldsymbol{\tau}}\mid \mathbf{Z}_{j 
\boldsymbol{\tau}},\sigma_{\epsilon j}^2\sim N(\mathbf{Z}_{j \boldsymbol{\tau}},
\sigma_{\epsilon j}^2\mathbf{I}_T)$, for $j=1,\ldots,C$.
To avoid redundant expressions, we include the term $\boldsymbol{\pi}_{C\boldsymbol{\tau}}$ 
and set it always equal to 1. As for the covariance structure, we study the joint probability 
of the repeated measurements on the same subject at time $\tau$ and $\tau^{\prime}$ taking 
category $j$ and $j^{\prime}$. Exploiting the conditional independence structure across 
the categories, 
\begin{equation}
\begin{split}
	&\text{Pr}(Y_{\tau}=j,Y_{\tau^{\prime}}=j^{\prime}\mid\{\mathbf{Z}_{\boldsymbol{j \tau}}\},\{\sigma_{\epsilon j}^2\})\\
	&=\left\{\begin{aligned} & \text{E}(\pi_{j\tau}\pi_{j\tau^{\prime}}\mid \mathbf{Z}_{j \boldsymbol{\tau}},\sigma_{\epsilon j}^2)\prod_{k\neq j}\text{E}[(1-\pi_{k\tau})(1-\pi_{k\tau^{\prime}})\mid \mathbf{Z}_{k \boldsymbol{\tau}},\sigma_{\epsilon k}^2] & j=j^{\prime}\\ & \text{E}[\pi_{j\tau}(1-\pi_{j\tau^{\prime}})\mid \mathbf{Z}_{j \boldsymbol{\tau}},\sigma_{\epsilon j}^2] \, 
 \text{E}[(1-\pi_{j^{\prime}\tau})\pi_{j^{\prime}\tau^{\prime}}\mid \mathbf{Z}_{j^{\prime} \boldsymbol{\tau}},\sigma_{\epsilon j^{\prime}}^2] & \\
 &\times \prod_{k\neq j,j^{\prime}}\text{E}[(1-\pi_{k\tau})(1-\pi_{k\tau^{\prime}})\mid \mathbf{Z}_{k \boldsymbol{\tau}},\sigma_{\epsilon k}^2] & j\neq j^{\prime} \end{aligned}\right..
\end{split}
	\label{eq:jointprobmult}
\end{equation}  
Hence, we can explore the covariance of the two ordinal responses 
$\mathbf{Y}_{\tau},\mathbf{Y}_{\tau^{\prime}}$ by studying the pairwise covariance 
for each entry.

The continuation-ratio logits structure is also key to efficient model implementation. It implies 
a sequential mechanism, such that the ordinal response is determined through a sequence of binary 
outcomes. Starting from the lowest category, each binary outcome indicates whether the ordinal 
response belongs to that category or to one of the higher categories. This mechanism inspires 
a novel perspective on the model implementation. That is, we can re-organize the original data 
set containing longitudinal ordinal responses to create $C-1$ data sets with longitudinal 
binary outcomes. Then, fitting model (\ref{eq:fddsmulti}) to the original data set is equivalent 
to fitting the model of Section \ref{subsec:standardmodel} separately on the $C-1$ re-organized 
data sets. The procedure is elaborated below.

Denote the set of all possible subject and time indices by $\boldsymbol{\mathcal{I}}_1$, 
that is, $\boldsymbol{\mathcal{I}}_1=\{(i,t):i=1,\ldots,n, \, t=1,\ldots,T_i\}$. To build the first 
re-organized data set with binary outcomes, we create binary indicators $Y^{(1)}_{it}$, such 
that $Y^{(1)}_{it}=1$ if $Y_{i1t}=1$ and $Y^{(1)}_{it}=0$ if $Y_{i1t}=0$. The first 
data set is then $\boldsymbol{\mathcal{D}}_1 = \{Y^{(1)}_{it}: (i,t)\in\boldsymbol{\mathcal{I}}_1\}$.  
Moving to the second data set, we first filter out the observations that are already 
categorized into the smallest scale, and denote the remaining indices set 
by $\boldsymbol{\mathcal{I}}_2=\boldsymbol{\mathcal{I}}_1\setminus\{(i,t):Y_{i1t}=1\}$. 
This is the set of indices with original ordinal responses belonging to categories higher 
than or equal to the second smallest scale. Then, we create new binary indicators $Y^{(2)}_{it}$, 
such that $Y^{(2)}_{it}=1$ if $Y_{i2t}=1$, and $Y^{(2)}_{it}=0$ if $Y_{i2t}=0$. 
The second data set is obtained as 
$\boldsymbol{\mathcal{D}}_2 = \{Y^{(2)}_{it}: (i,t)\in\boldsymbol{\mathcal{I}}_2\}$. 
The process is continued until we obtain the $(C-1)$-th data set, 
$\boldsymbol{\mathcal{D}}_{C-1}=\{Y^{(C-1)}_{it}: (i,t)\in\boldsymbol{\mathcal{I}}_{C-1}\}$, 
where $\boldsymbol{\mathcal{I}}_{C-1}$ is the indices set such that the original ordinal 
responses belong to either category $C-1$ or $C$. Notice that every re-organized data 
set $\boldsymbol{\mathcal{D}}_{j}$, for $j=1,\ldots,C-1$, contains longitudinal binary 
outcomes for which the model of Section \ref{subsec:standardmodel} is directly applicable. 
Provided the priors placed on each ordinal response category's parameters are independent,
it is straightforward to verify that fitting separately the model for binary responses to the 
re-organized data sets $\{\boldsymbol{\mathcal{D}}_{j}: j=1,\ldots,C-1\}$ is
equivalent to fitting model (\ref{eq:fddsmulti}) to the original data set.
We formalize the conclusion in the following proposition.

\begin{proposition}
\label{prop:fitseparate}
Fitting the ordinal responses model in (\ref{eq:fddsmulti}) is equivalent to fitting 
the model for binary responses separately, $C-1$ times to the data sets 
$\{\boldsymbol{\mathcal{D}}_{j}: j=1,\ldots,C-1\}$.
\end{proposition}

Based on Proposition \ref{prop:fitseparate}, the posterior simulation algorithm for the ordinal 
responses model can be parallelized and implemented on separate cores. In applications where 
the number of response categories is moderate to large, such a parallel computing scheme 
is especially beneficial. Also, since the binary responses model serves as the backbone for 
modeling ordinal responses, the prior specification strategy and the posterior simulation 
method described in Section \ref{subsec:modelapply} can be readily extended to 
model (\ref{eq:fddsmulti}). Finally, from (\ref{eq:probmult}) and (\ref{eq:jointprobmult}), 
it is clear that the posterior samples obtained from the $C-1$ separate models suffice to 
obtain full posterior inference for the ordinal response process.

\subsection{Data illustration}
\label{subsec:orddataexample}

As an illustration example, we consider the PAM arousal score on the original scale, which is 
obtained from the same EMA study discussed in Section \ref{sec:realapp}. PAM arousal is a $-2$ 
to 2 (excluding 0) score. We examine the same cohort of students on the same study period as 
described in Section \ref{subsec:datarealapp}. Over all observations, the distribution of 
arousal scores involves $16.6\%$ for level $-2$, $27.7\%$ for level $-1$, $12.6\%$ for 
level 1, and $12\%$ for level 2, while $31.1\%$ of the observations are missing.

To implement model (\ref{eq:fddsmulti}), we follow the procedure outlined above 
Proposition \ref{prop:fitseparate}. We re-organize the original data into separate data sets 
$\{\boldsymbol{\mathcal{D}_j}: j=1,\ldots,3\}$, each of them containing the binary responses 
indicating whether the arousal scores are at level $j$ or a higher level. Then, the proposed 
model is fitted to the three data sets in parallel. 

\begin{figure}[t!]
\centering
\includegraphics[width=16cm,height=10cm]{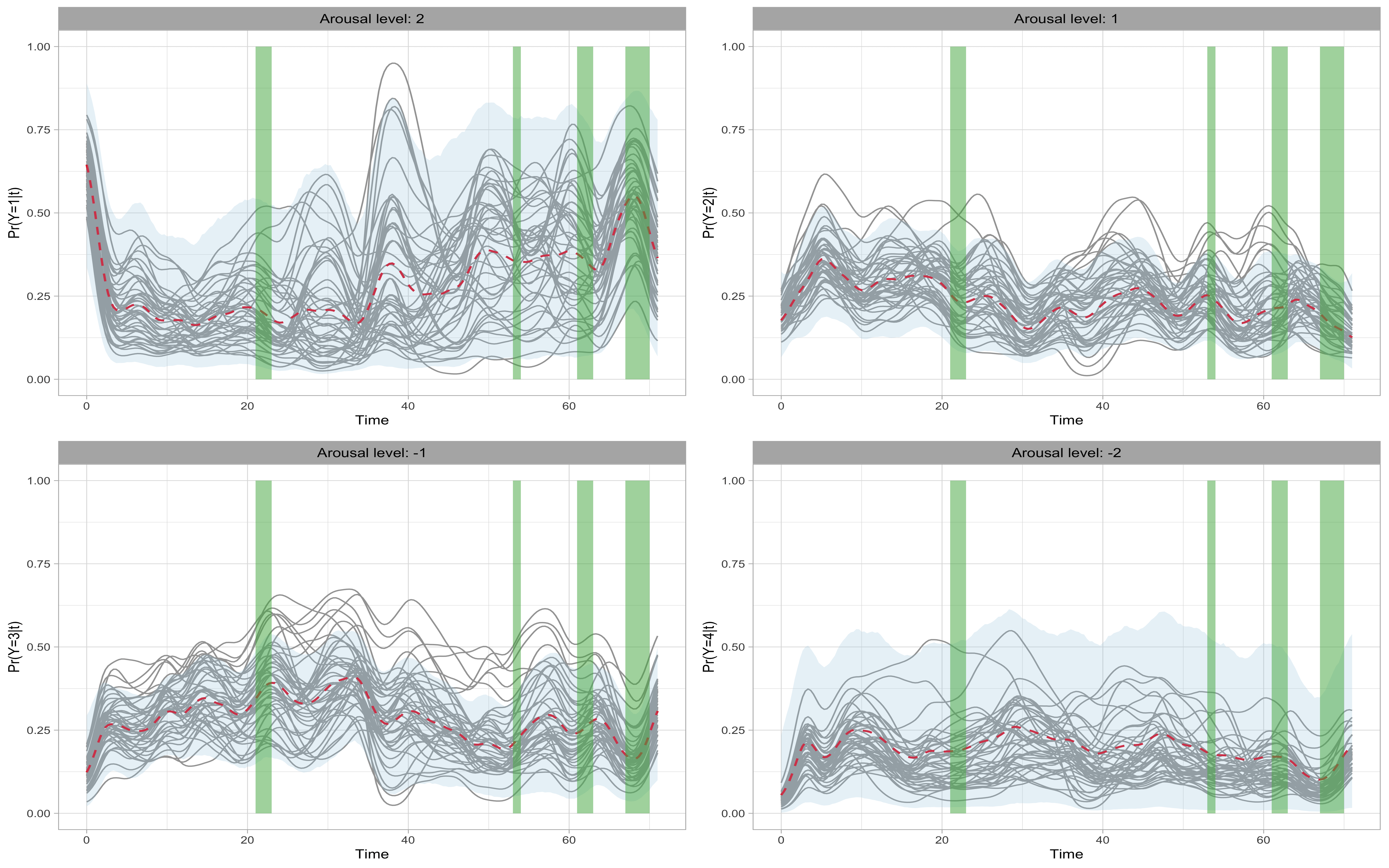}
\caption{Four levels arousal score data. Posterior mean (dashed line) and 95\% interval 
estimate (shaded region) of probability response curve for an out-of-sample subject. 
The posterior mean estimates for the probability response curves of in-sample subjects 
are given by the solid lines. The vertical shaded regions correspond to the four 
special time periods (see Section \ref{subsec:datarealapp}).}
\label{fig:quadaroprob}
\end{figure}

The primary inference focus is on the change of arousal scores as the term progresses, which 
is depicted by the probability response curve of each response level. We display posterior 
point and interval estimates of $\mathbf{P}_{\mathbf{j}\boldsymbol{\tau}}$ (defined 
in (\ref{eq:probmult})) in Figure \ref{fig:quadaroprob}. The probability of the highest arousal 
level drops dramatically as the term begins, indicating that the excitement of a new quarter 
may vanish within a week. The Boston marathon bombing slightly triggers higher probability 
for moderately low to low arousal level. There is a drop of the probability for moderately 
high to high arousal level after the Green Key festival and the Memorial Day holiday. 
The exams may have a significant impact on the arousal level. We observe peaks of arousal 
at the beginning of the final exam period, and also the middle of the term, which corresponding 
to the midterm exam period. Since the students are taking different courses, the midterm exam 
times vary, resulting in some curves with lead or lag peaks compared to the majority. This 
pattern is not clear in the analysis of binary arousal scores. Hence, examining the finer 
ordinal scale enables us to discover subtle changes of the students activation states.  
We have also investigated the temporal covariance structure of the ordinal responses, with 
details presented in the Supplementary Material.  

\section{Discussion}
\label{sec:summary}

We have developed a novel Bayesian hierarchical model for analyzing longitudinal binary 
data. We approach the problem from a functional data analysis perspective, resulting in 
a method that is suitable for either regularly or irregularly spaced longitudinal data. 
The modeling approach achieves flexibility and computational efficiency in full posterior 
inference. With regard to the former, the key model feature is the joint and nonparametric 
modeling of the mean and covariance structure. As illustrated by the data application, 
our approach enables interpretable inference with coherent uncertainty quantification, 
and provides improvement over the GLMM approach.  
The model formulation enables a natural extension to incorporate ordinal responses, 
which is accomplished by leveraging the continuation-ratio logits representation of 
the multinomial distribution. This representation leads to a factorization of the 
multinomial model into separate binomial models, on which the modeling approach for 
binary responses can be applied. The computational benefit is retained, since we can 
utilize parallel computing across response categories. 

The proposed methodology for modeling longitudinal binary and ordinal responses can be elaborated 
in different directions. We have focused on stationary specifications for the hierarchical GP 
prior. Nonstationary model components can be incorporated through a time-varying mean function 
$\mu_0$ and/or a nonstationary covariance kernel $\Psi_{\boldsymbol{\phi}}$. Moreover, 
longitudinal studies typically have predetermined covariates associated with each subject, 
or time-varying covariates corresponding to each observation. The predetermined covariates 
can be incorporated in the model through the prior placed on the mean function of the signal
process. Using the functional linear model may be a possible strategy for the more challenging 
task of accounting for time-varying covariates.

%
%

\section*{Supplementary material}

The Supplementary Material includes: details for the MCMC algorithm; proofs of the propositions; 
the simulation study; 
results from MCMC diagnostics and prior sensitivity analysis;
and, additional results for the real data examples. 



\section*{Acknowledgements}
The research was supported in part by the National Science Foundation under award SES 1950902.
The authors wish to thank two reviewers for several useful comments.

\section*{Declarations}
{\bf Conflict of interest} The authors have no conflicts of interest to declare that are 
relevant to the content of this article.




\bibliographystyle{jasa3}
\bibliography{sample}

\pagebreak
\vspace*{50pt}
\begin{center}
\textbf{\LARGE Supplementary Material: Flexible Bayesian
Modeling for Longitudinal Binary and
Ordinal Responses}
\end{center}
\bigskip
\bigskip
\bigskip
\bigskip
\setcounter{equation}{0}
\setcounter{figure}{0}
\setcounter{table}{0}
\setcounter{page}{1}
\setcounter{section}{0}
\makeatletter
\renewcommand{\theequation}{S\arabic{equation}}
\renewcommand{\thefigure}{S\arabic{figure}}
\renewcommand{\thetable}{S\arabic{table}}
\renewcommand{\thesection}{S\arabic{section}}
\renewcommand{\bibnumfmt}[1]{[S#1]}
\renewcommand{\citenumfont}[1]{S#1}

\section{MCMC posterior simulation details}
\label{sec:smmcmcdetail}

Based on the joint posterior distributions derived from (\ref{eq:jointpostmodelbin}), we design the MCMC sampling algorithm for the proposed model with binary responses. This process can be achieved entirely with Gibbs updates, by iterating the following steps. For notation simplicity, we let $(\phi\mid -)$ denote the posterior full conditional distribution for parameter $\phi$. 

\begin{description}
   \item[Step 1:] For $i=1,\ldots,n$ update $\boldsymbol{\mathcal{Z}}_i$ from $N(\mathbf{m}_i,\boldsymbol{\mathcal{V}}_i)$, where $\boldsymbol{\mathcal{V}}_i=(\Omega_i+(1/\sigma_{\epsilon}^2)\mathbf{I})^{-1}$, and $\mathbf{m}_i=\boldsymbol{\mathcal{V}}_i(\boldsymbol{\lambda}_i+(1/\sigma_{\epsilon}^2)\mathbf{Z}_i)$. Here $\Omega_i$ denote the diagonal matrix of $\boldsymbol{\xi}_i$, and $\boldsymbol{\lambda}_i=(Y_{i1}-1/2,\ldots,Y_{iT_i}-1/2)^{\top}$.   
   
   \item[Step 2:] Update the Pólya-Gamma random variables $\xi_{it}$ by sample from $PG(1,\mathcal{Z}_{it})$, for $i=1,\ldots,n$ and $t=1,\ldots,T_i$. 
   
   \item[Step 3:]  Update $\sigma_{\epsilon}^2$ by sample from $IG(a_{\epsilon}+\sum_{i=1}^nT_i/2,b_{\epsilon}+\sum_{i=1}^n(\boldsymbol{\mathcal{Z}}_i-\mathbf{Z}_i)^{\top}(\boldsymbol{\mathcal{Z}}_i-\mathbf{Z}_i)/2)$.
   
   \item[Step 4:] Update $\tilde{\mathbf{Z}}_i$ for $i=1,\ldots,n$,
   \begin{itemize}
   	\item In the case that all the subjects having observations on a common grid, $\mathbf{Z}_i^*$ vanishes and $\tilde{\mathbf{Z}}_i=\mathbf{Z}_i$. It has full conditional distribution $\mathbf{Z}_i\mid -\sim N(\tilde{\boldsymbol{\mu}}_i,\tilde{\mathbf{V}}_i)$, where $\tilde{\mathbf{V}}_i=((1/\sigma_{\epsilon}^2)\mathbf{I}+\boldsymbol{\Sigma}^{-1})^{-1}$, and $\tilde{\boldsymbol{\mu}}_i=\tilde{\mathbf{V}}_i((1/\sigma_{\epsilon}^2)\boldsymbol{\mathcal{Z}_i}+\boldsymbol{\Sigma}^{-1}\boldsymbol{\mu})$.
   
   	\item  In the case that the repeated measurements for the subjects are collected on uncommon grids, we first update $\mathbf{Z}_i^*$ from $N(\boldsymbol{\mu}_i^*,\boldsymbol{V}_i^*)$, where
   	\begin{equation*}
\begin{split}
&\boldsymbol{\mu}_i^*=\mu(\boldsymbol{\tau}^*_i)+\Sigma(\boldsymbol{\tau}^*_i,\boldsymbol{\tau}_i)\Sigma(\boldsymbol{\tau}_i,\boldsymbol{\tau}_i)^{-1}(\mathbf{Z}_i-\mu(\boldsymbol{\tau}_i))=\mathbf{B}_i\mathbf{Z}_i-\mathbf{u}_i\\
&\boldsymbol{V}_i^*=\Sigma(\boldsymbol{\tau}^*_i,\boldsymbol{\tau}^*_i)-\Sigma(\boldsymbol{\tau}^*_i,\boldsymbol{\tau}_i)\Sigma(\boldsymbol{\tau}_i,\boldsymbol{\tau}_i)^{-1}\Sigma(\boldsymbol{\tau}_i,\boldsymbol{\tau}^*_i)
\end{split}
\end{equation*}
with $\mathbf{B}_i=\Sigma(\boldsymbol{\tau}^*_i,\boldsymbol{\tau}_i)\Sigma(\boldsymbol{\tau}_i,\boldsymbol{\tau}_i)^{-1}$ and $\mathbf{u}_i=\mathbf{B}_i\mu(\boldsymbol{\tau}_i)-\mu(\boldsymbol{\tau}^*_i)$.

Then, to update $\mathbf{Z}_i$, we sample from $N(\tilde{\boldsymbol{\mu}}_i,\tilde{\mathbf{V}}_i)$, where
\begin{equation*}
\begin{split}
&\tilde{\mathbf{V}}_i=[(1/\sigma_{\epsilon}^2)\mathbf{I}+\Sigma(\boldsymbol{\tau_i},\boldsymbol{\tau_i})^{-1}+\mathbf{B}_i^T(\boldsymbol{V}_i^*)^{-1}\mathbf{B}_i]^{-1}\\
&\tilde{\boldsymbol{\mu}}_i=\tilde{\mathbf{V}}_i[(1/\sigma_{\epsilon}^2)\boldsymbol{\mathcal{Z}}_i+\Sigma(\boldsymbol{\tau_i},\boldsymbol{\tau_i})^{-1}\mu(\mathbf{\tau}_i)+\mathbf{B}_i^T(\boldsymbol{V}_i^*)^{-1}(\mathbf{u}_i+\mathbf{Z}_i^*)]
\end{split}
\end{equation*}
   \end{itemize}
 
   \item[Step 5:] Update $\boldsymbol{\mu}$ and $\boldsymbol{\Sigma}$ jointly by sample from $N(\boldsymbol{\mu}^*,\boldsymbol{\Sigma}/\kappa^*)$ and $IW(\nu^*,\boldsymbol{\Psi}^*)$, respectively, with
   \begin{equation*}
\begin{split}
&\boldsymbol{\mu}^*=\frac{\kappa}{\kappa+n}\boldsymbol{\mu}_0+\frac{n}{\kappa+n}\tilde{\mathbf{Z}}^m,\quad
\kappa^*=n+\kappa,\quad \nu^*=n+\nu\\
&\boldsymbol{\Psi}^*=\boldsymbol{\Psi}+S+\frac{n\kappa}{n+\kappa}(\tilde{\mathbf{Z}}^m-\boldsymbol{\mu}_0)(\tilde{\mathbf{Z}}^m-\boldsymbol{\mu}_0)^T,\quad
S=\sum^{n}_{i=1}(\tilde{\mathbf{Z}}_i-\tilde{\mathbf{Z}}^m)(\tilde{\mathbf{Z}}_i-\tilde{\mathbf{Z}}^m)^T
\end{split}
\end{equation*}
 
   \item[Step 6:] Update $\mu_0$ from $N(a_{\mu}^*,b_{\mu}^*)$, where $b_{\mu}^*=[\mathbf{1}^T\boldsymbol{\Sigma}^{-1}\mathbf{1}+\frac{1}{b_{\mu}}]^{-1}$, and $a_{\mu}^*=b_{\mu}^*[\mathbf{1}^T\boldsymbol{\Sigma}^{-1}\boldsymbol{\mu}+\frac{a_{\mu}}{b_{\mu}}]$.
   
   \item[Step 7:] Update $\sigma^2$ from $\text{Gamma}(a_{\sigma}+\frac{(\nu+|\boldsymbol{\tau}|-1)|\boldsymbol{\tau}|}{2},b_{\sigma}+\frac{1}{2}tr(\boldsymbol{\Psi}_{\rho}\boldsymbol{\Sigma}^{-1}))$. Here $\boldsymbol{\Psi}_{\rho}$ denotes the correlation matrix $\boldsymbol{\Psi}_{\boldsymbol{\phi}}/\sigma^2$.
   
   \item [Step 8:] Using the Griddy-Gibbs sampler by \citet{Ritter1992}, update $\rho$ from 
   \begin{equation*}
   	P(\rho=g_l\mid -)=\frac{|\boldsymbol{\Psi}_{g_l}|^{(\nu+|\boldsymbol{\tau}|-1)/2}\exp(-\frac{1}{2}tr(\boldsymbol{\Psi}_{g_l}\boldsymbol{\Sigma}^{-1}))}{\sum_{l=1}^G|\boldsymbol{\Psi}_{g_l}|^{(\nu+|\boldsymbol{\tau}|-1)/2}\exp(-\frac{1}{2}tr(\boldsymbol{\Psi}_{g_l}\boldsymbol{\Sigma}^{-1}))}
   \end{equation*}
   where $g_1,\ldots,g_G$ are grid points and $\boldsymbol{\Psi}_{g_l}$ denotes the correlation matrix when $\rho$ taking the value $g_l$.
   
   \item [Step 9:] Using the Griddy-Gibbs sampler, update $\nu$ from
   \begin{equation*}
   	P(\nu=g_l\mid -)=\frac{N(\boldsymbol{\mu}\mid\boldsymbol{\mu}_0,(g_l-3)\boldsymbol{\Sigma})IW(\boldsymbol{\Sigma}\mid g_l+|\boldsymbol{\tau}|-1,\boldsymbol{\Psi}_{\boldsymbol{\phi}})}{\sum_{l=1}^GN(\boldsymbol{\mu}\mid\boldsymbol{\mu}_0,(g_l-3)\boldsymbol{\Sigma})IW(\boldsymbol{\Sigma}\mid g_l+|\boldsymbol{\tau}|-1,\boldsymbol{\Psi}_{\boldsymbol{\phi}})}
   \end{equation*}

 \end{description}

\section{Proofs}

\subsection*{Proof of Proposition \ref{prop:meancovcondsignal}}
\label{subsec:proofmeancovcond}

\begin{proof}
Firstly, the probability response curve can be obtained by 
\begin{equation}
    \begin{split}
       \text{Pr}(\mathbf{Y}_{\boldsymbol{\tau}}=\mathbf{1}\mid\mathbf{Z}_{\boldsymbol{\tau}},\sigma_{\epsilon}^2)&=\int p(\mathbf{Y}_{\boldsymbol{\tau}}=\mathbf{1}\mid\boldsymbol{\mathcal{Z}}_{\boldsymbol{\tau}},\mathbf{Z}_{\boldsymbol{\tau}},\sigma_{\epsilon}^2)p(\boldsymbol{\mathcal{Z}}_{\boldsymbol{\tau}}\mid\mathbf{Z}_{\boldsymbol{\tau}},\sigma_{\epsilon}^2)d\boldsymbol{\mathcal{Z}}_{\boldsymbol{\tau}}\\
       &=\int \varphi(\boldsymbol{\mathcal{Z}}_{\boldsymbol{\tau}})N(\boldsymbol{\mathcal{Z}}_{\boldsymbol{\tau}}\mid\mathbf{Z}_{\boldsymbol{\tau}},\sigma_{\epsilon}^2\mathbf{I})d\boldsymbol{\mathcal{Z}}_{\boldsymbol{\tau}}=\text{E}(\varphi(\boldsymbol{\mathcal{Z}}_{\boldsymbol{\tau}})\mid \mathbf{Z}_{\boldsymbol{\tau}},\sigma_{\epsilon}^2).
    \end{split}
    \label{eq:probcurvecondsignal}
\end{equation}

Then, to find the diagonal and off-diagonal elements for the covariance matrix of $\mathbf{Y}_{\boldsymbol{\tau}}$, we use the law of total variance/covariance. For the diagonal elements,
\begin{equation}
    \begin{split}
        \text{Var}(Y_{\tau}\mid\mathbf{Z}_{\boldsymbol{\tau}},\sigma_{\epsilon}^2)&=\text{Var}[\text{E}(Y_{\tau}\mid \mathcal{Z}_{\boldsymbol{\tau}})\mid\mathbf{Z}_{\boldsymbol{\tau}},\sigma_{\epsilon}^2]+\text{E}[\text{Var}(Y_{\tau}\mid \mathcal{Z}_{\boldsymbol{\tau}})\mid\mathbf{Z}_{\boldsymbol{\tau}},\sigma_{\epsilon}^2]\\
        &=\text{Var}[\varphi(\mathcal{Z}_{\tau})\mid\mathbf{Z}_{\boldsymbol{\tau}},\sigma_{\epsilon}^2]+\text{E}[\varphi(\mathcal{Z}_{\tau})(1-\varphi(\mathcal{Z}_{\tau}))\mid \mathbf{Z}_{\boldsymbol{\tau}},\sigma_{\epsilon}^2]\\
        &=\text{E}[\varphi(\mathcal{Z}_{\tau})\mid \mathbf{Z}_{\boldsymbol{\tau}},\sigma_{\epsilon}^2]-\text{E}^2[\varphi(\mathcal{Z}_{\tau})\mid \mathbf{Z}_{\boldsymbol{\tau}},\sigma_{\epsilon}^2].
    \end{split}
    \label{eq:varcondsignal}
\end{equation}
As for the off-diagonal entries, similarly,
\begin{equation}
    \begin{split}
        \text{Cov}(Y_{\tau},Y_{\tau^{\prime}}\mid\mathbf{Z}_{\boldsymbol{\tau}},\sigma_{\epsilon}^2)&=\text{Cov}[\text{E}(Y_{\tau}\mid \boldsymbol{\mathcal{Z}}_{\boldsymbol{\tau}}),\text{E}(Y_{\tau^{\prime}}\mid \boldsymbol{\mathcal{Z}}_{\boldsymbol{\tau}})\mid\mathbf{Z}_{\boldsymbol{\tau}},\sigma_{\epsilon}^2]+\text{E}[\text{Cov}(Y_{\tau},Y_{\tau^{\prime}}\mid \boldsymbol{\mathcal{Z}}_{\boldsymbol{\tau}})\mid\mathbf{Z}_{\boldsymbol{\tau}},\sigma_{\epsilon}^2]\\
        &=\text{Cov}[\varphi(\mathcal{Z}_{\tau}),\varphi(\mathcal{Z}_{\tau^{\prime}})\mid\mathbf{Z}_{\boldsymbol{\tau}},\sigma_{\epsilon}^2]
    \end{split}
    \label{eq:covcondsignal}
\end{equation}
\end{proof}

\subsection*{Proof of Proposition \ref{prop:deltaapproxlogitnormal}}
\label{subsec:proofdeltaapprox}

\begin{proof}
Write $\mathbf{Z}=\boldsymbol{\mu}+\boldsymbol{\zeta}$, where $\boldsymbol{\zeta}\sim N(0,\boldsymbol{\Sigma})$.
By Taylor expansion around the mean, 
\begin{equation}
    \varphi(Z_i)\approx \varphi(\mu_i)+\zeta_i\varphi^{\prime}(\mu_i)+\frac{\zeta^2}{2}\varphi^{\prime\prime}(\mu_i).
\label{eq:taylorlogitnomralmean}
\end{equation}
Then taking expectation yields $\text{E}(\varphi(Z_i))\approx \varphi(\mu_i)+\frac{\sigma^2}{2}\varphi^{\prime\prime}(\mu_i)$, $i=1,2$.

The expectation of $\varphi^2(Z_i)$ can be derived using the same technique,
\begin{equation}
    \varphi^2(Z_i)\approx \varphi^2(\mu_i)+2\zeta_i\varphi(\mu_i)\varphi^{\prime}(\mu_i)+\zeta_i^2[(\varphi^{\prime}(\mu_i))^2+\varphi(\mu_i)\varphi^{\prime\prime}(\mu_i)].
    \label{eq:taylorlogitnormalvar}
\end{equation}
Taking expectation with respect to $\zeta_i$ again, we arrive at the result. 

As for $E(\varphi(Z_1)\varphi(Z_2))$, consider the function $f(\mathbf{Z})=\varphi(Z_1)\varphi(Z_2)$, using the bivariate version of Taylor expansion,
\begin{equation}
f(\mathbf{Z})\approx f(\boldsymbol{\mu})+\bigtriangledown f(\boldsymbol{\mu})^{\top}\boldsymbol{\zeta}+\frac{1}{2}\boldsymbol{\zeta}^{\top}\bigtriangledown^2f(\boldsymbol{\mu})\boldsymbol{\zeta}.
\label{eq:taylorlogitnormalcov}    
\end{equation}
Similarly, taking expectation with respect to $\boldsymbol{\zeta}$, we can obtain the result. 

\end{proof}

\subsection*{Proof of Proposition \ref{prop:marginalsignal}}
\label{subsec:proofmarginalsignal}

\begin{proof}
The result is proved by considering the corresponding f.d.d.s. on any finite grid 
$\boldsymbol{\tau}$. Let the bold letter denote the corresponding process evaluated 
at $\boldsymbol{\tau}$. From the model assumption, 
\begin{equation}
    p(\mathbf{Z})=\int\int p(\mathbf{Z}\mid \boldsymbol{\mu},\boldsymbol{\Sigma})p(\boldsymbol{\mu}\mid \boldsymbol{\Sigma})p(\boldsymbol{\Sigma})d\boldsymbol{\mu}d\boldsymbol{\Sigma}
    \label{eq:marginalsignal}
\end{equation}
We first notice that marginalizing over its mean $\boldsymbol{\mu}$,  $\mathbf{Z}\sim N(\boldsymbol{\mu}_0,(\nu-2)\boldsymbol{\Sigma})$. Based on that, 
\begin{equation}
    \begin{split}
        p(\mathbf{Z})&=\int p(\mathbf{Z}\mid \boldsymbol{\Sigma})p(\boldsymbol{\Sigma})d\boldsymbol{\Sigma}\\
        &\propto \int \frac{\exp\{-\frac{1}{2}\text{Tr}[(\boldsymbol{\Psi}_{\boldsymbol{\phi}}+\frac{(\mathbf{Z}-\boldsymbol{\mu}_0)(\mathbf{Z}-\boldsymbol{\mu}_0)^{\top}}{\nu-2})\boldsymbol{\Sigma}^{-1}]\}}{|\boldsymbol{\Sigma}|^{(\nu+|\boldsymbol{\tau}|+1)/2}}d\boldsymbol{\Sigma}\\
        &\propto [1+\frac{(\mathbf{Z}-\boldsymbol{\mu}_0)^{\top}\boldsymbol{\Psi}_{\boldsymbol{\phi}}^{-1}(\mathbf{Z}-\boldsymbol{\mu}_0)}{\nu-2}]^{-(\nu+|\boldsymbol{\tau}|)/2},
   \end{split}
   \label{eq:marginalizescalematrix}
\end{equation}
which can be recognized as the kernel of a MVT distribution, thus providing the result.

\end{proof}

\section{Synthetic data examples}
\label{sec:simstudy}

The principal goal of analyzing longitudinal data is to estimate the mean and covariance structure 
of the subject's repeated measurements. We conduct simulation studies to evaluate the proposed 
method on fulfilling this goal. In particular, Section \ref{subsec:simmean} evaluates 
the proposed model's capacity to capture the fluctuation of the mean structure, and 
Section \ref{subsec:simcov} explores its performance in estimating within subject covariance structure. 
To address a comment from a reviewer, in Section \ref{subsec:irregular}, we evaluate 
model performance on a scenario where the observations are made on irregular time points.
Unless otherwise specified, the posterior analyses in this section are based on 5000 posterior 
samples collected every 4 iterations from a Markov chain of 30000 iterations, with the first 
10000 samples being discarded.

\subsection{Estimating mean structure}
\label{subsec:simmean}

Consider a generic process of generating longitudinal binary responses,
\begin{equation}
\begin{split}
    &\mathbf{Y}_i = Y_i(\boldsymbol{\tau}_{i}) \mid \mathcal{Z}_i(\boldsymbol{\tau}_{i})
    \stackrel{i.i.d.}{\sim} Bin(1,\eta(\mathcal{Z}_i(\boldsymbol{\tau}_{i}))),
    \quad \boldsymbol{\tau}_i=(\tau_{i1},\ldots,\tau_{iT_i}),\quad i=1,\ldots,n,\\
    & \mathcal{Z}_i(\boldsymbol{\tau}_{i})=\boldsymbol{\mathcal{Z}}_i = f(\boldsymbol{\tau}_i)+\boldsymbol{\omega}_i+\boldsymbol{\epsilon}_i\quad \boldsymbol{\epsilon}_i\stackrel{i.i.d.}{\sim} N(\mathbf{0},\sigma_{\epsilon}^2\mathbf{I}),
    \end{split}    
    \label{eq:datagensim}
\end{equation}
where $\eta(\cdot)$ is a generic link function mapping $\mathbb{R}$ to $(0,1)$, $f(\tau)$ is a 
signal function, and 
$\boldsymbol{\omega}_i$ is a realization from a mean zero continuous stochastic process that depicts the temporal covariance within subject. The objective is twofold. First, to estimate the subject's probability response curve, which is defined as the probability of obtaining positive response, as a function of time. Second, to estimate the true underlying signal function.

We consider three data generating processes. The specific choice of $\eta(\cdot)$, $f(\tau)$ and $\boldsymbol{\omega}_i$ for each generating process is summarized as follows:
\begin{itemize}
\item Case 1: $\eta_1(\cdot)=\varphi(\cdot)$, where $\varphi(\cdot)$ is the expit function, $f_1(\tau)=0.3+3\sin(0.5\tau)+\cos(\tau/3)$, and $\boldsymbol{\omega}_i\stackrel{i.i.d.}{\sim} N(\mathbf{0},K_1(\boldsymbol{\tau},\boldsymbol{\tau}))$, with covariance kernel $K_1(\tau_{t},\tau_{t^{\prime}})=\exp(-|\tau_t-\tau_{t^{\prime}}|^2)$.
\item Case 2: $\eta_2(\cdot)=\Phi(\cdot)$, where $\Phi(\cdot)$ denotes the CDF of standard normal distribution, $f_2(\tau)=0.1+2\sin(0.25\tau)+\cos(0.25\tau)$, and $\boldsymbol{\omega}_i\stackrel{i.i.d.}{\sim} MVT(5,\mathbf{0},K_2(\boldsymbol{\tau},\boldsymbol{\tau}))$, with covariance kernel $K_2(\tau_{t},\tau_{t^{\prime}})=\frac{1}{3}\exp(-|\tau_t-\tau_{t^{\prime}}|^2)$.
\item Case 3: a mixture of Case 1 and Case 2, with equal probability of generating data from each model. 
\end{itemize}

For $n=30$ subjects, we simulate $T=31$ binary observations at time $\tau=0,\ldots,30$, following the aforementioned data generating processes. To enforce an unbalanced study design, we randomly drop out a proportion of the simulated data. We term the drop out proportion sparsity level, for which we consider $10\%$, $25\%$ and $50\%$. 

\begin{figure}[t!]
\centering
\begin{subfigure}{\textwidth}
  \centering 
  \includegraphics[width=16cm,height=3.2cm]{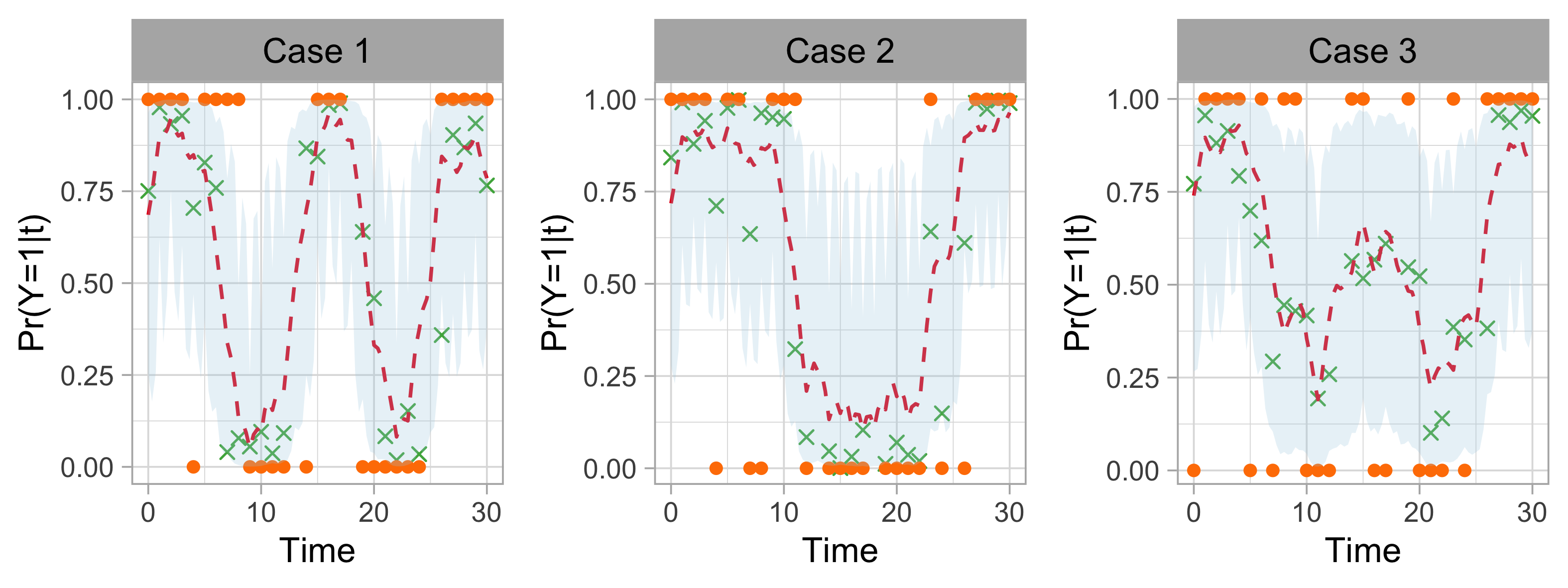}  
  \caption{Sparsity level at $10\%$.}
  \label{subfig:signalprobcurve90}
\end{subfigure}
\begin{subfigure}{\textwidth}
  \centering 
  \includegraphics[width=16cm,height=3.2cm]{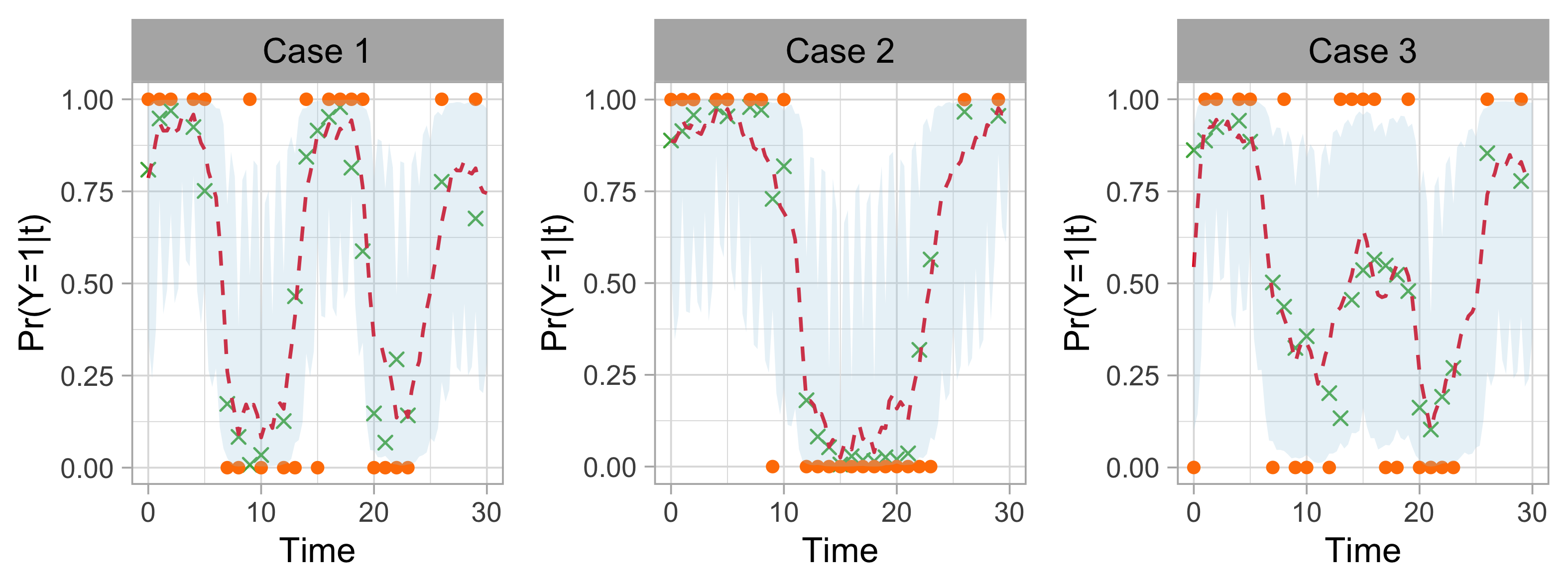}  
  \caption{Sparsity level at $25\%$.}
  \label{subfig:signalprobcurve75}
\end{subfigure}
\begin{subfigure}{\textwidth}
  \centering
  \includegraphics[width=16cm,height=3.2cm]{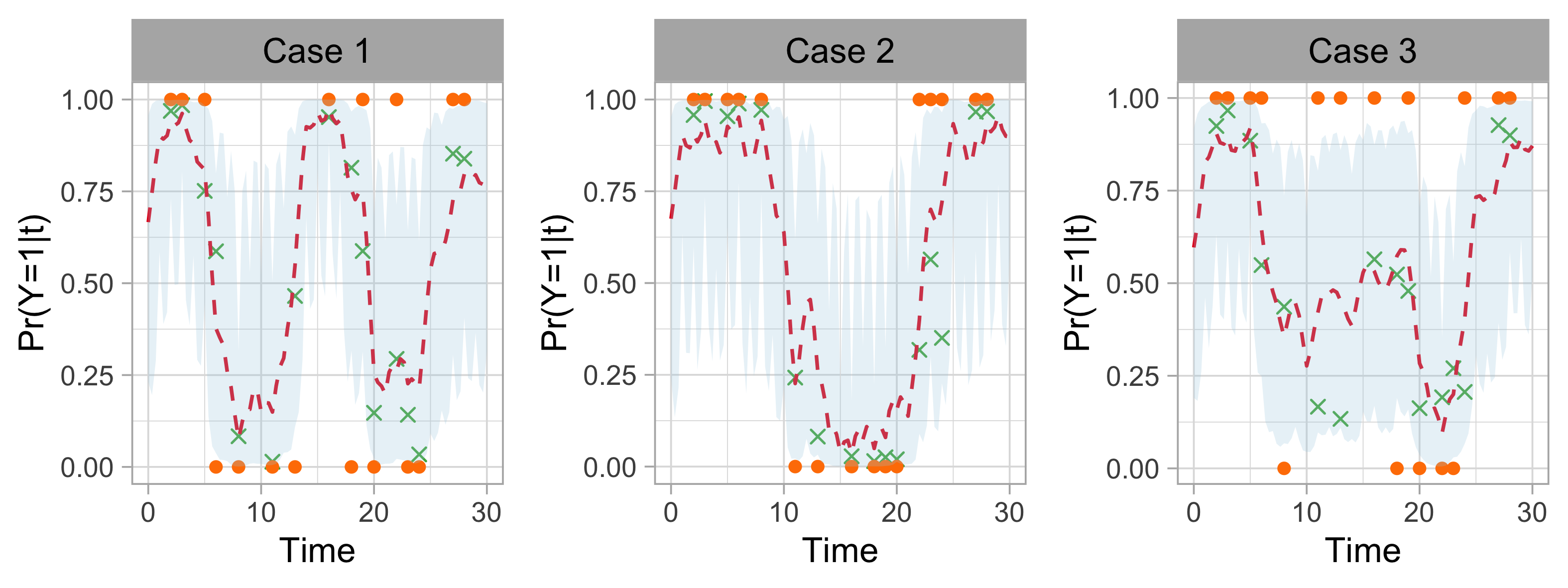}  
  \caption{Sparsity level at $50\%$.}
  \label{subfig:signalprobcurve50}
\end{subfigure}

\caption{Simulation study regarding the mean structure. Inference results for the probability response curve.  In each panel, the dashed line and shaded region correspond to the posterior mean and $95\%$ credible interval estimates, the (orange) dot is the original binary data, whereas the (green) cross denotes the true probability of generating that responses.}
\label{fig:signalprobcurvesub}
\end{figure}

The proposed hierarchical model is applied to the data, with a weakly informative prior placed on the mean structure. We obtain posterior inference of the probability response curve and the signal process on a finer grid $\boldsymbol{\tau}^+=(0,\frac{1}{3},\frac{2}{3},\ldots,30)$. Figure \ref{fig:signalprobcurvesub} plots posterior point and interval estimates of the subject's probability response curve for a randomly selected one in each case. Despite the data generating process and the sparsity level, the model can recover the evolution of the underlying probability used in generating binary responses. 
We observe shrinkage in the interval estimate at the set of grid points where at least one 
subject has observation, that is, $\boldsymbol{\tau}$. The increase in the credible interval width 
at $\check{\boldsymbol{\tau}}$ reflects the lack of information at those time grids.   

\begin{figure}[t!]
\centering
\includegraphics[width=16cm,height=4cm]{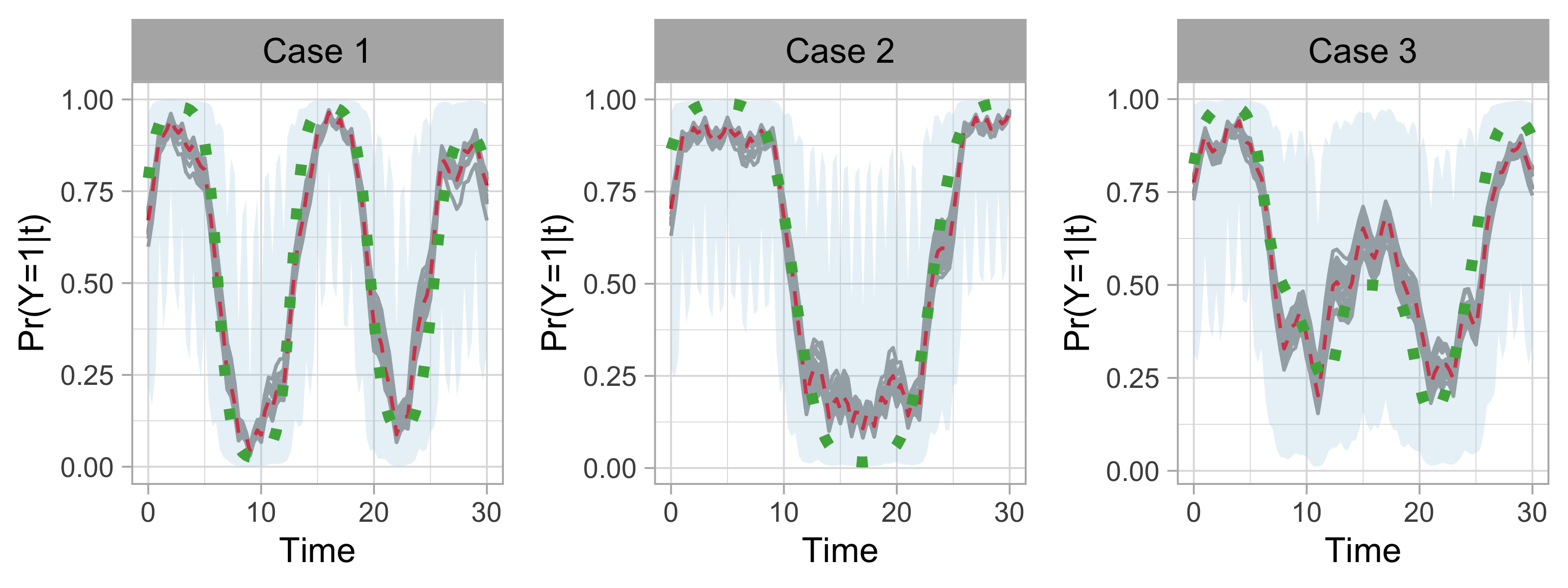}
\caption{Simulation study regarding the mean structure. Prediction of the probability response curve for a new subject. In each panel, the dashed line and shaded region show the posterior mean and $95\%$ interval estimates of probability response curve for a new subject. The solid lines are the posterior mean estimates of probability response curves for the in-sample subjects. The dotted line is the true probability function for generating binary responses.}
\label{fig:signalprobcurvenew}
\end{figure}

We further investigate the model's ability in out-of-sample prediction, by estimating the probability response curve for a new subject from the same cohort. Figure \ref{fig:signalprobcurvenew} shows the posterior point and interval estimates of $\text{Pr}(Y_{*}(\tau_{*t})=1)$, including, as a reference point, the posterior mean estimates of each subject's probability response curve $\text{Pr}(Y_{i}(\tau_{it})=1)$, $i=1,\ldots,n$. The true probability function that triggered the binary response, given as the signal transformed by the link function, is also shown in the figure. It is obtained with the simulated data with $10\%$ sparsity, while there is no major difference for the other two sparsity levels. The behavior of the probability response curve for the new subject is to be expected. It follows the overall trend depicted by the true underlying probability function, while suffers from a comparable level of measurement error with the observed subjects.  

\begin{figure}[t!]
\centering
\includegraphics[width=16cm,height=4cm]{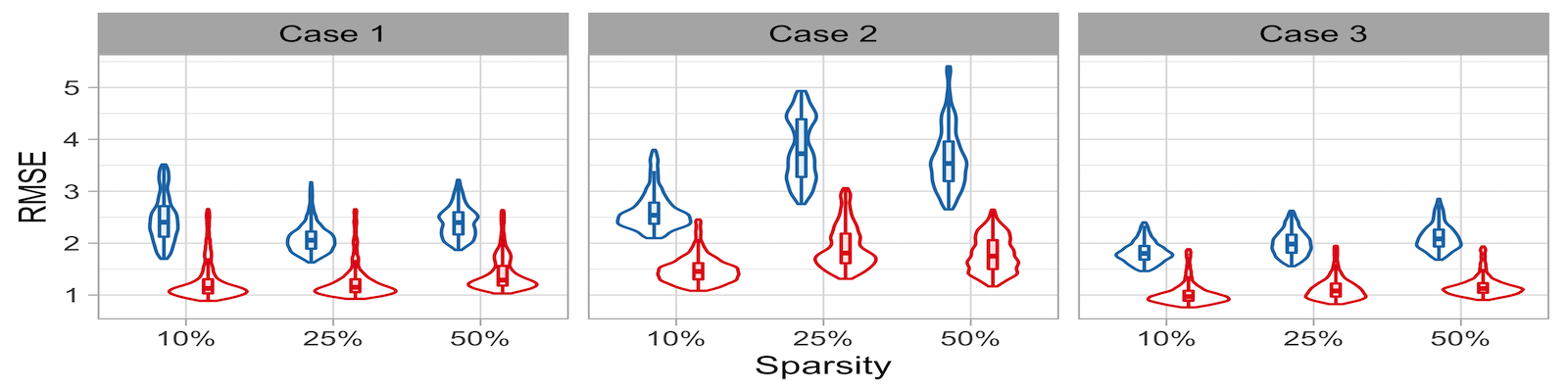}
\caption{Simulation study regarding the mean structure. Box and violin plots of the posterior samples of RMSE for different data generating process and sparsity level combinations. The red box corresponds to the proposed model while the blue box is for the simplified model. }
\label{fig:signalRMSE}
\end{figure}

It is also of interest to assess the model's ability in recovering the underlying continuous signal process, since the signal process describes the intrinsic behavior and is crucial to answer related scientific questions.  
In our proposed model, the signal process is modeled nonparametrically through a GP. To further emphasize the benefits of this model formulation, we compare the proposed model with its simplified backbone. The simpler model differs from the original one in modeling the mean function. 
Instead of modeling the mean function $\mu$ through a GP, we consider the parametric 
form $\mu(\tau)\equiv\mu_0$, with $\mu_0\sim N(a_{\mu},b_{\mu})$. 
The model's ability in capturing the signal process is summarized by the rooted mean square error (RMSE), which is defined by $\text{RMSE}^{\mathcal{M}}=\sqrt{\frac{1}{n}\sum_{i=1}^n\frac{1}{|\boldsymbol{\tau}^+|}\sum_{\tau\in\boldsymbol{\tau}^+}(\hat{Z}^{\mathcal{M}}_i(\tau)-f(\tau))^2}$. 
Here, $\hat{Z}^{\mathcal{M}}_i(\tau)$ denotes the estimated signal for subject $i$ evaluated at 
time $\tau$, under model $\mathcal{M}$, which can be obtained at every MCMC iteration. 
Figure \ref{fig:signalRMSE} explores the posterior distribution of the RMSE under the proposed model and its simplified version, for different data generating process and sparsity level combinations. Despite the scenario, the proposed model shows a notably smaller RMSE. Contrasting the performance with the simpler model highlights the practical utility of including the GP prior layer for the mean function 
in terms of effective estimation of the underlying continuous signal process.     

\subsection{Estimating covariance structure}
\label{subsec:simcov}

Since we emphasize the importance of modeling dependence in longitudinal data, we now explore 
how well our model works for estimating different covariance structures. Consider the data 
generating process in (\ref{eq:datagensim}), with expit link function and signal $f(\tau)=0.1+2\sin(0.5\tau)+\cos(0.5\tau)$. We examine a number of possible choices for generating $\boldsymbol{\omega}_i$, that imply covariance structures which would not be in the same form as the covariance kernel used in the proposed model. The primary interest is to exhibit the robustness of covariance kernel choice to different true covariance structures. We let $T_i=T$ and $\tau_{it}=\tau_t$, namely that all subjects are observed over the same time grids. For $n=100$ subjects, we generate sequences of length $T=11$ at time $\tau=0,\ldots,10$. We study the following options of generating $\boldsymbol{\omega}_i$:
         
\begin{itemize}
\item Case 1: $\boldsymbol{\omega}_i\stackrel{i.i.d.}{\sim} N(\mathbf{0},K_1(\boldsymbol{\tau},\boldsymbol{\tau}))$, with squared exponential kernel $K_1(\tau_{t},\tau_{t^{\prime}})=\exp(-|\tau_t-\tau_{t^{\prime}}|^2/(2\cdot 3^2))$. Each realized trajectory is infinitely differentiable. 
\item Case 2:  $\boldsymbol{\omega}_i\stackrel{i.i.d.}{\sim} N(\mathbf{0},K_2(\boldsymbol{\tau},\boldsymbol{\tau}))$, with exponential kernel $K_2(\tau_{t},\tau_{t^{\prime}})=\exp(-|\tau_t-\tau_{t^{\prime}}|/5)$. Each realization is effectively from a continuous-time AR(1) GP.
\item Case 3:  $\boldsymbol{\omega}_i\stackrel{i.i.d.}{\sim} MVT(5,\mathbf{0},K_3(\boldsymbol{\tau},\boldsymbol{\tau}))$, with compound symmetry kernel $K_3(\tau_{t},\tau_{t^{\prime}})=\mathbf{I}_{\{\tau_t=\tau_{t^{\prime}}\}}+0.4\mathbf{I}_{\{\tau_t\neq\tau_{t^{\prime}}\}}$. The covariance between two observations remains a constant, despite their distance.
\item Case 4: $\boldsymbol{\omega}_i\stackrel{i.i.d.}{\sim} MVT(5,\mathbf{0},K_4(\boldsymbol{\tau},\boldsymbol{\tau}))$, with kernel $K_4(\tau_{t},\tau_{t^{\prime}})=0.7K_2(\tau_{t},\tau_{t^{\prime}})+0.3K_3(\tau_{t},\tau_{t^{\prime}})$, a mixture of AR(1) and compound symmetry covariance structure.
\end{itemize}

\begin{figure}[t!]
\centering
\includegraphics[width=16cm,height=4cm]{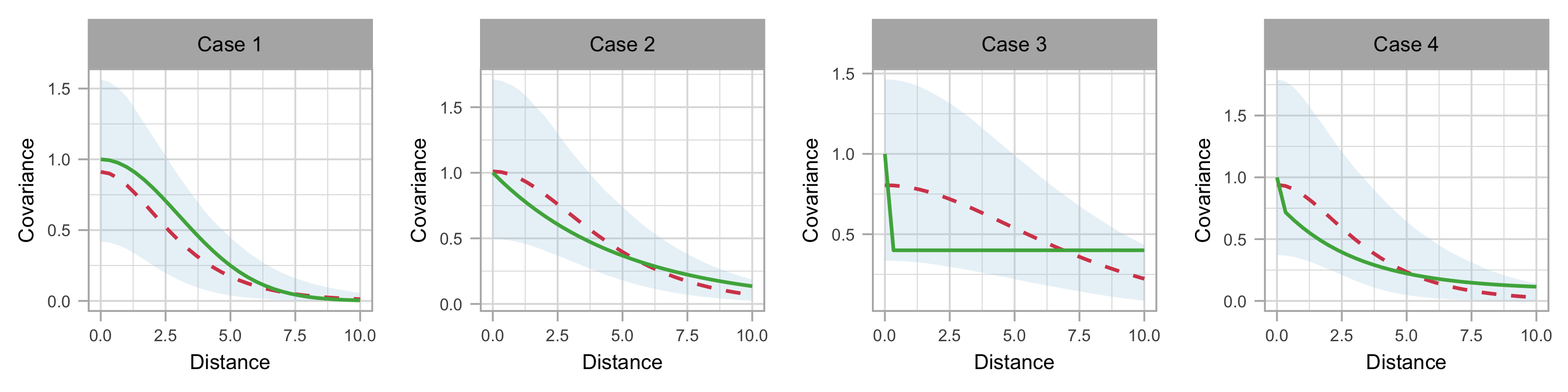}
\caption{Simulation study regarding the covariance structure. Inference results for the 
signal covariance kernels. In each panel, the dashed line and shaded region correspond 
to the posterior mean and 95\% credible interval estimates, whereas the solid line 
denotes the true covariance kernel.}
\label{fig:covariogram}
\end{figure}

In terms of longitudinal binary responses, the covariance structure can be elucidated in two senses, namely the covariance between the pair of binary data $(Y_i(\tau_{t}),Y_i(\tau_{t^{\prime}}))$ and between the pair of signal $(Z_i(\tau_t),Z_i(\tau_{t^{\prime}}))$. We consider the covariance structure of the signal process first. From Proposition \ref{prop:marginalsignal}, $\text{Cov}(Z_i(\tau_t),Z_i(\tau_{t^{\prime}}))=\Psi_{\boldsymbol{\phi}}(\tau_t,\tau_{t^{\prime}})$, $\forall i$, where the covariance function $\Psi_{\boldsymbol{\phi}}$ is defined in (\ref{eq:matern52covfun}). Hence, the signal covariance structure estimated from the model is also isotropic, facilitating a graphic comparison between the posterior estimate of  $\Psi_{\boldsymbol{\phi}}(\tau_d)$ versus the true covariance kernel $K(\tau_d)$, where $\tau_d=|\tau_t-\tau_{t^{\prime}}|$. The results are presented in Figure \ref{fig:covariogram}. 
The proposed model recovers the truth, despite the mis-specification of the covariance kernel. 
Comparing with the other three cases, the posterior point estimate of covariance kernel is less accurate in Case 3. This can be explained by noticing that the constant covariance in that case violates the model assumption. Nonetheless, the posterior interval still covers the truth. 

\begin{figure}[t!]
    \centering
    \begin{subfigure}[b]{0.24\textwidth}
            \includegraphics[width=\textwidth,height=4cm]{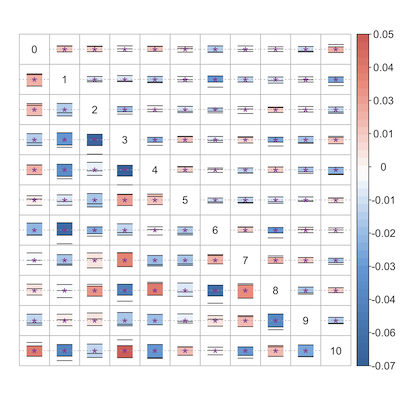}
            \caption{Case 1.}
    \end{subfigure}
    \begin{subfigure}[b]{0.24\textwidth}
            \includegraphics[width=\textwidth,height=4cm]{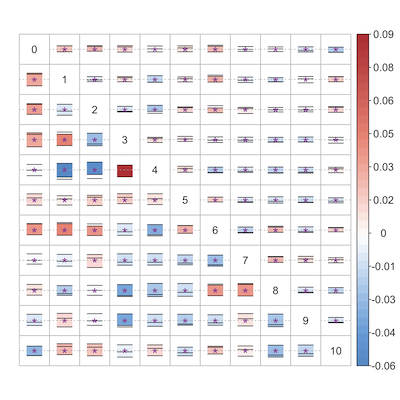}
            \caption{Case 2.}
    \end{subfigure}
    \begin{subfigure}[b]{0.24\textwidth}
            \includegraphics[width=\textwidth,height=4cm]{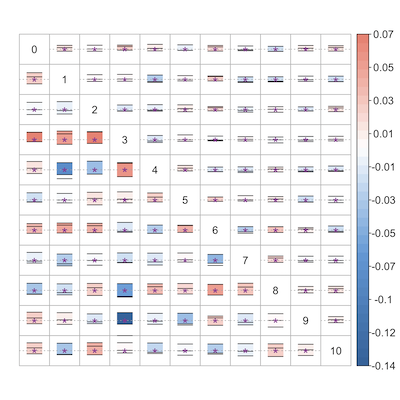}
            \caption{Case 3.}
    \end{subfigure}
    \begin{subfigure}[b]{0.24\textwidth}
            \includegraphics[width=\textwidth,height=4cm]{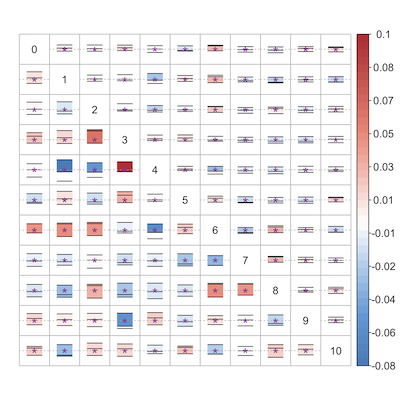}
            \caption{Case 4.}
    \end{subfigure}
    \caption{Simulation study regarding the covariance structure. Posterior interval 
    estimate of correlation coefficients (``box'') versus point estimate obtained from 
    the true data generating process (``$\star$''). In each panel, the upper triangle 
    and the lower triangle are for the Pearson and the  
    tetrachoric correlation coefficient, respectively.}
    \label{fig:bincorrCI}
\end{figure}

As for the covariance between the pair of binary responses, we consider two measurements, 
the Pearson correlation coefficient and the tetrachoric correlation coefficient. For a review 
of the definitions and properties of these two correlation coefficients, we refer to \citet{Joakim2011}. At each MCMC iteration, we predict a new sequence of binary responses of length $T$, denoted as $\{Y^{(s)}_{i^*}(\boldsymbol{\tau}):s=1,\ldots,S\}$. Correspondingly, we also obtain samples of binary sequences from the true data generating process, denoted by $\{\hat{Y}^{(s)}_{i^*}(\boldsymbol{\tau}):s=1,\ldots,S\}$. Both sets of binary sequences form $S/n$ datasets that mimic  the original samples. From the datasets comprised by posterior predictive samples $Y^{(s)}_{i^*}(\boldsymbol{\tau})$, we obtain interval estimates of the two correlation coefficients. In addition, for $\hat{Y}^{(s)}_{i^*}(\boldsymbol{\tau})$ that are generated from the truth, we obtain point estimates, which can be viewed as the correlation coefficients from the data, accounting for the variation in the data generating process. Notice that marginally the binary process is not guaranteed to be isotropic. Hence, the correlation coefficients should be calculated for every possible pair of $(\tau_t,\tau_{t^{\prime}})\in\boldsymbol{\tau}$. The resulting point and interval estimates of both types of correlation coefficients are displayed in Figure \ref{fig:bincorrCI}. All the posterior interval estimates cover the truth, indicating that the proposed model effectively captures the binary covariance structure.  

\begin{figure}[t!]
\centering
\includegraphics[width=16cm,height=6cm]{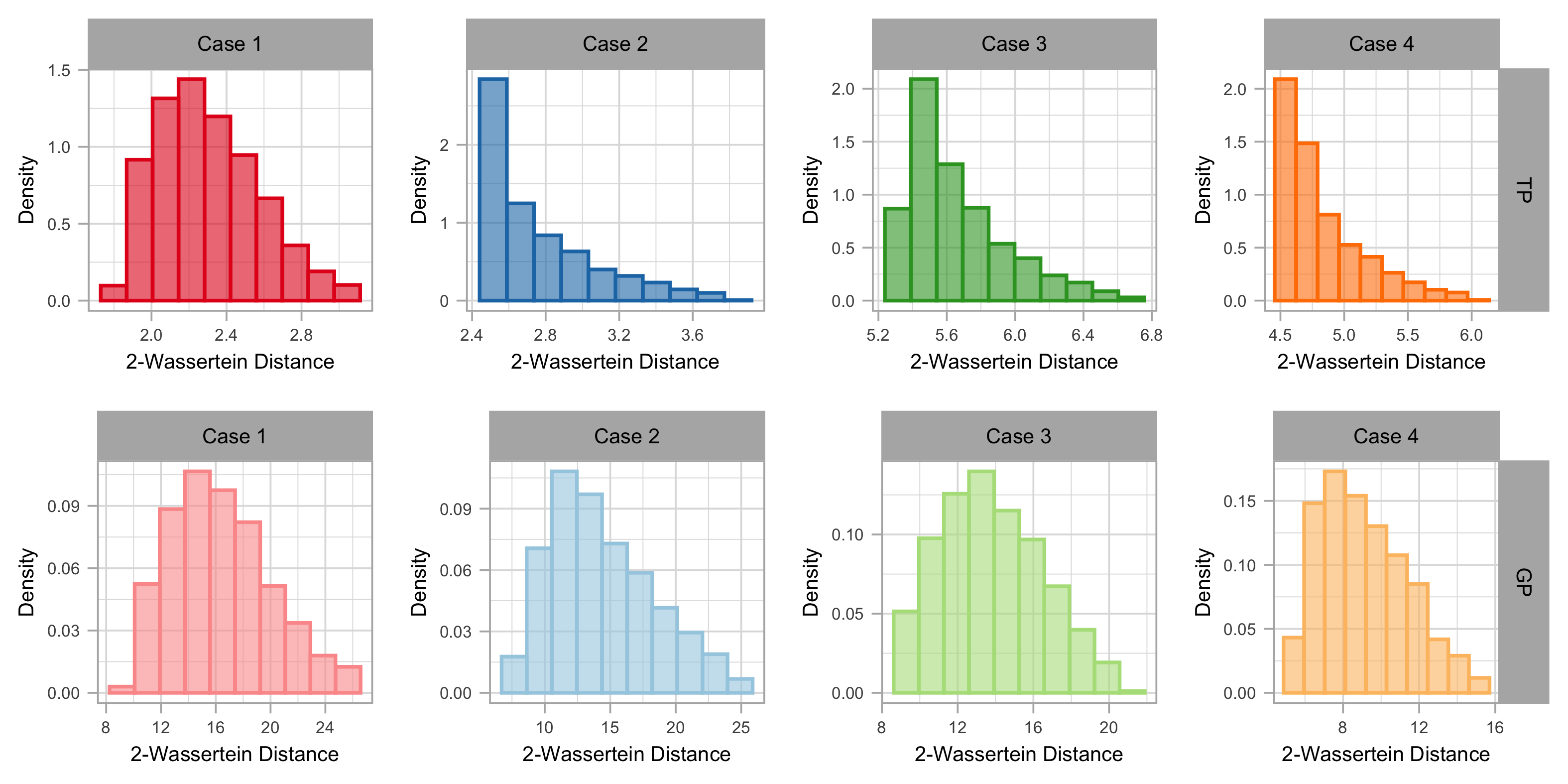}
\caption{Simulation study regarding the covariance structure. Histogram for the posterior samples of the 2-Wasserstein distance between the f.d.d.s. of the centralized signal process obtained from the proposed model (upper panel) and the simplified model (lower panel) to the truth.}
\label{fig:covwassd}
\end{figure}

The simulation studies have illustrated the benefits of our approach, which is avoiding 
possible bias in covariance structure estimation caused by mis-specification of the covariance 
kernel for the signal process. This model feature is driven by the IWP prior placed on the 
covariance function. To emphasize this point, we consider an alternative, simplified modeling 
approach, with $Z_i\mid \mu\stackrel{i.i.d.}{\sim}GP(\mu,\Psi_{\boldsymbol{\phi}})$, $\mu\sim GP(\mu_0,\Psi_{\boldsymbol{\phi}}/\kappa)$. That is, instead of modeling the covariance function nonparametrically, we assume a covariance kernel of certain parametric form, specified by $\Psi_{\boldsymbol{\phi}}$. We consider the centralized signal process $\omega_i=Z_i-\mu$ evaluated at a finite grid $\boldsymbol{\tau}$, denoted as $\boldsymbol{\omega}_i$. Under the proposed model, $\boldsymbol{\omega}_i\stackrel{i.i.d.}{\sim}MVT(\nu,\mathbf{0},\Psi_{\boldsymbol{\phi}}(\boldsymbol{\tau},\boldsymbol{\tau}))$, while under the simplified model, $\boldsymbol{\omega}_i\stackrel{i.i.d.}{\sim} N(\mathbf{0},(1+\frac{1}{\kappa})\Psi_{\boldsymbol{\phi}}(\boldsymbol{\tau},\boldsymbol{\tau}))$. We know the true distribution of $\boldsymbol{\omega}_i$ from the data generating process. Therefore, we can compute the 2-Wasserstein distance between the model estimated distribution of $\boldsymbol{\omega}_i$ to the truth. The usage of 2-Wasserstein distance is motivated by its straightforward interpretation: a 2-Wasserstein distance of $d$ means that coordinate-wise standard deviations differ by at most $d$ \citep[Thm.~3.4]{Huggins2020}. Iterating over the posterior samples of model parameters, we obtain the distributions of 2-Wasserstein distance between the model estimated distribution of $\boldsymbol{\omega}_i$ and the truth, which is shown in Figure \ref{fig:covwassd}. 
Clearly, for the proposed model, the 2-Wasserstein distances are substantially small. Contrasting 
the performance highlights the practical benefits of modeling the covariance structure nonparametrically.

\subsection{Model performance with irregular observing points}
\label{subsec:irregular}

\begin{figure}[t!]
\centering
\includegraphics[width=15.6cm,height=3.9cm]{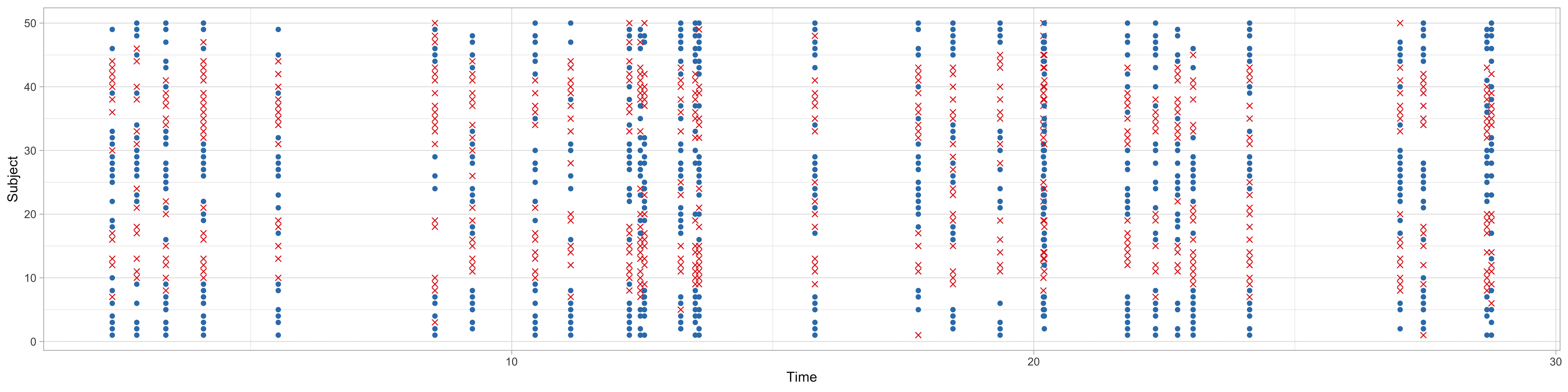}
\caption{Simulation study with irregular observing points. Visualization of the repeated measurements for each subject. The blue dot marks a positive response while the red cross represents a negative response.}
\label{fig:irrdata}
\end{figure}

The simulation studies discussed above focus on longitudinal settings with observations made 
at integer time points, which is the typical scenario in longitudinal studies. To further 
illustrate the practical benefit of adopting the functional data analysis perspective, we 
consider a synthetic scenario in which observations are made irregularly. Specifically, 
the pooled grid $\boldsymbol{\tau}$ consists of 30 grid points that are uniformly sampled on 
the interval $(0,30)$. We consider $n=50$ subjects. For each of them, we first generate repeated 
measurements on the pooled grid, following the scheme described in Section \ref{subsec:simmean} 
Case 1. The unbalanced setting is imposed by randomly dropping out 30$\%$ of the simulated observations. 
The observed data are visualized in Figure \ref{fig:irrdata}, which shows heavily irregular pattern.

\begin{figure}[t!]
\centering
\includegraphics[width=15.6cm,height=3.9cm]{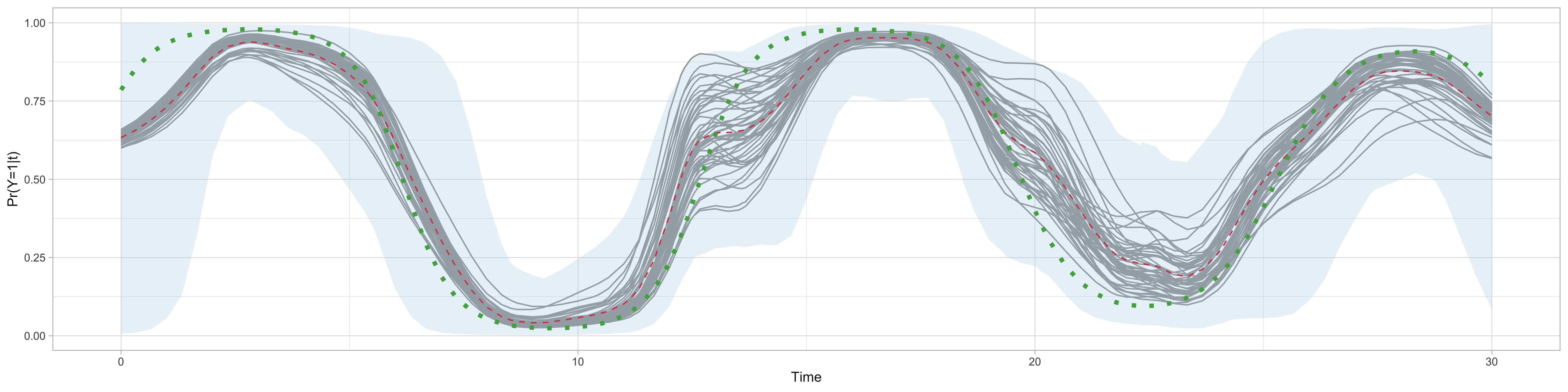}
\caption{Simulation study with irregular observing points. Posterior inference of a new subject's 
probability response curve. The dashed line and shaded region show the posterior mean and 95\% 
interval estimates of probability response curve for a new subject. The dotted line is the true 
probability function for generating binary responses. As a reference, the solid lines are the 
posterior mean estimates of probability response curves for the in-sample subjects.}
\label{fig:irrjointinf}
\end{figure}

To assess the model's performance in out-of-sample prediction, we plot posterior point and 
interval estimates of a new subject's probability response curve in Figure \ref{fig:irrjointinf}, 
including the posterior mean estimate of each in-sample subjects' probability response curve. 
Similar to the scenarios discussed in Section \ref{subsec:simmean}, the predicted mean captures 
the true probability function well. Comparing to the cases with more regular observed time points, 
the shrinkage of the credible interval at observed points is less prominent. Nonetheless, the 
intervals are shorter at the region where observing points are more concentrated, which is to 
be expected.

Moreover, we compare our model with a traditional approach, which postulates a GLMM structure. 
Specifically, the model used for comparison is formulated as follows:
\begin{equation*}
		Y_{it}\mid \mathcal{Z}_{it} \stackrel{ind.}{\sim} Bin(1,\varphi(\mathcal{Z}_{it})), \,\,
		\mathcal{Z}_{it}=\tilde{\boldsymbol{\tau}}_{it}^{\top}\boldsymbol{\beta}+\sum_{k=1}^KS_{itk}b_{k}+\mu_i+\epsilon_{it},\,\, 
		t=1,\ldots,T_i, \,\, i=1,\ldots, n.
\end{equation*}  
The components of this model are set similar to the modeling approach described in 
Section \ref{subsec:comparerealapp} of the main manuscript, except that here the cubic 
B-spline basis functions have 4 inner knots that separate the whole observing period 
into 5 equal length intervals. We perform model comparison using the posterior predictive 
loss criterion and CRPS, with the results summarized in Table \ref{tab:compsim}. Our model 
is favored by both criteria. The key distinction between the two models is that 
we adopt a flexible, functional data analysis modeling approach, which appears to be beneficial,
especially when the observing time points are highly irregular.

\begin{table}[t!] \centering
\small
\caption{Simulation study with irregular observing points. Comparison between the proposed model 
and the generalized linear mixed effects model using two different criteria. The values 
in bold correspond to the model favored by the particular criterion.} 
\label{tab:compsim}
\begin{tabular}{ccccc}
\hline
\hline
\multirow{2}{*}{Model} & \multicolumn{3}{c}{Posterior predictive loss} & \multirow{2}{*}{CRPS} \\
\cline{2-4}  & $G(\mathcal{M})$ & $P(\mathcal{M})$ & $G(\mathcal{M})+P(\mathcal{M})$ & \\
\hline
\hline
Proposed & \textbf{125.78} & \textbf{152.33} & \textbf{278.11} & \textbf{0.12}\\
GLMM & 150.55 & 154.16 & 304.71 & 0.14 \\
\hline
\hline
\end{tabular}
\end{table}

\section{Model implementation details}
\label{sec:algocheck}

\subsection{Prior sensitivity analysis}
\label{subsec:priorsen}

We have proposed a prior specification strategy in Section \ref{subsec:modelapply}, which 
relies on the model properties. The prior specification regarding parameter $\nu$ is less 
structured, despite a suggested range which is found conservative in practice. 
To evaluate the potential influence of the prior choice for $\nu$, 
we conduct sensitivity analysis, using Case 1 of the second set of simulation studies 
as an illustration. We choose this case on purpose, because each individual signal process 
is a realization from a GP, whereas the model assumes that it follows a TP.

\begin{figure}[t!]
\centering
\includegraphics[width=16cm,height=9cm]{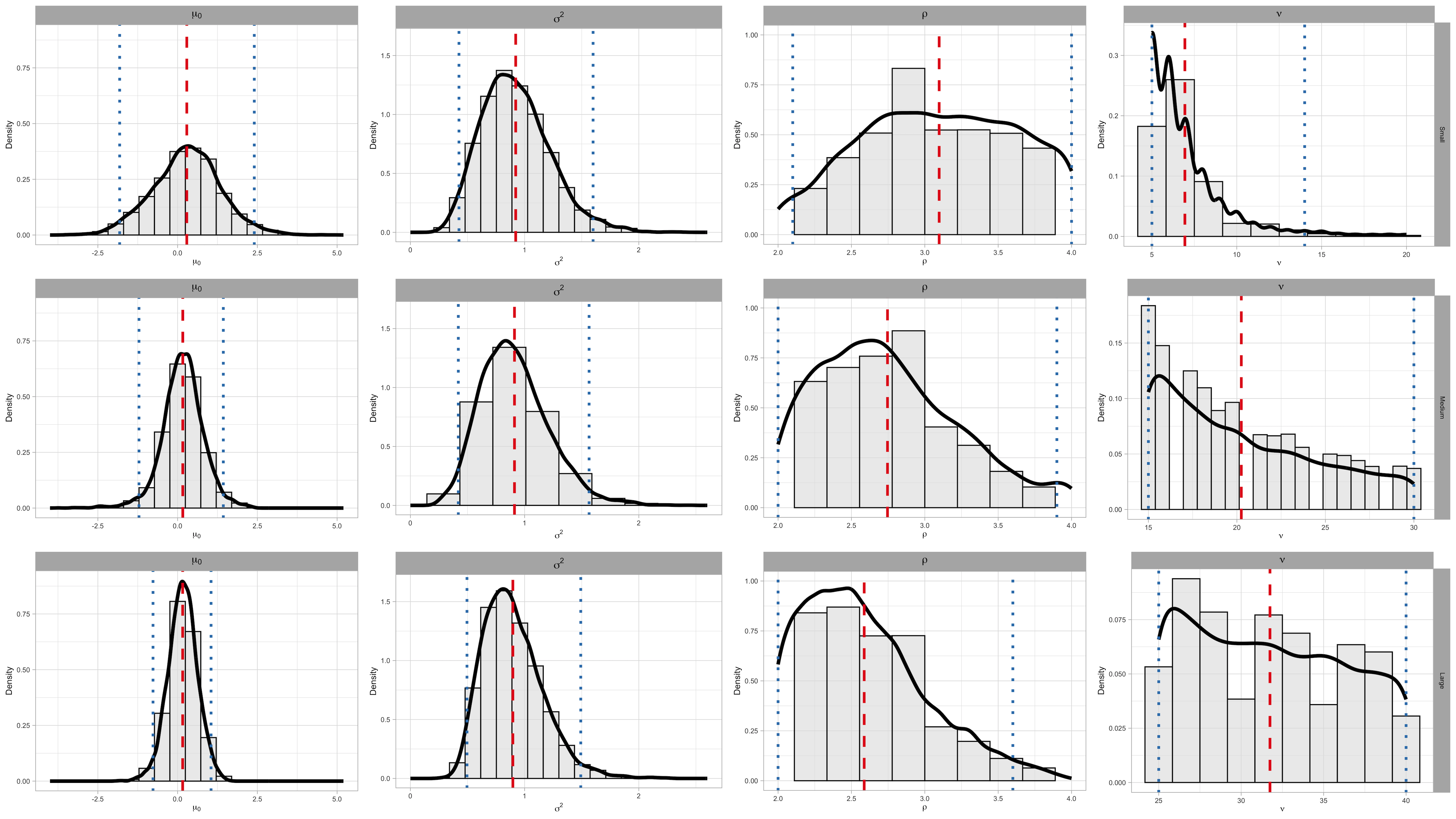}
\caption{Prior sensitivity analysis regarding a simulation scenario. Histogram of the posterior 
samples for model parameters $\mu_0$, $\sigma^2$, $\rho$, and $\nu$. In each panel, the solid line 
depicts the kernel density estimate. The dashed line corresponds to the mean and the dotted lines 
represent the $2.5\%$ and $97.5\%$ percentile, respectively.}
\label{fig:simpriorsen}
\end{figure}

We consider three specifications for the prior of $\nu$, namely, the ``small'' 
case $\nu\sim Unif(5,20)$, the ``medium'' case $\nu\sim Unif(15,30)$, and the ``large'' 
case $\nu\sim Unif(25,40)$. We start with checking the posterior distribution of the 
parameters $\mu_0$, $\sigma^2$, $\rho$, and $\nu$, under the three specifications. 
Figure \ref{fig:simpriorsen} shows the posterior distribution of them, obtained from 
5000 posterior samples after burn in and thinning.
Although the posterior distribution for $\nu$ changes substantially (as expected 
given the priors with different support), there is a smaller effect on the posterior 
distribution of the other TP prior hyperparameters.

\begin{figure}[t!]
\centering
\includegraphics[width=16cm,height=4cm]{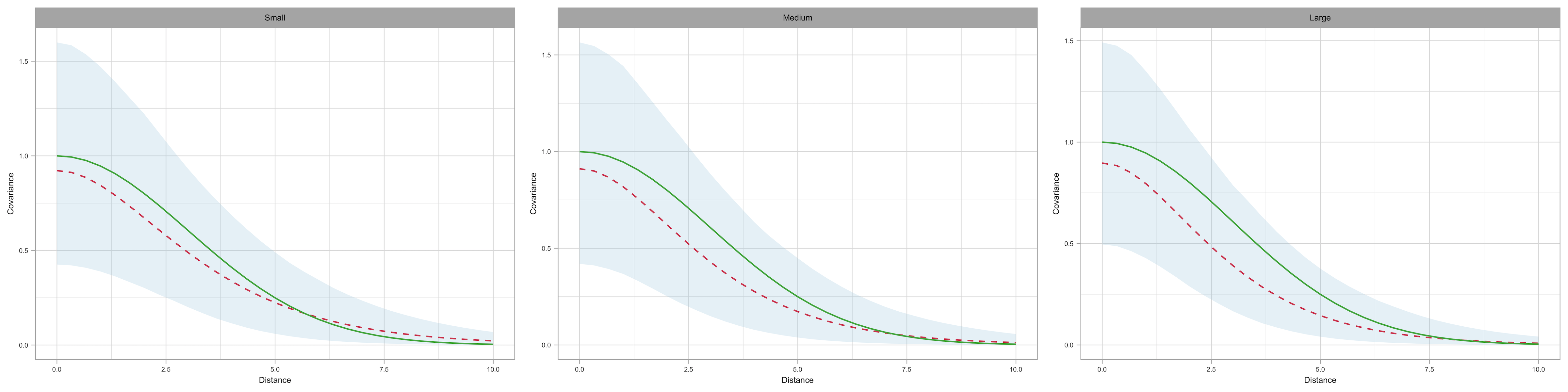}
\caption{Prior sensitivity analysis regarding a simulation scenario. Posterior inference of the
signal covariance kernel. In each panel, the dashed line and shaded region correspond to
the posterior mean and 95\% credible interval estimates, respectively, whereas the solid line denotes the
true covariance kernel.}
\label{fig:simsencov}
\end{figure}

\begin{figure}[t!]
\centering
\includegraphics[width=16cm,height=4cm]{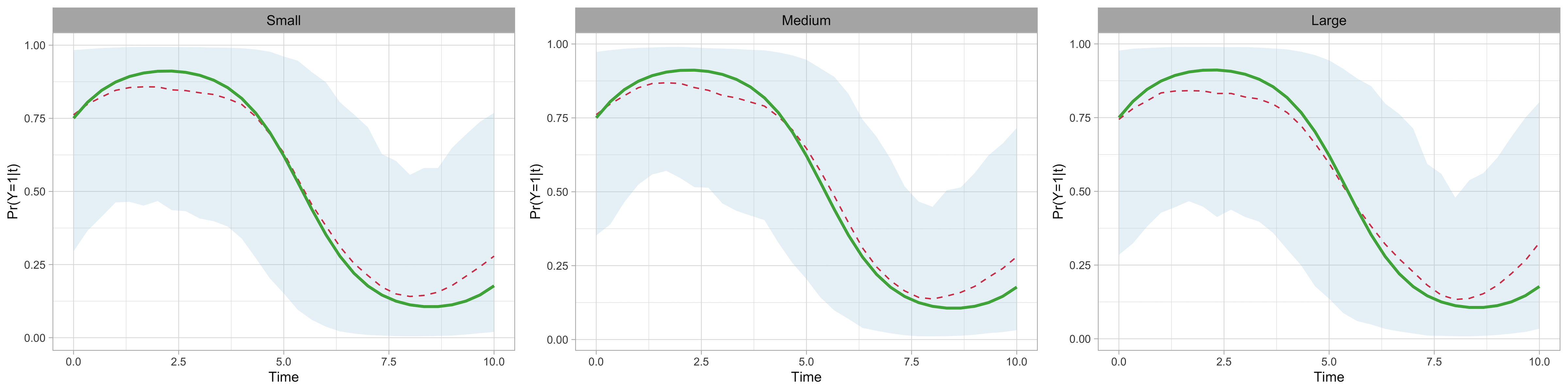}
\caption{Prior sensitivity analysis regarding a simulation scenario. Posterior inference of a
new subject's probability response curve. In each panel, the dashed line and shaded region 
correspond to the posterior mean and 95\% interval estimates of the probability response curve, 
respectively. The solid line is the true probability function used to generate the binary responses.}
\label{fig:simsensignal}
\end{figure}

It is more directly relevant to assess the effect of the prior choice for $\nu$ on posterior
estimation of key functionals pertaining to the main inference objectives. To this end, we 
plot the posterior point and interval estimates of the signal covariance kernel 
(Figure \ref{fig:simsencov}) and of a new subject's probability response curve 
(Figure \ref{fig:simsensignal}). Both sets of inferences are similar indicating that 
the hyperprior choice of $\nu$ has negligible influence on the posterior inference of 
the temporal trend and the covariance structure.

\begin{figure}[t!]
\centering
\includegraphics[width=16cm,height=4cm]{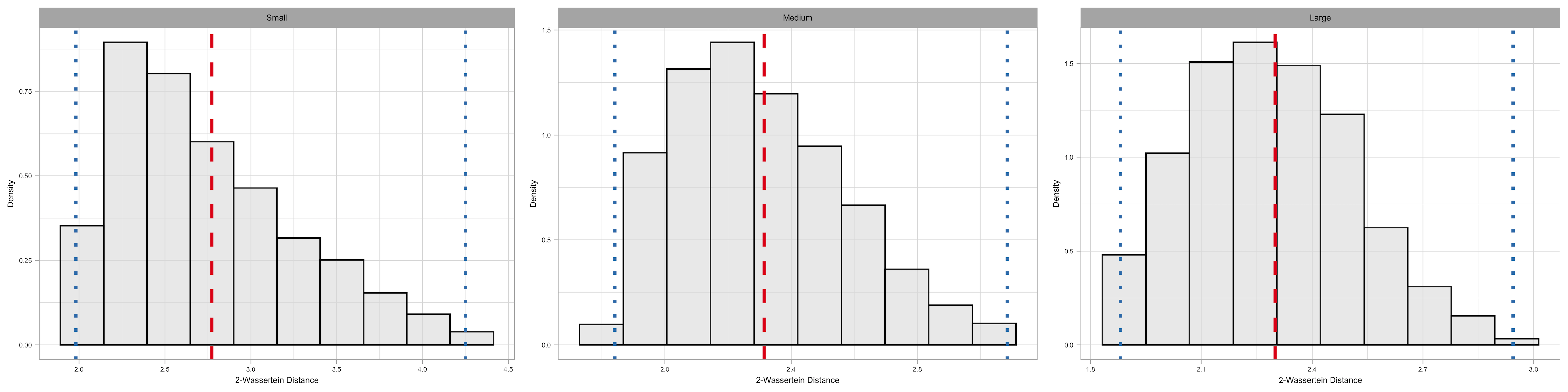}
\caption{Prior sensitivity analysis regarding a simulation scenario. Histogram of the posterior 
samples for the 2-Wasserstein distance between the f.d.d.s. of the centralized signal process estimated
under the proposed model and the truth. In each panel, the dashed line corresponds to the mean and 
the dotted lines represent the $2.5\%$ and $97.5\%$ percentile, respectively.}
\label{fig:simsenwassd}
\end{figure}

The effect of different prior choices for $\nu$ emerges when we consider the full distribution 
of the signal process. Specifically, the 2-Wasserstein distances between the model estimated 
distribution of $\boldsymbol{\omega}_i$ and the truth, calculated under the three priors for $\nu$, 
are shown in Figure \ref{fig:simsenwassd}. The distributions of the 2-Wasserstein distances 
are comparable under the medium and large case, in both cases taking smaller values than the 
distribution under the small case. This pattern is to be expected, because $\nu$ controls the 
tail behavior and a larger value is closer to the true data generating process.

In conclusion, posterior inference regarding the temporal trend and covariance structure 
of the probability response curve is robust with respect to the prior specification of $\nu$. 
However, the model can sacrifice accuracy in estimating the distribution of the probability 
response curve, if the prior of $\nu$ does not span a sufficiently wide range. In practice, 
the suggested default prior for $\nu$ is a safe (conservative) option.

\subsection{MCMC diagnostics}
\label{subsec:mcmcdiag}

Here, we provide some results from assessing convergence of the MCMC algorithm. 
For simplicity in the presentation of results, we focus on the real data example with 
binary arousal responses. For this example, a total of 5000 MCMC samples were taken 
from a chain of length 30000. The first 10000 were discarded as burn in, and the 
remaining draws were thinned to reduce autocorrelation.

\begin{figure}[t!]
\centering
\includegraphics[width=15.5cm,height=3.1cm]{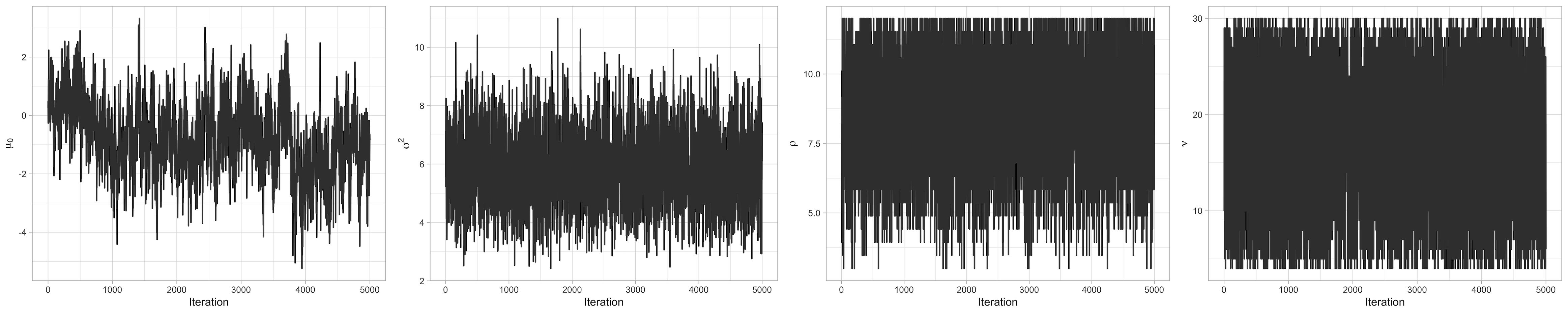}
\caption{MCMC diagnostics for the real data example with binary arousal responses. 
Trace plots of the 5000 remaining posterior samples for the TP prior hyperparameters.}
\label{fig:mcdiaghyper}
\end{figure}

\begin{figure}[t!]
\centering
\includegraphics[width=15.5cm,height=3.1cm]{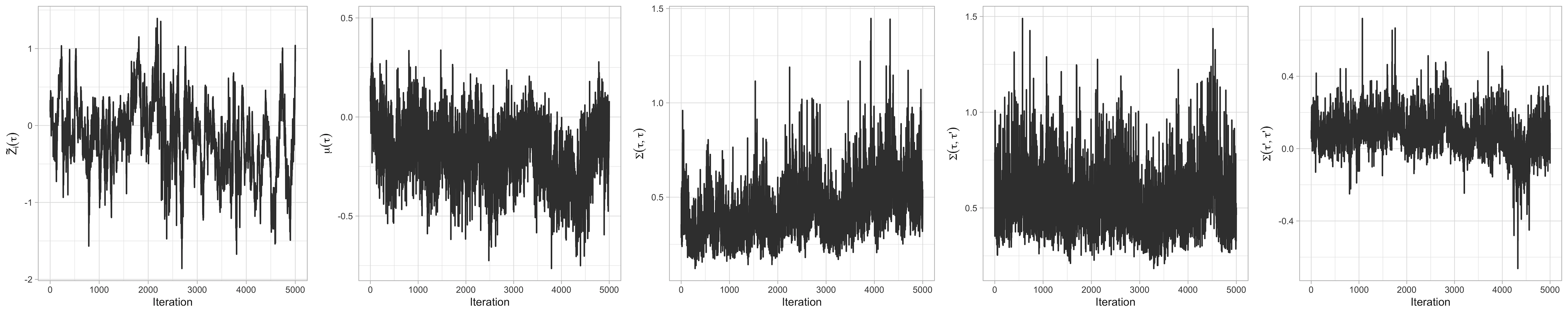}
\caption{MCMC diagnostics for the real data example with binary arousal responses. 
Trace plots of the 5000 remaining posterior samples for randomly selected 
parameters $\boldsymbol{\tilde{Z}}_i(\tau)$, $\mu(\tau)$, $\Sigma(\tau,\tau)$,
$\Sigma(\tau,\tau^{\prime})$ and $\Sigma(\tau^{\prime},\tau^{\prime})$.}
\label{fig:mcdiagpar}
\end{figure}

Figure \ref{fig:mcdiaghyper} shows the trace plots of the remaining MCMC samples 
for the TP prior hyperparameters.
To facilitate presentation for other model parameters, which include multiple components,
we randomly select a subject $i$ and time points $\tau$, $\tau^{\prime}$ from the pooled 
grid $\boldsymbol{\tau}$. We then focus on the posterior samples for 
$\boldsymbol{\tilde{Z}}_i(\tau)$, $\mu(\tau)$, $\Sigma(\tau,\tau)$, 
$\Sigma(\tau,\tau^{\prime})$ and $\Sigma(\tau^{\prime},\tau^{\prime})$. Their trace 
plots, corresponding to the 5000 MCMC samples collected after burn in and thinning, are 
displayed in Figure \ref{fig:mcdiagpar}. Overall, such results suggest convergence of 
the Markov chains.

\section{Additional results for data examples}
\label{sec:dataaddresults}

\subsection{Binary responses from \textit{Studentlife} study}

We follow the prior specification strategy discussed in Section \ref{subsec:modelapply}
of the main paper. We suggest the default hyperprior for $\mu_0$ and $\nu$ as $\mu_0\sim N(0,100)$ 
and $\nu\sim Unif(4,30)$. Parameters $\sigma^2$ and $\rho$ control the covariance structure.  
Their prior hyperparameters can be determined by exploring the covariance structure of the data. 
On the other hand, the hyperprior for $\sigma_{\epsilon}^2$ depends on the belief about the range 
and the degree of freedom of the measurement error. Hence, it is useful to perform prior sensitivity 
analysis with respect to the hyperprior on $\sigma_{\epsilon}^2$, especially for the real data 
analyses.

In general, the measurement error reflects the remaining variability of the underlying 
continuous process, whose major change has been captured by the signal process. Hence, 
it should have small probability of taking large values. For the analysis conducted in 
Section \ref{subsec:resultsrealapp} of the main paper, we consider the measurement error range 
to be small, and pick a moderate value for the error degree of freedom. Specifically, we take 
$R=0.1$ and $\upsilon=10$. Then, using the method described in Section \ref{subsec:modelapply} 
of the main paper, we obtain the $IG(5,0.001)$ distribution as the hyperprior for 
$\sigma_{\epsilon}^2$ (referred to as the original hyperprior).

\begin{figure}[t!]
\centering
\begin{subfigure}{\textwidth}
  \centering 
  \includegraphics[width=16cm,height=6cm]{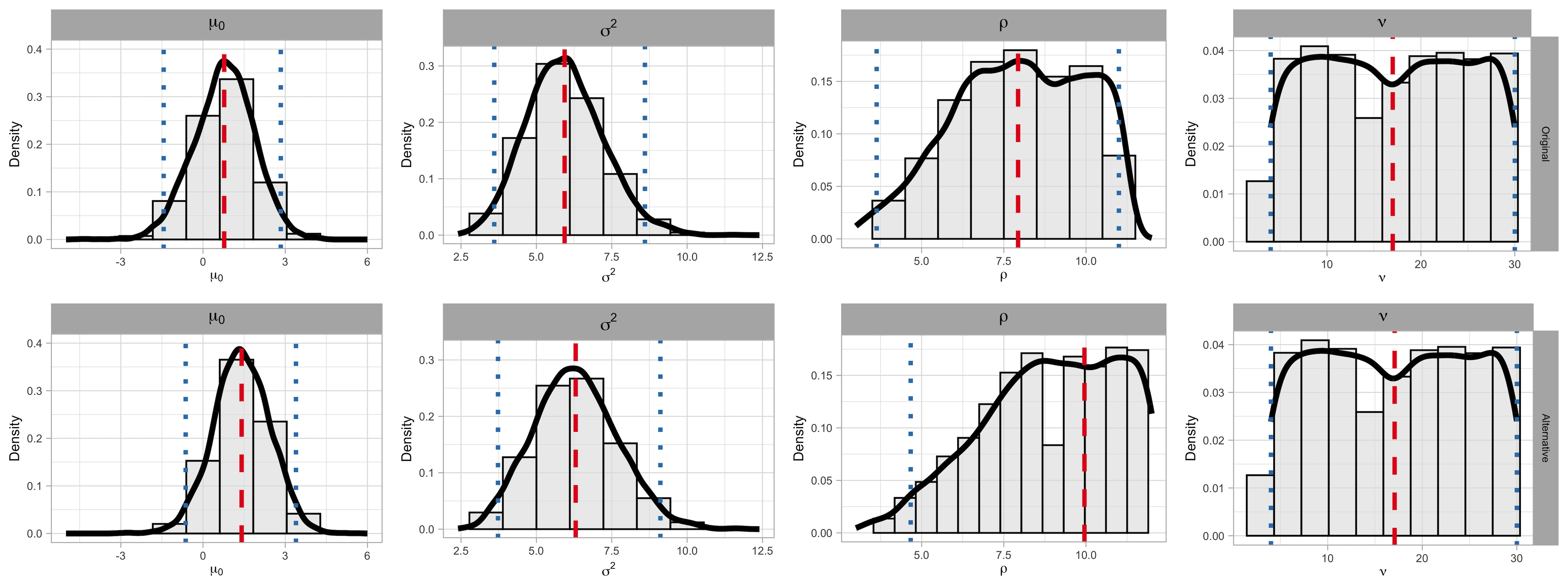}  
  \caption{Prior sensitivity analysis on the valence data.}
  \label{subfig:valpriorsen}
\end{subfigure}
\begin{subfigure}{\textwidth}
  \centering 
  \includegraphics[width=16cm,height=6cm]{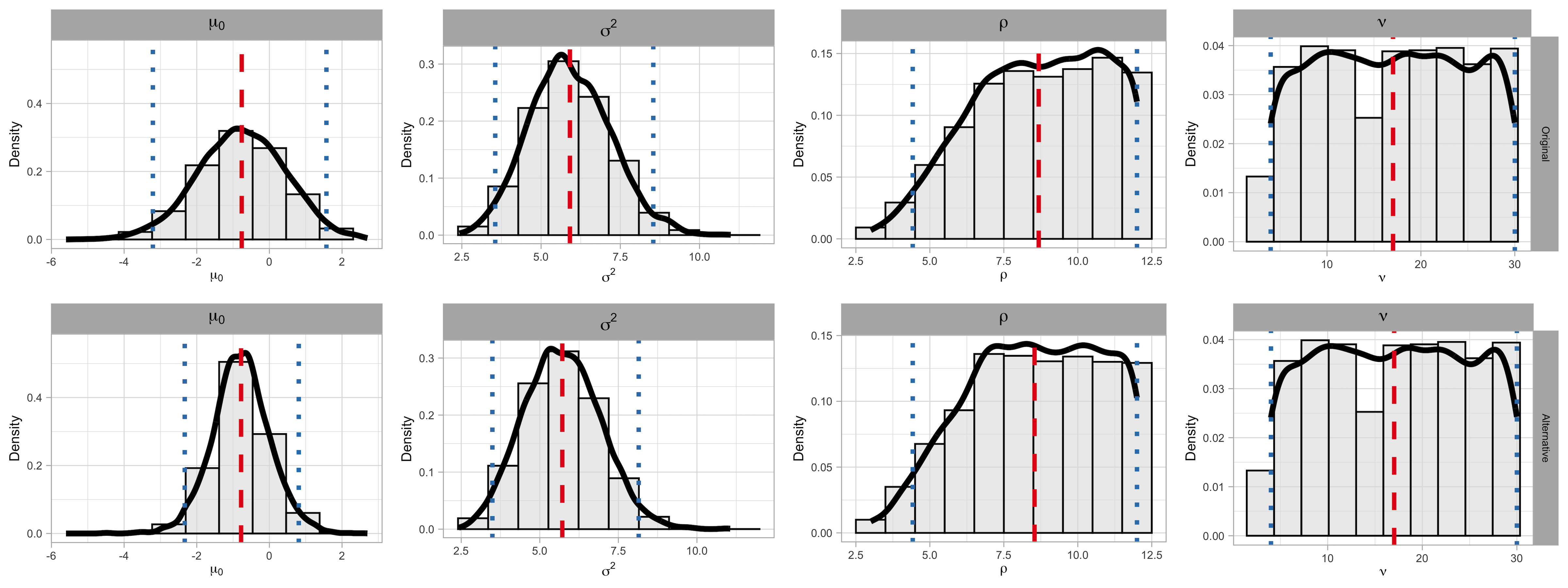}  
  \caption{Prior sensitivity analysis on the arousal data.}
  \label{subfig:aropriorsen}
\end{subfigure}
\caption{\textit{Studentlife} data. Histogram of the posterior samples for the TP prior 
parameters $\mu_0$, $\sigma^2$, $\rho$, and $\nu$. The solid line depicts the kernel density 
estimate. The dashed line corresponds to the mean and the dotted lines represent the $2.5\%$ 
and $97.5\%$ percentile, respectively.}
\label{fig:priorsen}
\end{figure}

To perform prior sensitivity analysis, we assume an alternative hyperprior for
$\sigma_{\epsilon}^2$. In the case of valence score, we assume a larger measurement 
error range $R=0.5$, resulting in the hyperprior $\sigma_{\epsilon}^2\sim IG(5,0.02)$. 
As for the arousal score, we assume the error distribution has a heavier tail, achieved 
by setting $\upsilon=6$. The hyperprior in this case is $\sigma_{\epsilon}^2\sim IG(3,0.0007)$.
We focus on the posterior distribution of the TP prior parameters, $\mu_0$, $\sigma^2$, $\rho$, 
and $\nu$, because they determine the signal process, which is the target of 
primary inferential interest. Results are shown in Figure \ref{fig:priorsen}. The posterior 
distributions of the particular model parameters are similar, suggesting robustness with 
respect to the error variance prior.

Moreover, we elaborate here on the missing-at-random assumption for the particular 
real data analysis.
The main reason to treat the missingness as random is that the 
responses are from an ecological momentary assessment (EMA) study (mentioned explicitly 
in the original publication \citep{StudentLife2014} and the description of the corresponding 
R package). To reduce the potential bias caused by non-random missing responses, at the 
design stage, EMA studies place a premium on obtaining high levels of subject compliance 
with the assessment protocol \citep{Shiffman2008}. As a result, one can assume the occurrence 
of missing values is driven by a 
random process \citep{EMABook2018}, and therefore is ignorable 
\citep[see e.g.][]{Hedeker2009,Shiffman2009}.

\begin{figure}[t!]
    \centering
    \begin{subfigure}[b]{0.48\textwidth}
            \includegraphics[width=7.2cm,height=5.4cm]{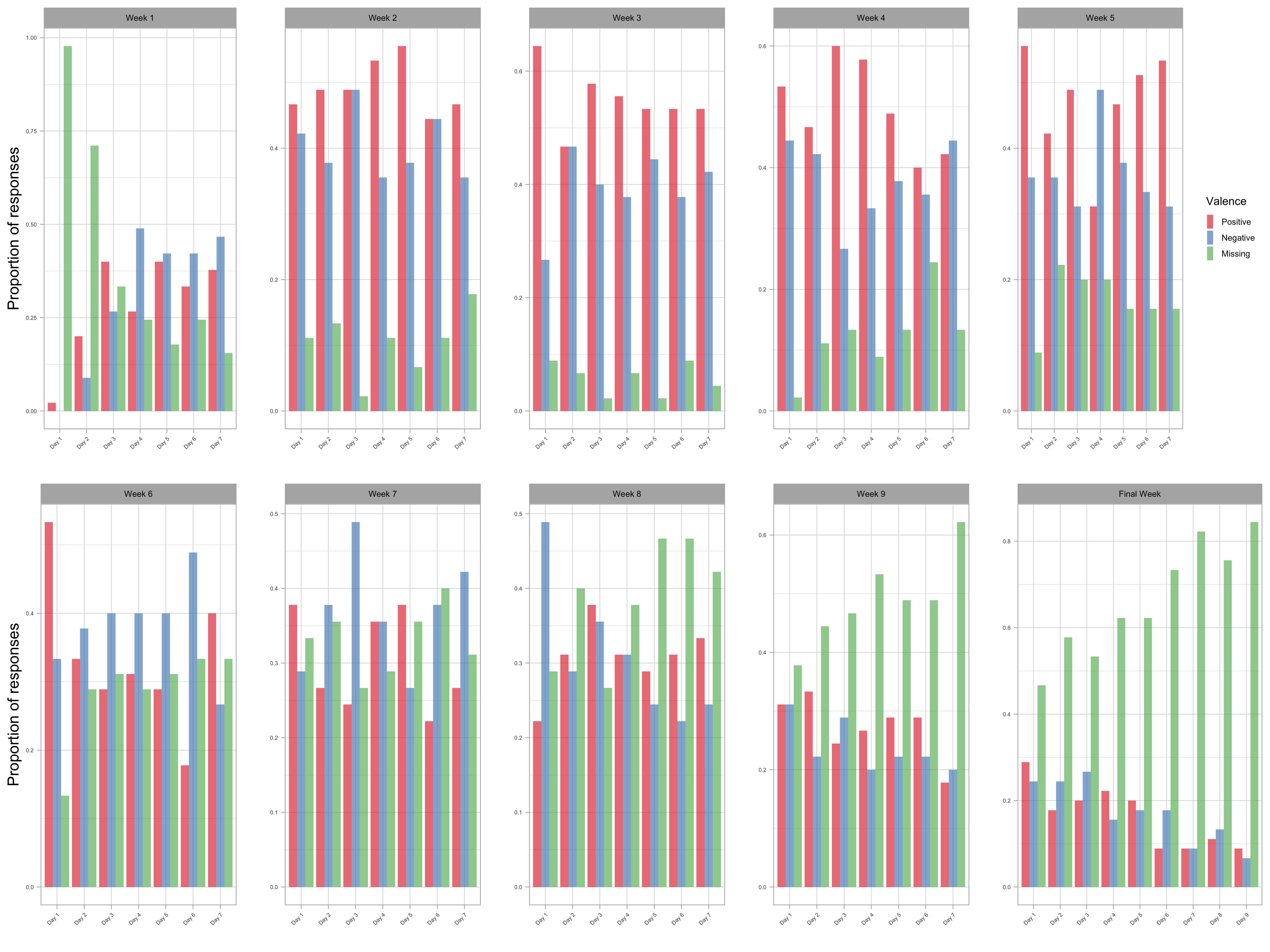}
            \caption{{\small Valence score}}
    \end{subfigure}
    \begin{subfigure}[b]{0.48\textwidth}
            \includegraphics[width=7.2cm,height=5.4cm]{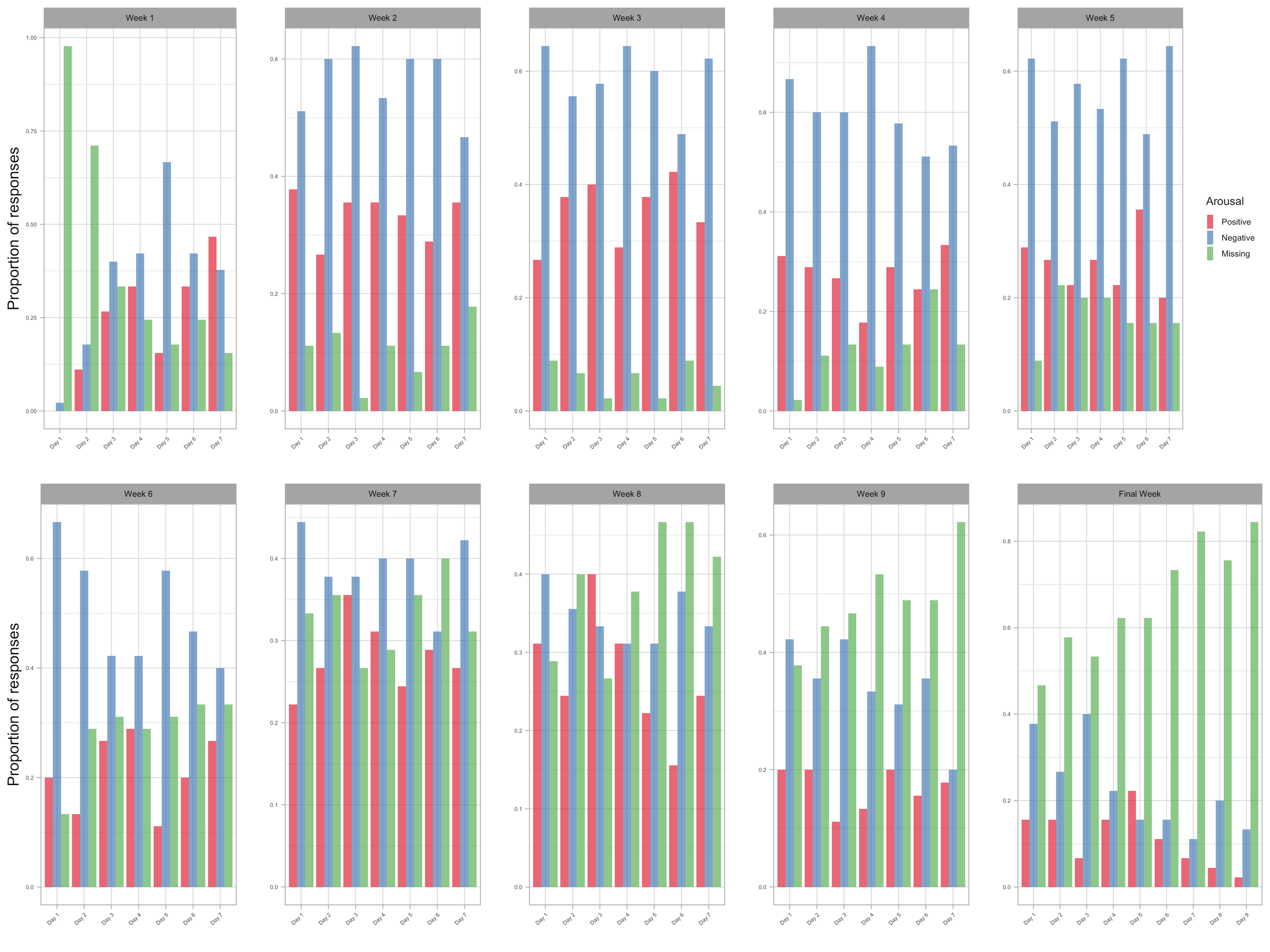}
            \caption{{\small Arousal score}}
    \end{subfigure}
\caption{\textit{Studentlife} data. Proportion of three types of response (positive, 
negative, and missing) over time, for valence and arousal scores.}
    \label{fig:bindatprop}
\end{figure}

For an example of empirical evaluation for the specific data, 
we plot the proportion of the three types of responses (positive, negative, and missing) over 
time, for valence and arousal scores, aggregated over the subjects. The corresponding plot 
is displayed in Figure \ref{fig:bindatprop}. The plots show no 
strong pattern of missingness 
over time, apart from that more missing responses appear at the beginning and toward the end 
of the study. Combining with the feature from the design of EMA studies, 
the missing-at-random assumption is arguably plausible for our illustrative data analyses.

\subsection{Four levels arousal score data}

\begin{figure}[t!]
\centering
\includegraphics[width=16cm,height=10cm]{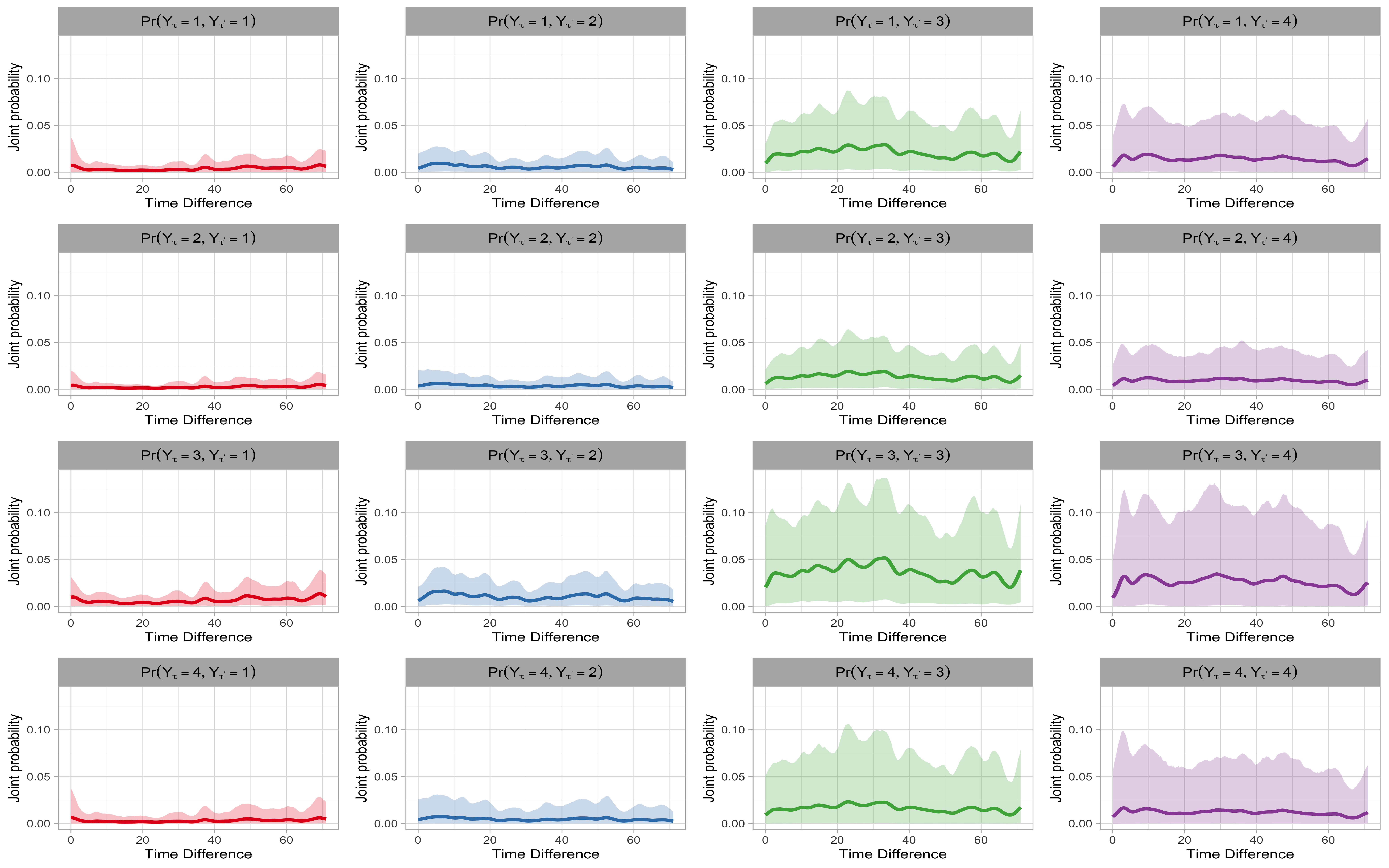}
\caption{Four levels arousal score data. Posterior mean (dashed line) and 95\% interval estimate (shaded region) of the joint probability of the observations on the same subject made at time $\tau$ and $\tau^{\prime}$.}
\label{fig:quadarojointprob}
\end{figure}

Particular to the ordinal responses, we assess the time dependence through the 
joint probability $\text{Pr}(\mathbf{Y}_{\tau}=j,\mathbf{Y}_{\tau^{\prime}}=j^{\prime}\mid\{\mathbf{Z}_{j\boldsymbol{\tau}}\},\{\sigma^2_{\epsilon j}\})$, for which  
inference can be obtained by evaluating Equation (\ref{eq:jointprobmult}) of the main paper
with the posterior samples of model parameters. Figure \ref{fig:quadarojointprob} displays 
posterior point and interval estimates for all possible pairs of the joint probabilities. 

\end{document}